\documentclass[fleqn,usenatbib]{mnras}
\usepackage{newtxtext,newtxmath}
\usepackage{floatrow}
\usepackage{caption}
\usepackage{multirow}
\usepackage{comment}
\usepackage[T1]{fontenc}
\usepackage{orcidlink}
\DeclareRobustCommand{\VAN}[3]{#2}
\let\VANthebibliography\thebibliography
\def\thebibliography{\DeclareRobustCommand{\VAN}[3]{##3}\VANthebibliography}
\usepackage{graphicx}	
\usepackage{amsmath}	
\usepackage{float}
\floatstyle{plaintop}
\restylefloat{table}

\newcommand{\grizy}{\protect\hbox{$grizy$} }
\newcommand{\omatter}{\ensuremath{\Omega_{\mathrm{M}}}}

\newcommand{\tardis}{\textsc{Tardis}}
\newcommand{\caiihk}{\ensuremath{\mathrm{Ca}\,\textsc{ii}\,\mathrm{H\&K}}}
\newcommand{\mgii}{\ensuremath{\mathrm{Mg}\,\textsc{ii}}}
\newcommand{\cii}{\ensuremath{\mathrm{C}\,\textsc{ii}}}
\newcommand{\kms}{\,km\,s$^{-1}$}
\newcommand{\nai}{\ensuremath{\mathrm{Na}\,\textsc{i}}}

\title[UV spectra of 03fg-like SNe]{The Ultraviolet Spectra of 2003fg-like Type Ia Supernovae}

\author[Bhattacharjee et al.]
{Snehasish~Bhattacharjee$^{1}$\thanks{E-mail: snehasish@astro.ncu.edu.tw}\orcidlink{0000-0002-7350-7043}
Yen-Chen~Pan$^{1}$\thanks{E-mail: ycpan@astro.ncu.edu.tw}\orcidlink{0000-0001-8415-6720}
Hao-Yu~Miao$^{2,3}$\orcidlink{0000-0003-2736-5977},
Charles~D.~Kilpatrick$^{4}$\orcidlink{0000-0002-5740-7747},
\newauthor
Willem~B.~Hoogendam$^{5}$\orcidlink{0000-0003-3953-9532},
Katie~Auchettl$^{6,7}$\orcidlink{0000-0002-4449-9152},
Aaron~Do$^{8}$\orcidlink{0000-0003-3429-7845}, and
Yossef~Zenati$^{9,10,11}$\orcidlink{0000-0002-0632-8897}
\\
$^{1}$Graduate Institute of Astronomy, National Central University, 300 Zhongda Road, 32001 Zhongli, Taiwan\\
$^{2}$Institute of Space Sciences (ICE-CSIC), Campus UAB, Carrer de Can Magrans, s/n, E-08193 Barcelona, Spain\\
$^{3}$Institut d'Estudis Espacials de Catalunya (IEEC), 08860 Castelldefels (Barcelona), Spain\\
$^{4}$Center for Interdisciplinary Exploration and Research in Astrophysics (CIERA), Northwestern University, 1800 Sherman Ave, Evanston, IL 60201, USA\\
$^{5}$Institute for Astronomy, University of Hawai'i, Honolulu, HI 96822, USA\\
$^{6}$School of Physics, University of Melbourne, VIC 3010, Australia\\
$^{7}$Department of Astronomy and Astrophysics, University of California, Santa Cruz, CA 95064, USA\\
$^{8}$Institute of Astronomy and Kavli Institute for Cosmology, Madingley Road, Cambridge CB3 0HA, UK\\
$^{9}$Physics and Astronomy Department, Johns Hopkins University, Baltimore, MD 21218, USA\\
$^{10}$Astrophysics Research Center of the Open University (ARCO), The Open University of Israel, Ra’anana 4353701, Israel\\
$^{11}$Department of Natural Sciences, The Open University of Israel, Ra'anana 4353701, Israel\\
}

\pubyear{2025}

\begin{document}
\label{firstpage}
\pagerange{\pageref{firstpage}--\pageref{lastpage}}
\maketitle

\begin{abstract}
2003fg-like Type Ia supernovae (03fg-like SNe~Ia) are rare subtype of SNe~Ia, photometrically characterized by broader optical light curves and bluer ultraviolet (UV) colors compared to normal SNe~Ia. In this work, we study four 03fg-like SNe~Ia using \textit{Swift} UltraViolet and Optical Telescope (UVOT) grism observations to understand their unique UV properties and progenitor scenario(s). We report 03fg-like SNe~Ia to have similar UV features and elemental compositions as normal SNe~Ia, but with higher UV flux relative to optical. Previous studies have suggested that the UV flux levels of normal SNe~Ia could be influenced by their progenitor properties, such as metallicity, with metal-poor progenitors producing higher UV flux levels. While 03fg-like SNe were previously reported to occur in low-mass and metal-poor host environments, our analysis indicates that their UV excess cannot be explained by their host-galaxy parameters. Instead, we demonstrate that the addition of a hot blackbody component, likely arising from the interaction with the circumstellar material (CSM), to the normal SN~Ia spectrum, can reproduce their distinctive UV excess. This supports the hypothesis that 03fg-like SNe~Ia could explode in a CSM-rich environment.
\end{abstract}

\begin{keywords}
supernovae: general - supernovae
\end{keywords}



\section{Introduction}

Type Ia supernovae (SNe~Ia) are among the most important cosmic explosions, as their light curves exhibit a well-known correlation between peak brightness and the post-peak decline rate \citep[e.g.,][]{phillips1993,phillips1999,kattner2012}, a relationship that allows SNe~Ia to be used as an extragalactic distance indicator, leading to the discovery of the accelerating expansion of the Universe \citep{reiss1998,perlmutter1999}. These luminous transients are observed in all types of galaxies and are spectroscopically characterized by broad Si, Ca, and Fe lines, along with the absence of both H and He lines \citep[e.g.,][]{livio2018,ruiter2024}.

Ultraviolet (UV) observations of SNe~Ia offer a valuable approach to exploring the physics of their explosions and progenitor systems. For example, the UV spectra of SNe Ia are dominated by densely packed absorption lines from iron-group elements (IGEs), which significantly influence the opacity at UV wavelengths \citep{baron1996,pinto2000b}. UV photons undergo multiple cycles of absorption and re-emission, often scattering to longer wavelengths before escaping the expanding SN ejecta. This process is highly sensitive to both the explosion dynamics and the composition of the progenitor white dwarfs (WD), providing insight into the composition of the optically transparent outer layers of the SN \citep{sauer2008,foley2016,derKacyet2020,derKacyet2023}. This has also been supported by recent studies, which found mild correlations between SN UV colors and host-galaxy properties, such as stellar mass and metallicity, in the sense that SNe~Ia in more metal-poor galaxies tend to show higher UV flux levels (\citealp{pan2020} though this result has been contested in \citealp{brown2020}). 

UV observations are essential not only for normal SNe~Ia but also for understanding peculiar subtypes like 2003fg-like SNe~Ia \citep[03fg-like~SNe~Ia;][]{howell2006,jha2019,taubenberger2019,ashall2021}. These SNe have slower-declining light curves and are generally more luminous than typical SNe~Ia \citep[e.g.,][]{taubenberger2017, hsiao2020, ashall2021}. The optical spectra of 03fg-like SNe~Ia often show \cii\ lines at 4745 \AA\ and at 7234 \AA\ at early epochs, along with the \cii\ feature at 6580 \AA\ \citep[e.g.,][]{branch2003,prieto2006,hicken2007,parrent2016}, which is only seen in a handful of normal SNe~Ia \citep[e.g.,][]{thomas2011}. In particular, they are distinctive in the UV, often appearing more luminous and bluer, with different color evolution compared to normal SNe~Ia \citep[e.g.,][]{brown2014, hoogendam2024}. They were also referred to as super-Chandrasekhar mass (super-$\rm M_{ch}$) SNe~Ia due to some producing large amounts of $^{56}$Ni, suggesting progenitors with masses exceeding the Chandrasekhar limit \citep[e.g.,][]{howell2006, tanaka2010, scalzo2010, taubenberger2011, hsiao2020}. However, other 03fg-like SNe have inferred progenitor masses below $\rm M_{ch}$, indicating a diverse range of observed phenomena \citep{hicken2007, chakradhari2014, chen2019, lu2021}. These SNe~Ia are also frequently found in low-mass, metal-poor galaxies with high specific star-formation rates \citep{howell2006, childress2011, khan2011, taubenberger2011, chakradhari2014, hsiao2020, ashall2021, lu2021}.

Several theoretical models have been proposed to explain the unique characteristics of 03fg-like SNe~Ia, such as their potentially super-Chandrasekhar mass progenitors and overluminous nature relative to normal SNe~Ia. These include the merger of two white dwarfs (WDs) with a combined mass exceeding $\rm M_{ch}$ \citep{howell2006, scalzo2010,dutta2022,dimitriadis2023,ohora2024}, a merger of a CO WD with the core of an AGB star or a WD explosion within a carbon-enriched circumstellar material \citep[CSM;][]{hoeflich1996,hachinger2012,noebauer2016,nagao2017, nagao2018,hsiao2020,jiang2021,lu2021,ashall2021,siebert2023,siebert2024,kwok2024}, head-on collisions of massive WDs \citep{sharon2024}, or a super-$\rm M_{ch}$ WD explosion supported by rapid rotation or magnetic fields \citep{yoon2005,das2013}. However, no consensus on the most likely scenario has been reached. Alternatively, it has been reported that the UV flux levels of SNe~Ia should correlate with their progenitor metallicities \citep{lentz2000,walker2012}, suggesting that host-galaxy properties such as metallicity could also contribute to their unique UV properties.

In this work, we investigate the UV spectroscopic properties of 03fg-like SNe with Neil Gehrels Swift Observatory ({\it Swift}) UltraViolet and Optical Telescope (UVOT) grism observations \citep{gehrels2004,roming2004}. Our sample consists of four 03fg-like SNe: SN~2009dc, SN~2011aa, SN~2012dn, and SN~2020hvf. We examined their UV grism spectra and compared these with a sample of normal SNe~Ia that also have UV spectra. The article is organized as follows: In Section \ref{sec:sample}, we describe our sample and data. Section \ref{sec:compare} presents the spectroscopic analysis. The discussion and conclusions are presented in Sections~\ref{sec:discussion} and \ref{sec:conclude}, respectively. Throughout this paper, we assume $\mathrm{H_0}=70$\,km\,s$^{-1}$\,Mpc$^{-1}$ and a flat universe with $\omatter=0.3$.

\section{Data}\label{sec:sample}

\subsection{Swift observations of 03fg-like SN sample}
\label{sec:03fg-sample}

Our sample contains all the 03fg-like SNe~Ia observed with Swift UVOT grism: SN~2009dc \citep{puckett2009}, SN~2011aa \citep{puckett2011}, SN~2012dn \citep{bock2012}, and SN~2020hvf \citep{tonry2020a}. These SNe have been classified as 03fg-like SNe by previous studies based on both their photometric and spectroscopic properties \citep[e.g.,][]{taubenberger2011,brown2014,jiang2021,siebert2023}. Among these, the UV spectra of SN~2009dc, SN~2011aa, and SN~2012dn were previously included in \citet{brown2014}. All of the 03fg-like SNe, except for SN~2020hvf, have only single-epoch observations. For SN~2020hvf, two epochs of UV spectra before peak luminosity are available. We adopt the $\Delta m_{15}(B)$ and host-galaxy $E(B-V)$ values reported by earlier studies \citep{taubenberger2011,brown2014,taubenberger2019,jiang2021} for each SN in the analysis. These host-galaxy $E(B-V)$ values are all estimated from the \nai\,D lines in the spectrum \citep{turatto2003,poznanski2012,phillips2013}, as they cannot be adequately derived with conventional light-curve fitters used for normal SNe~Ia \citep[e.g.,][]{ashall2020}. Given that the host-galaxy $E(B-V)$ of these 03fg-like SNe are derived using a method fundamentally different from that used for our normal SN~Ia comparison sample (see Section~\ref{sec:normal-sample}; via light-curve fitting), we have discussed the potential impact of this in Section~\ref{sec:discussion}. The basic properties of these 03fg-like SNe are summarized in Table~\ref{tab:03fg}.

The UV spectroscopic observations were carried out using the {\it Swift} UVOT grism \citep{gehrels2004,roming2004,kuin2015}. Due to its slitless design, contamination from nearby background sources (e.g., their host galaxies) often affects the spectra. \citet{pan2018} developed a data reduction pipeline to mitigate this background contamination, enabling users to construct background spectra and better estimate the flux beneath the target by adjusting aperture size and offsets from the target spectrum. In this work, we further improve the pipeline by making it more efficient and user-friendly. The improved version enables interactive background flux estimation, thereby optimizing background subtraction. The apertures used for extracting both target and background spectra for each 03fg-like SN are shown in Appendix~\ref{cross:cut}.

\begin{figure*}
\centering
\includegraphics[width=\columnwidth]{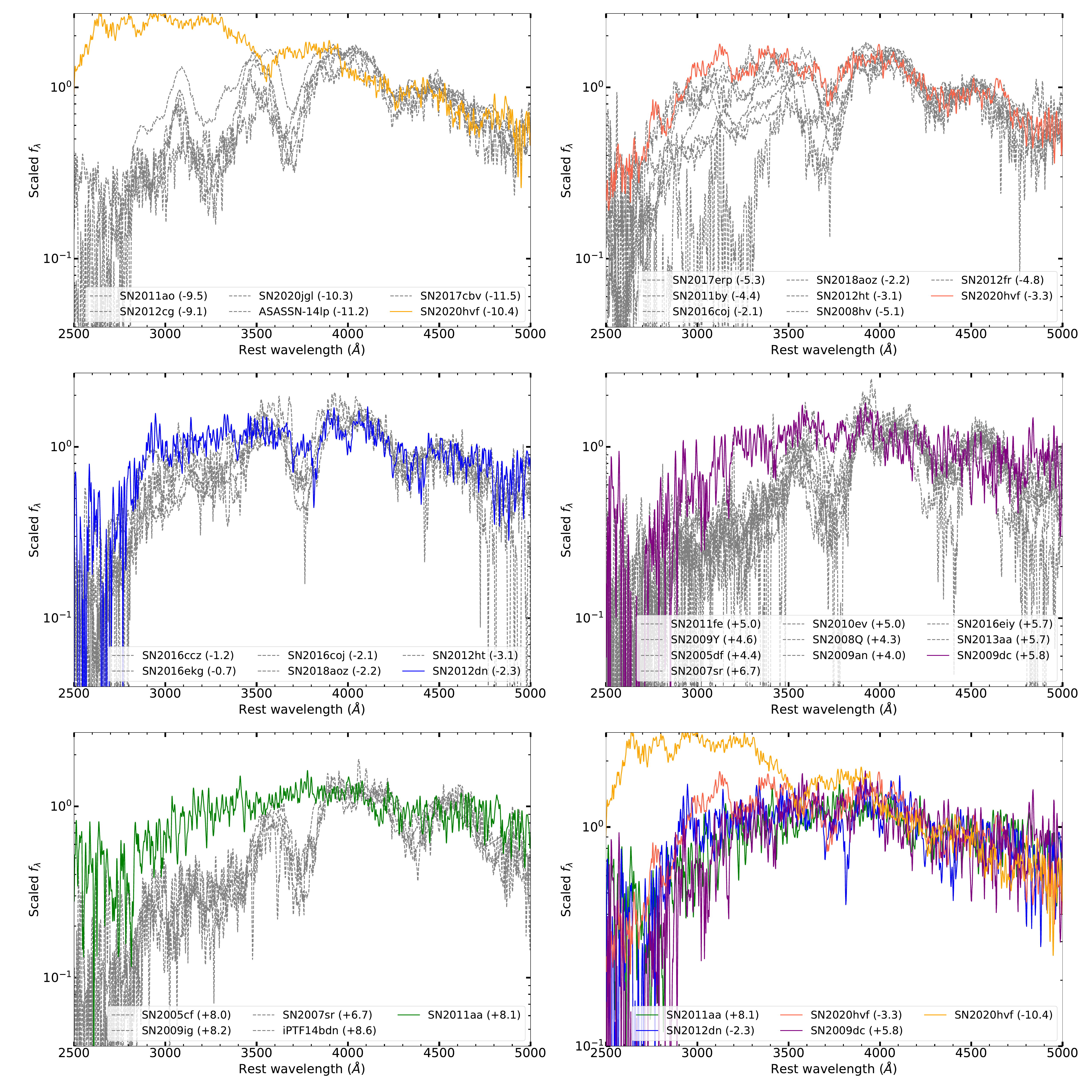}
\caption{Spectral comparisons between 03fg-like SNe and normal SNe~Ia (in grey) of phase differences within 2 days. The panels are organized in ascending order of the phases of 03fg-like UV spectra. The bottom-right panel shows the UV spectra of all the 03fg-like objects in our sample. All the spectra are normalized with the median flux at 4000-4500\,\AA.}\label{spec:compare}
\end{figure*}

After the data reduction, we correct the spectra for foreground Galactic reddening using the calibrations of \citet{sf2011}. We also correct the spectra for host-galaxy reddening using the host-galaxy $E(B-V)$ reported in Table~\ref{tab:03fg}. Here we adopt $R_{V} = 3.1$ and a \citet*[][CCM]{ccm89} reddening law. The final spectra of all the 03fg-like SNe studied in this work are shown in Figure~\ref{spec:compare}

\subsection{Normal SN~Ia sample}
\label{sec:normal-sample}

We use the 26 spectroscopically normal SNe~Ia from \citet{pan2020} as our comparison sample. Additionally, our comparison sample of normal SNe~Ia also include five recently observed SNe~Ia presented for the first time in this work, including SN~2019np \citep{itagaki2019}, SN~2020ue \citep{itagaki2020}, SN~2020ftl \citep{villi2020}, SN~2020jgl \citep{tonry2020b}, and SN~2020uxz \citep{itagaki2020}. We adopt the $\Delta m_{15}(B)$ and host-galaxy $E(B-V)$ from previous studies for some of these SNe \citep{burke2022,tinyanont2021,peterson2023,deckers2023,sai2022}. If not available, we derive them by fitting the optical light curves with the \texttt{SNooPY} package \citep{burns2011}. The optical light curves were obtained from the Lulin One-meter Telescope (LOT) and the one-meter telescope from Las Cumbres Observatory Global Telescope \citep[LCOGT;][]{brown2013}. The light curves and \texttt{SNooPY} fitting can be found in appendix~\ref{snpy:app}. The \texttt{EBV\_model} method is adopted in the fitting. The details of these newly added SNe can be found in Table~\ref{tab:sne}.

\begin{figure}
\centering
\includegraphics[scale=0.3]{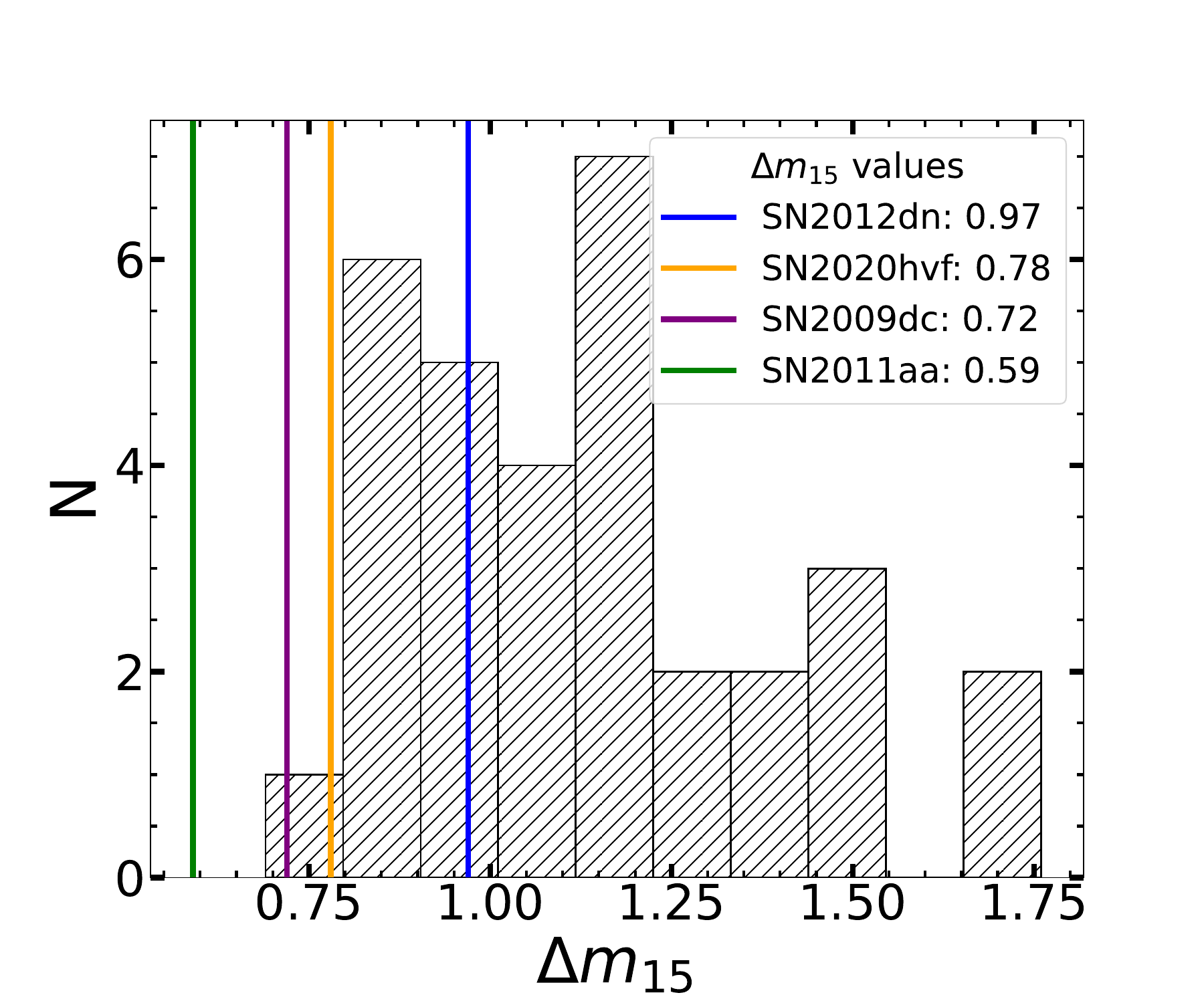}
\caption{Distribution of $\Delta m_{15}(B)$ of the normal SNe~Ia (black histogram) with the colored vertical lines representing the $\Delta m_{15}(B)$ of the 03fg-like SNe~Ia in our sample.}\label{hist:dm15}
\end{figure}

\subsection{Host-galaxy properties}
This work also explores the relationship between UV spectra and host-galaxy properties. The host-galaxy properties of our newer sample are derived by fitting the multicolor photometry of the host galaxy with the photometric redshift code \texttt{z-peg} \footnote{\url{http://www2.iap.fr/pegase/}} \citep{borgne2002}. We adopt the same configuration as that used in \citet{pan2020} to ensure consistency. The \textsc{PYTHON} package \texttt{HostPhot}\footnote{\url{https://github.com/temuller/hostphot}.} \citep{hostphot2022} is utilized for measuring the host-galaxy photometry from the PS1 \grizy\ imaging data \citep{chambers2016}. An aperture is applied to the imaging data of each filter to measure the host-galaxy photometry. The size and shape of the aperture are optimized by \texttt{HostPhot} with Kron flux parameters in the \textsc{PYTHON} library for Source Extraction and Photometry \citep[\texttt{SEP};][]{Bertin_Arnouts_SEXtrator,Barbary2016_SEP}, and are manually modified if needed. The photometry is then corrected for foreground Milky Way reddening with $R_{V} = 3.1$ and a CCM reddening law. \texttt{z-peg} then fits the observed galaxy photometry with various galaxy spectral energy distribution (SED) templates and gives the best-fit host-galaxy parameters. A \citet{salpeter1955} initial-mass function (IMF) is assumed in the fitting. More details about \texttt{z-peg} fitting can be found in \citet{pan2020}. 

\begin{figure}
\centering
\includegraphics[scale=0.28]{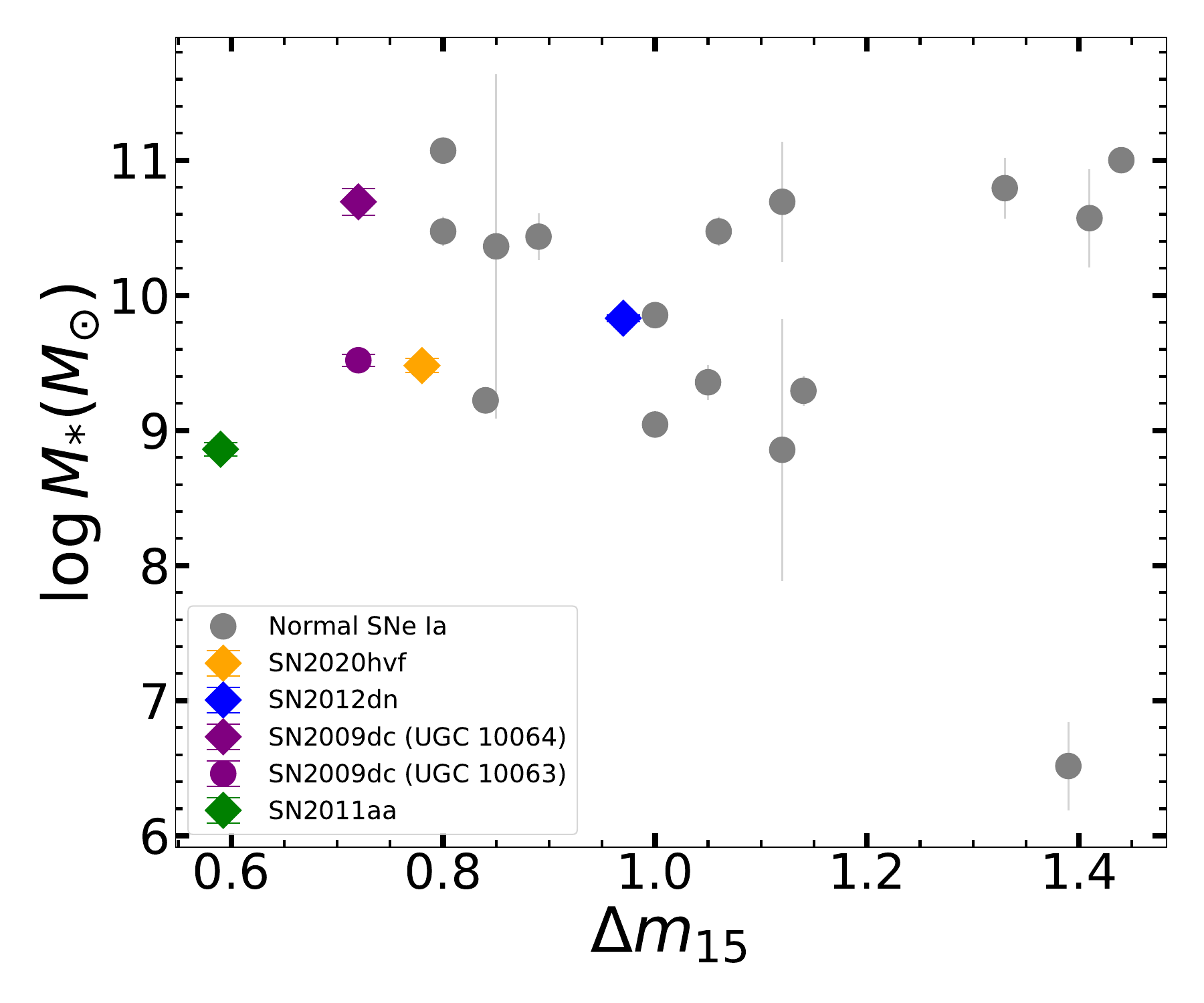}
\caption{Host-galaxy stellar mass as a function of $\Delta m_{15}(B)$ for our sample. Normal SNe~Ia are shown in grey, while 03fg-like SNe are highlighted in color. For SN~2009dc, which is likely located in an interacting system (consisting of UGC~10064 and UGC~10063), both host masses are shown.}\label{mass:dm15}
\end{figure}

In Figures \ref{hist:dm15} and \ref{mass:dm15}, we compare the $\Delta m_{15}(B)$ and host-galaxy stellar mass of 03fg-like SNe~Ia with that of normal SNe~Ia. 03fg-like SNe~Ia tend to cluster towards the lower end of $\Delta m_{15}(B)$ distribution (i.e., having slower decline rates). While previous studies \citep[e.g.,][]{brown2014,ashall2021} have noticed that 03fg-like SNe tend to reside in lower-mass galaxies compared to normal SNe~Ia, \citet{taubenberger2011} pointed out that SN~2009dc is likely located in an interacting system consisting of two galaxies (UGC~10064 and UGC~1006). In this case, the SN~2009dc may be associated with either a massive ($\log (M_{*}/M_{\odot})>10$) or low-mass galaxies. To account for this ambiguity, we include the masses of both host candidates in Table~\ref{tab:03fg} and Figure~\ref{mass:dm15}.

\begin{table*}
    \centering
    \caption{Properties of the 03fg-like SNe~Ia sample in this work.}
    \label{tab:03fg}
    \small
    \begin{tabular}{cccccccc}
        \hline
        \hline
        SN Name & Redshift & Phase & $\Delta m_{15}(B)^{a}$  & Host galaxy & $E(B-V)_{host}^{b}$ & $E(B-V)_{MW}^{c}$ & Host $M_{stellar}^{d}$ \\
         & ($z$) & (Days) & (mag) &  & (mag) & (mag) & ($\log (M_{*}/M_{\odot})$) \\
        \hline
        SN~2009dc & $0.0214$ & $+5.8$ & $0.71 \pm 0.03$ (1)  & UGC 10064 & $0.023 \pm 0.022$ (1) & $0.150 \pm 0.020$  & $10.693^{+0.009}_{-0.187}$ \\
          &  &  &  & UGC 10063 &  &   & $9.520^{+0.010}_{-0.080}$ (5)\\
        \hline
        SN~2011aa & $0.0124$ & $+8.1$ & $0.59 \pm 0.07$ (2) & UGC 03906 & $0$ (2) & $0.023 \pm 0.001$ & $8.861^{+0.054}_{-0.044}$ \\
        \hline
        SN~2012dn & $0.0102$  & $-2.3$ & $0.97 \pm 0.00$ (3) & ESO 462-016 & $0.044 \pm 0.013$ (3) & $0.064 \pm 0.006$  & $9.831^{+0.036}_{-0.010}$ \\
        \hline
        SN~2020hvf & $0.0057$ & $-10.4$, $-3.3$ & $0.78 \pm 0.00$ (4) & NGC 3643 & $0$ (4) & $0.035 \pm 0.001$ &$9.480^{+0.036}_{-0.068}$  \\
        \hline
    \end{tabular}
    \vspace{6pt}
    \begin{flushleft}
         Ref: (1)  \citet{ashall2021}, (2) \citet{brown2014}, (3) \citet{taubenberger2019}, (4) \citet{jiang2021}, (5) \citet{taubenberger2011}.\\
         \textsuperscript{a} The decline of $B$-band magnitude 15~days after peak brightness.\\
         \textsuperscript{b} Host-galaxy color excess.\\
        \textsuperscript{c} Milky-Way color excess from \citet{sf2011}.\\
        \textsuperscript{d} Host-galaxy stellar mass.\\
         
    \end{flushleft}
\end{table*}

\begin{table*}
    \centering
    \caption{Properties of the five normal SNe~Ia in this work not sourced from \citet{pan2020}.}
    \label{tab:sne}
    \small
    \begin{tabular}{ccccccc}
        \hline
        \hline
        SN Name & Redshift & Phase & $\Delta m_{15}(B)$   & Host galaxy & $E(B-V)_{host}$ & $E(B-V)_{MW}$ \\
         & ($z$) & (Days) & (mag) &  & (mag) & (mag)   \\
        \hline
        SN~2019np & $0.0045$ & $-6.9$ & $0.98 \pm 0.01$ (1) & NGC 3254 & $0.110 \pm 0.066$ (6) & $0.017 \pm  0.001$ \\
        \hline
       SN~2020ue & $0.0033$  & $-6.5$ & $1.09 \pm  0.01$ (2) & NGC 4636 & $0.000 \pm 0.009$ (3) & $0.024 \pm  0.001$  \\
        \hline
        SN~2020ftl & $0.0073$ & $-7.8$, $-6.0$ & $1.14 \pm 0.03$ (3) & NGC 4277 & $0.000 \pm 0.011$ (3) & $0.016 \pm 0.001$  \\
        \hline
        SN~2020jgl & $0.0067$ & $-12.5$, $-10.3$, $-3.7$ & $1.20 \pm 0.07$ (4)  & PGC 26905 & $0.040 \pm 0.016$ (3)  & $0.059 \pm 0.002$  \\
        \hline
        SN~2020uxz & $0.0082$ & $-12.6$ & $0.99 \pm 0.01$ (5) & NGC 0514 & $0.000 \pm 0.005$ (3) & $0.033 \pm 0.001$ \\
        \hline
        \hline
    \end{tabular}
    \vspace{6pt}
    \begin{flushleft}
         Ref: (1) \cite{burke2022} (2) \citet{tinyanont2021}  (3) This work (4) \citet{peterson2023} (5)  \citet{deckers2023} (6) \citet{sai2022}
    \end{flushleft}
\end{table*}

\section{Results}\label{sec:compare}

\subsection{UV Spectra of 03fg-like SNe}
\label{sec:uvspec}
Figure~\ref{spec:compare} shows the spectral comparison of 03fg-like objects with normal SNe~Ia. We only compare the spectra of phase differences within 2~days. This phase criterion ensures that the comparison sample is at similar phases while maintaining a reasonable sample size. A stricter phase cut does not alter our conclusions, but it reduces the robustness of our statistical analysis. The phases of our 03fg-like SNe range from $-10$ to $+8$\,days. All the spectra are normalized to the median flux in the range of 4000-4500\,\AA. It is apparent (with eye inspection) that the 03fg-like SNe tend to be bluer in the UV than the majority of normal SNe~Ia in our comparison sample, regardless of the phases investigated in this work. We see a potential trend between the UV flux level and phase of the spectrum (see the bottom-right panel of Figure~\ref{spec:compare}), with earlier epochs having higher UV flux levels than those of later epochs. More quantitative analyses can be found in Section~\ref{section:flux-ratio}. 

The UV spectra of 03fg-like SNe exhibit fewer strong features compared to normal SNe~Ia, but still show some similarities, such as \caiihk\ between 3500-4000\,\AA\ and \mgii\ between 2500-3000\,\AA. The flux between 3000-3500\,\AA\ is likely contributed by the IGEs, including Ni, Co, and Fe. These lines are strongly blended with each other. Detailed spectral modeling for the 03fg-like SNe, including spectral element decomposition for the UV region, is provided in Section~\ref{sec:tardis}. Notably, SN~2020hvf has a much higher \caiihk\ velocity than that of normal SNe~Ia at the same phase. This is consistent with the high \ion{Si}{II} velocity observed in \citet{jiang2021}. Using the method described in \citet{pan2024}, we measure a \caiihk\ velocity of $30036\pm207$\,\kms\ at $-10.4$\,days. This velocity decreases significantly to $17184\pm 173$\,\kms\ at $-3.3$\,days, though both velocities remain higher than typical \caiihk\ velocities of normal SNe~Ia. \citep[e.g.,][]{foley2011}.

\subsection{UV flux ratio}
\label{section:flux-ratio}
The flux ratios of the UV spectra have been commonly used to investigate the UV spectral properties of SNe~Ia \citep[e.g.,][]{foley2016,pan2020}. Following those studies, we compute the flux ratios $f_{2550}$ and $f_{3025}$ for the 03fg-like SNe in our sample. The flux ratio $f_{2550}$ is defined as the median flux at 2450-2650\,\AA\ (in the mid-UV region) divided by the median flux at 4000-4500\,\AA, while $f_{3025}$ is defined as the median flux at 2900-3150\,\AA\ (in the near-UV region) divided by the median flux at 4000-4500 {\AA}. These flux ratios are essentially analogous to UV colors measured through photometry. Uncertainties from flux, wavelength, and host-galaxy reddening are combined through standard error propagation to estimate the uncertainties in flux ratios. Since most of our sample do not have error spectra, the flux uncertainties are estimated from the scatter in the residuals between the observed and smoothed spectra. The resulting uncertainties on the flux ratios are quantified by performing a Monte Carlo experiment using this scatter as the standard deviation of the distribution. To account for wavelength uncertainties, we perform another Monte Carlo experiment by introducing small random shifts (with a standard deviation of $\sim 20$\,\AA) in wavelength and evaluating their impact on the flux ratios. Finally, we incorporate the uncertainty in host-galaxy reddening by performing the same experiment using the uncertainty of host-galaxy reddening.

\begin{figure*}
\centering
\includegraphics[scale=0.25]{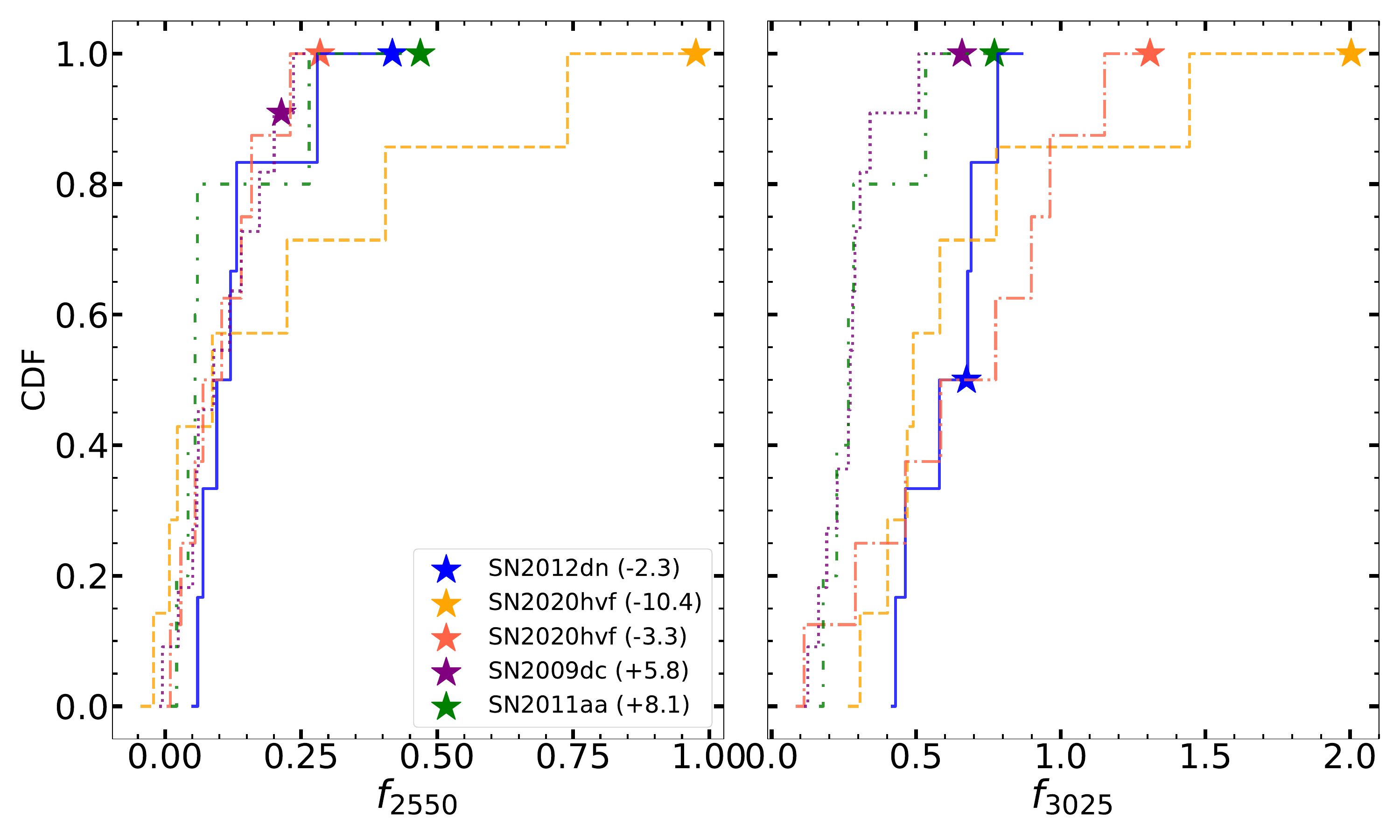}
\includegraphics[scale=0.25]{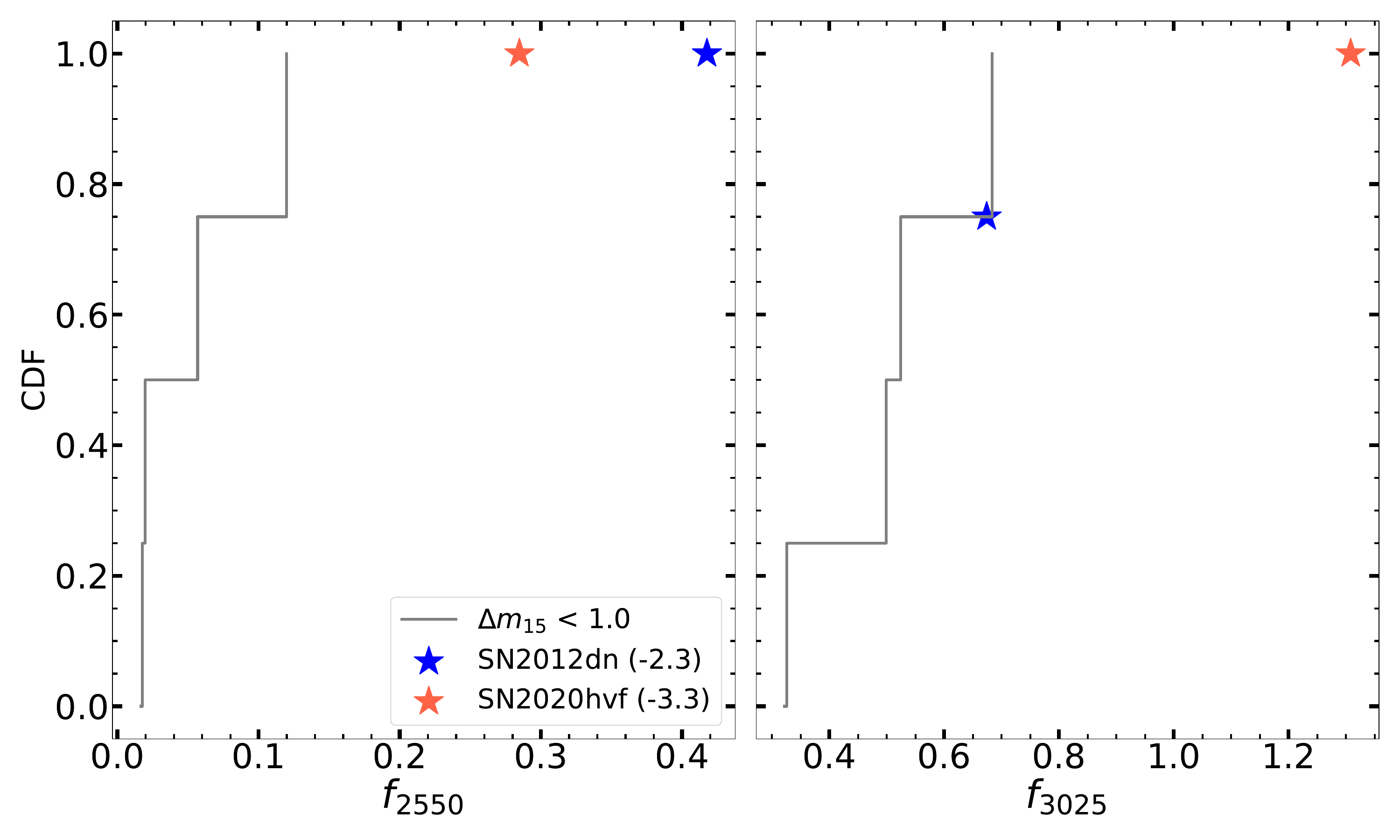}
\caption{\textit{Upper left:} The cumulative distribution functions (CDFs) of the flux ratio $f_{2550}$ are shown for SNe at phases within 2 days of each 03fg-like SN, including both normal and 03fg-like SNe~Ia. The color of each CDF matches the color of the corresponding star marker, which represents the interpolated CDF value for the $f_{2550}$ of the 03fg-like SN. \textit{Upper right:} The same as the left panel, but with $f_{3025}$ instead.
\textit{Lower left:} CDF of $f_{2550}$ for the normal SNe~Ia at near peak with $\Delta m_{15}(B) < 1$ (in black). The star markers represent the interpolated CDF values corresponding to the $f_{2550}$ of the near peak 03fg-like SNe. \textit{Lower right:} The same as the left panel, but with $f_{3025}$ instead.}\label{cdf}
\end{figure*}

The upper panels of Figure~\ref{cdf} show the cumulative distribution functions (CDFs) of flux ratios $f_{2550}$ and $f_{3025}$ for the normal SNe~Ia at different phases. where the star markers represent the interpolated CDF values corresponding to the $f_{2550}$ and $f_{3025}$ of 03fg-like SNe. As mentioned in Section~\ref{sec:uvspec}, only the spectra of normal SNe~Ia within a phase difference of 2~days from that of 03fg-like SNe are compared. There is a clear trend that the UV flux ratios of the 03fg-like SNe tend to be higher than the majority of the normal SNe~Ia. The only exception is the spectrum of SN~2012dn taken at $-2.3$\,day from the peak, with a $f_{3025}$ higher than only 50\% of the comparison sample. The low $f_{3025}$ of SN~2012dn may be related to its relatively high $\Delta m_{15}(B)$ (0.97; the highest among all the 03fg-like SNe in our sample), since previous studies have shown that the $f_{3025}$ is sensitive to the decline rate of SNe~Ia, with faster declined SNe~Ia having lower $f_{3025}$ \citep[e.g.,][]{pan2020}. 

The lower panels of Figure~\ref{cdf} show the CDFs of $f_{2550}$ and $f_{3025}$ for near-peak normal SNe~Ia with $\Delta m_{15}(B) < 1$. These are compared with the interpolated CDF values corresponding to the flux ratios of our 03fg-like SNe with near-peak measurements (i.e., SN~2012dn and SN~2020hvf). Unlike the upper panels of Figure~\ref{cdf}, we do not extend the analysis to other phases, as the sub-sample of normal SNe~Ia with $\Delta m_{15}(B) < 1$ would become too small for meaningful comparison. Despite limiting the comparison to those near-peak slow decliners ($\Delta m_{15}(B) < 1$), we find that 03fg-like SNe consistently exhibit higher flux ratios than normal SNe~Ia.

We list the flux ratios of 03fg-like SNe and their differences from the mean flux ratios of our comparison sample (i.e., from the normal SNe~Ia shown in Figure~\ref{spec:compare}) in Table~\ref{tab:sig_best_fit}. Consistent with the trend found in Section~\ref{sec:uvspec}, there is a trend that the UV flux ratios of 03fg-like SNe are significantly higher than the mean flux ratios of normal SNe~Ia at similar phases. 

\begin{figure*}
\centering
\includegraphics[scale=0.116]{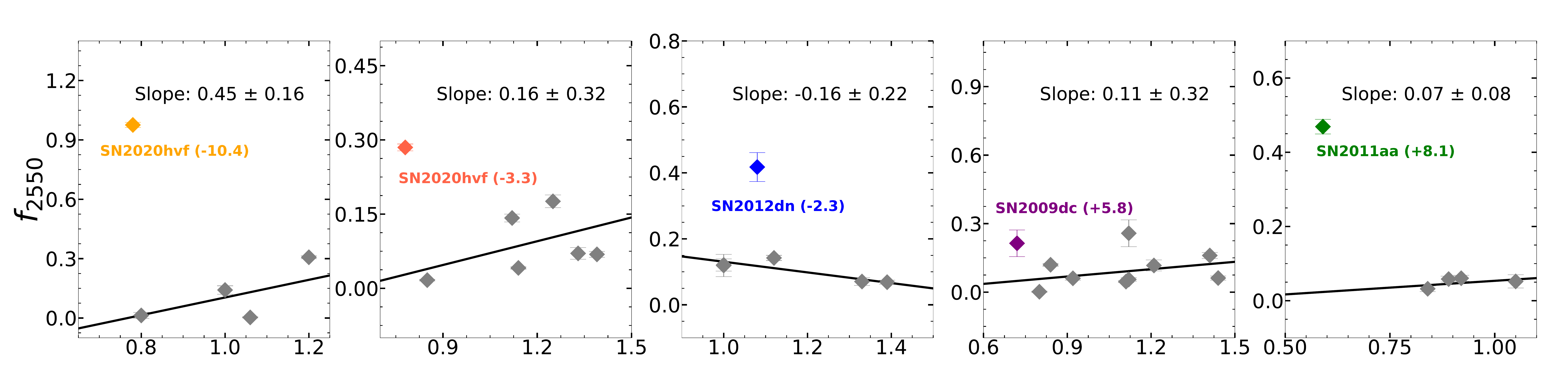}
\includegraphics[scale=0.116]{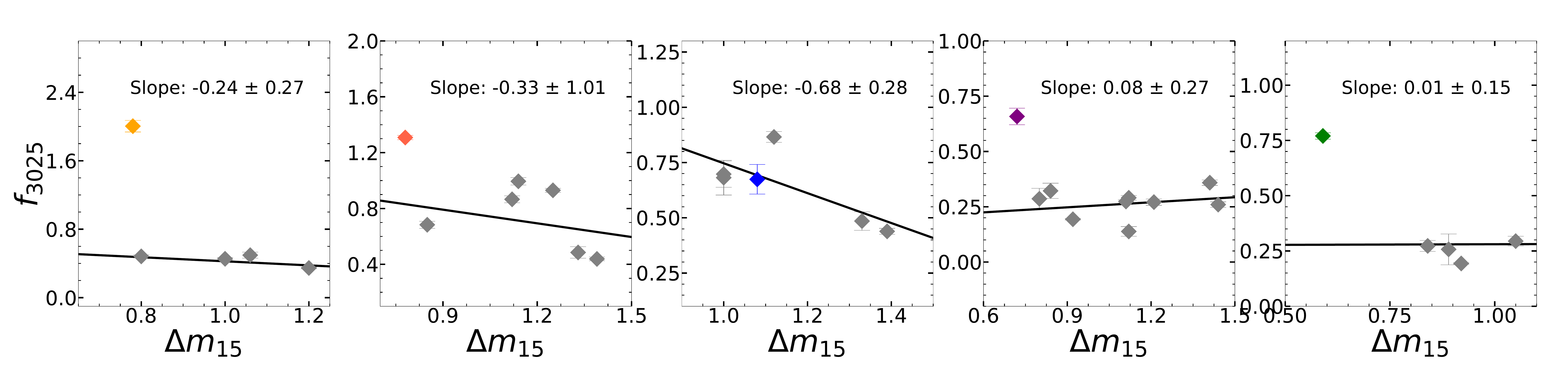}
\caption{\textit{Upper:} The flux ratio $f_{2550}$ as a function of $\Delta m_{15}$. Normal SNe~Ia are shown in grey, while 03fg-like SNe are highlighted and labeled in color. A linear fit to all normal SNe~Ia is represented by a grey solid line, with the slope displayed in the panel. \textit{Lower:} The same as the upper panels, but with $f_{3025}$ instead.}\label{ratio:2550_3025:dm15}
\end{figure*}

Next, we investigate if the UV excess found for 03fg-like SNe stays significant compared to the normal SNe~Ia of the same decline rates. The UV flux ratios $f_{2550}$ and $f_{3025}$ as a function of $\Delta m_{15}$ are shown in the upper and lower panels of Figure~\ref{ratio:2550_3025:dm15}, respectively. Again, only spectra with a phase difference of 2~days or less are used for comparison. We perform linear fitting with the Monte Carlo Markov Chain (MCMC) \textsc{PYTHON} package \texttt{LINMIX} \citep{kelly2007} on our normal SN~Ia sample (i.e., the gray line in each panel) to determine the difference of UV flux ratios between 03fg-like SNe and the predicted ratios from the normal SNe~Ia (through the linear fitting). In general, we find a significant trend that the UV flux ratios of 03fg-like SNe tend to be higher than those of normal SNe~Ia even at similar decline rates (except for the $f_{3025}$ of SN~2012dn, as discussed earlier in the section). This is based on the assumption of a linear relationship between the flux ratios and $\Delta m_{15}$, which is a reasonable approximation for SN~Ia near the peak luminosity as reported in \citet{pan2020}. All relevant statistics can be found in Table~\ref{tab:sig_best_fit}.

\begin{figure*}
\centering
\includegraphics[scale=0.27]{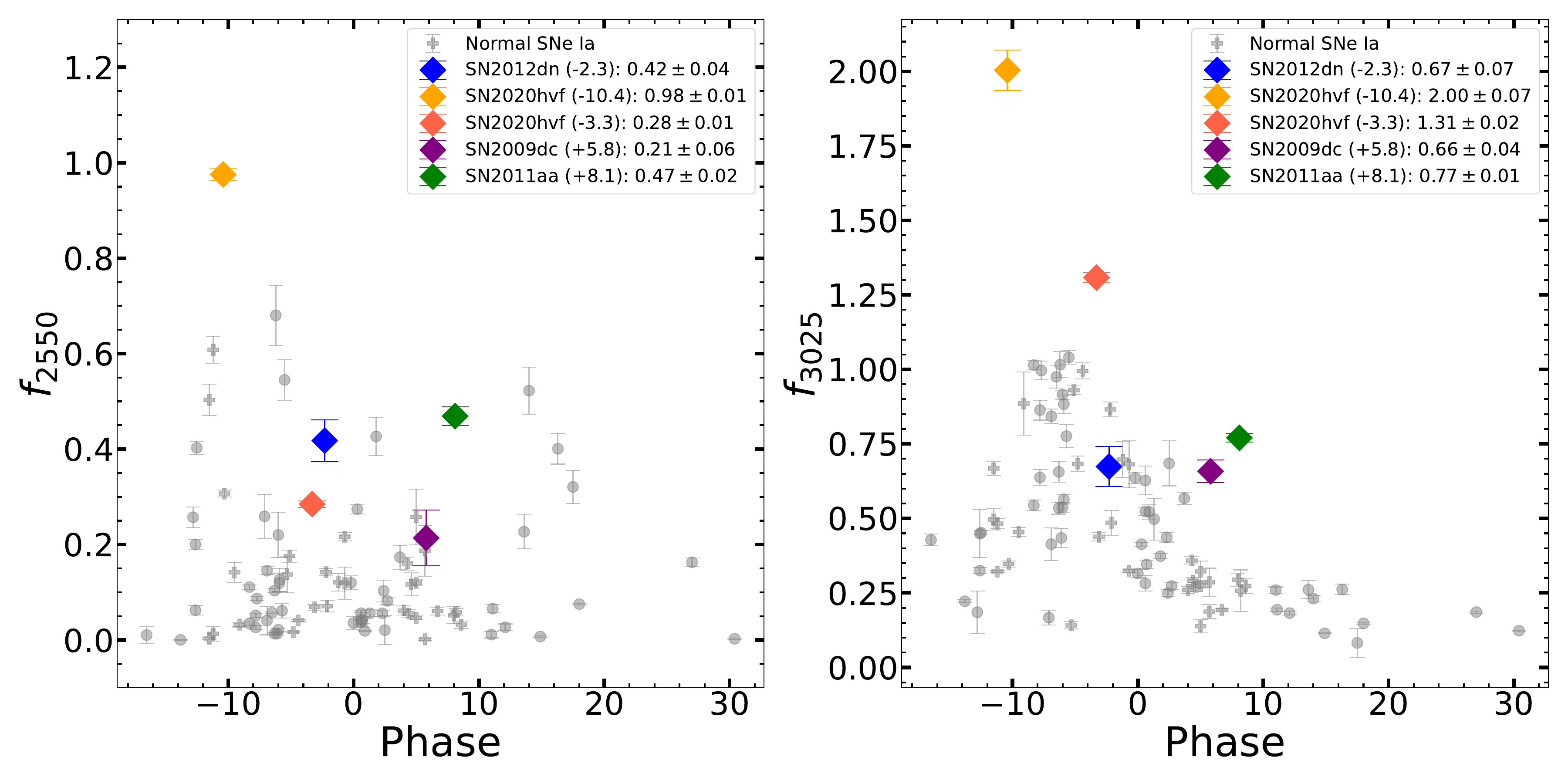}
\caption{The Left panel shows the flux ratio $f_{2550}$ as a function of SN phase, while the right panel shows the same for the flux ratio $f_{3025}$. Normal SNe~Ia are shown in grey, and 03fg-like SNe are highlighted in color.}\label{phase}
\end{figure*}

The temporal evolution of the UV flux ratios is shown in Figure~\ref{phase}. There are possible trends with both $f_{2550}$ and $f_{3025}$ for 03fg-like SNe, in the sense that higher UV flux ratios tend to be found at earlier phases (at least from the phase of $-10$\,days). In contrast, $f_{2550}$ of normal SNe~Ia shows no clear trend, while $f_{3025}$ increases after the explosion, peaking at $\sim$1 week before maximum light. However, the trend of 03fg-like SNe for $f_{2550}$ may be due to the high flux ratio of SN~2020hvf at $-10.4$\,day. The trend vanishes if we exclude that data point from the analysis. A larger sample of 03fg-like SNe is needed for further investigation. 

\begin{table*}
    \centering
    \caption{Comparison of flux ratios of 03fg-like SNe with the mean flux ratios and predictions from the best fit of normal SNe~Ia, within a phase difference of 2~days.}
    \label{tab:sig_best_fit}
    \small
    \begin{tabular}{cccccccccccc}
        \hline
        \hline
        SN Name & Phase & $f_{2550}$ & $f_{2550} - \overline{f_{2550}}$  & $f_{2550} - \text{Best-fit}_{f_{2550}}$  & $f_{3025}$ & $f_{3025} - \overline{f_{3025}}$  & $f_{3025} - \text{Best-fit}_{f_{3025}}$  \\
        \hline
        \hline
        SN~2020hvf & $-10.4$ & 0.976 $\pm$ 0.008  & 0.755 ($71.0\sigma$) & 0.970 ($121.2\sigma$)  &  2.004 $\pm$ 0.089 & 1.505 ($16.8\sigma$) & 1.527 ($17.2\sigma$) \\
        \hline
        SN~2020hvf & $-3.3$ & 0.285 $\pm$ 0.011  & 0.133 ($11.0\sigma$) & 0.257 ($23.4\sigma$) &  1.308 $\pm$ 0.055 & 0.587 ($10.5\sigma$) & 0.477 ($8.7\sigma$) \\
        \hline
        SN~2012dn & $-2.3$ & 0.418 $\pm$ 0.034  & 0.237 ($6.9\sigma$) & 0.300 ($8.8\sigma$) & 0.674 $\pm$ 0.053 & $-0.006$ ($0.1\sigma$) & $-0.019$ ($0.4\sigma$) \\
        \hline
        SN~2009dc & $+5.8$ & 0.214 $\pm$ 0.064  & 0.107 ($1.6\sigma$) & 0.165 ($2.6\sigma$)&  0.658 $\pm$ 0.060 & 0.400 ($6.6\sigma$) & 0.424 ($7.1\sigma$)\\
        \hline
        SN~2011aa & $+8.1$ & 0.469 $\pm$ 0.013  & 0.418 ($28.3\sigma$) & 0.445 ($34.2\sigma$) & 0.770 $\pm$ 0.013 & 0.515 ($22.4\sigma$) & 0.492 ($37.8\sigma$) \\
        \hline
        \hline
    \end{tabular}
        \smallskip
    \parbox{\linewidth}{\small\textbf{Note:} $f_{2550} - \overline{f_{2550}}$ and $f_{3025} - \overline{f_{3025}}$ represent the deviations in flux ratios for 03fg-like SNe relative to the mean values of normal SNe~Ia at similar phases. Similarly, $f_{2550} - \text{Best-fit}_{f_{2550}}$ and $f_{3025} - \text{Best-fit}_{f_{3025}}$ represent the same but for the deviations of 03fg-like SNe flux ratios from the predictions for normal SNe~Ia, as determined by the linear fit to flux ratio and $\Delta m_{15}(B)$ relation. The values in parentheses represent the significance levels of the differences.}
\end{table*}

\subsection{Comparison with the mean spectra of normal SNe~Ia}\label{sec:mean}
Here, we investigate the UV properties of 03fg-like SNe by comparing their spectra with the mean spectra of normal SNe~Ia produced in \citet{pan2020}. Given that only the near-peak mean spectra are available, we limit our comparison to the spectra of similar phases within our 03fg-like SN sample, specifically SN~2012dn and SN~2020hvf at $-2.3$\,days and $-3.3$\,days, respectively. 

\begin{figure*}
\centering
\includegraphics[scale=0.19]{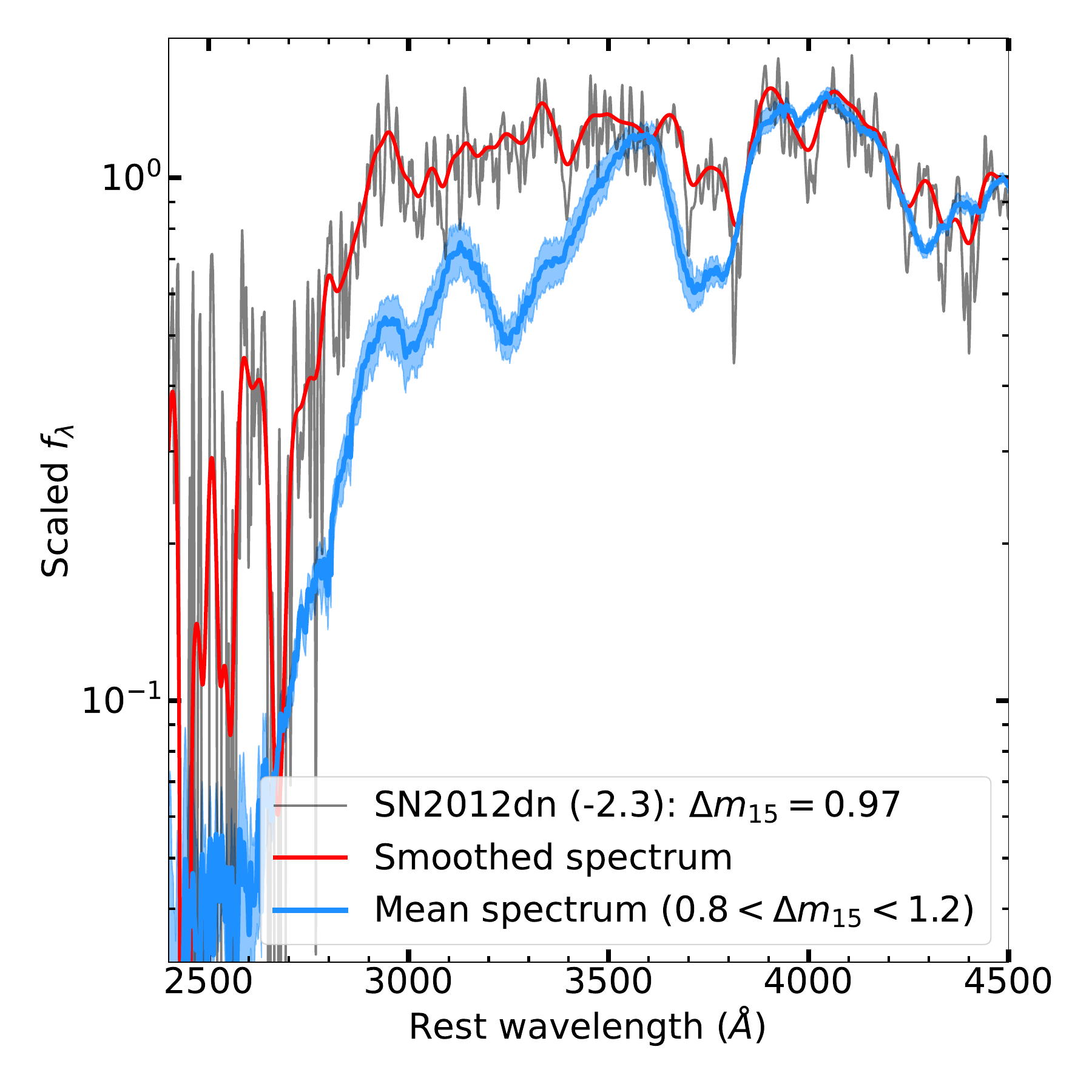}
\includegraphics[scale=0.19]{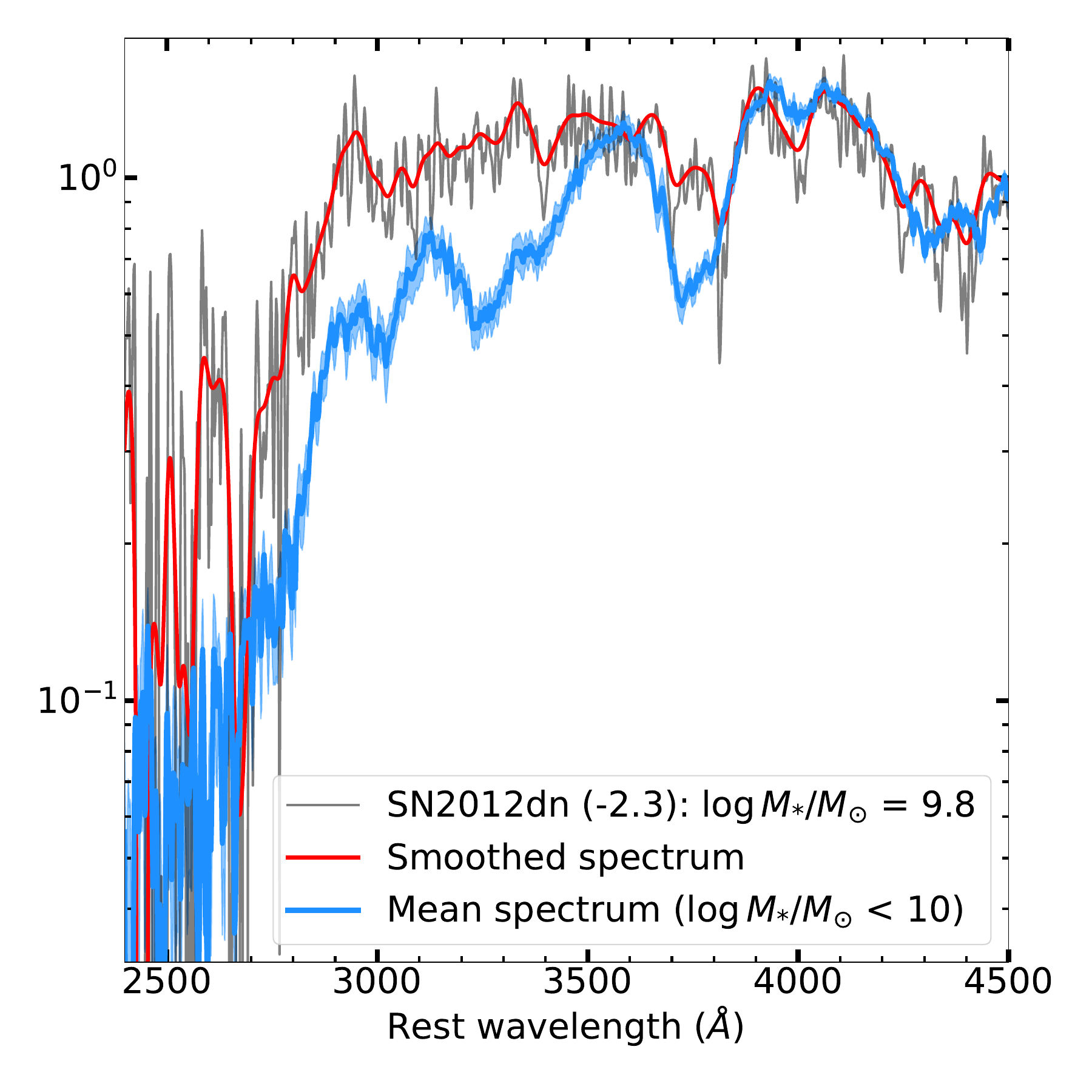}
\includegraphics[scale=0.19]{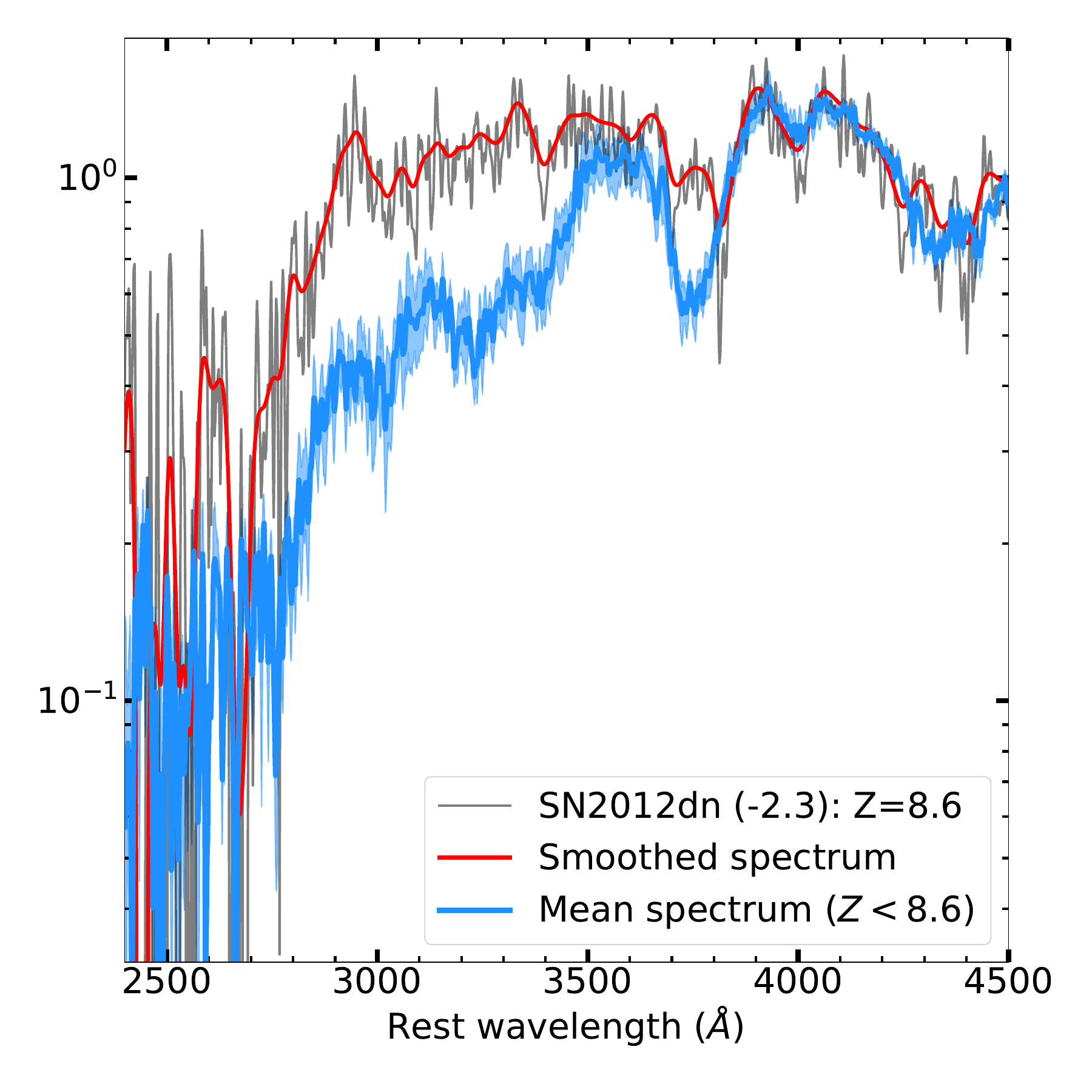}\\
\includegraphics[scale=0.19]{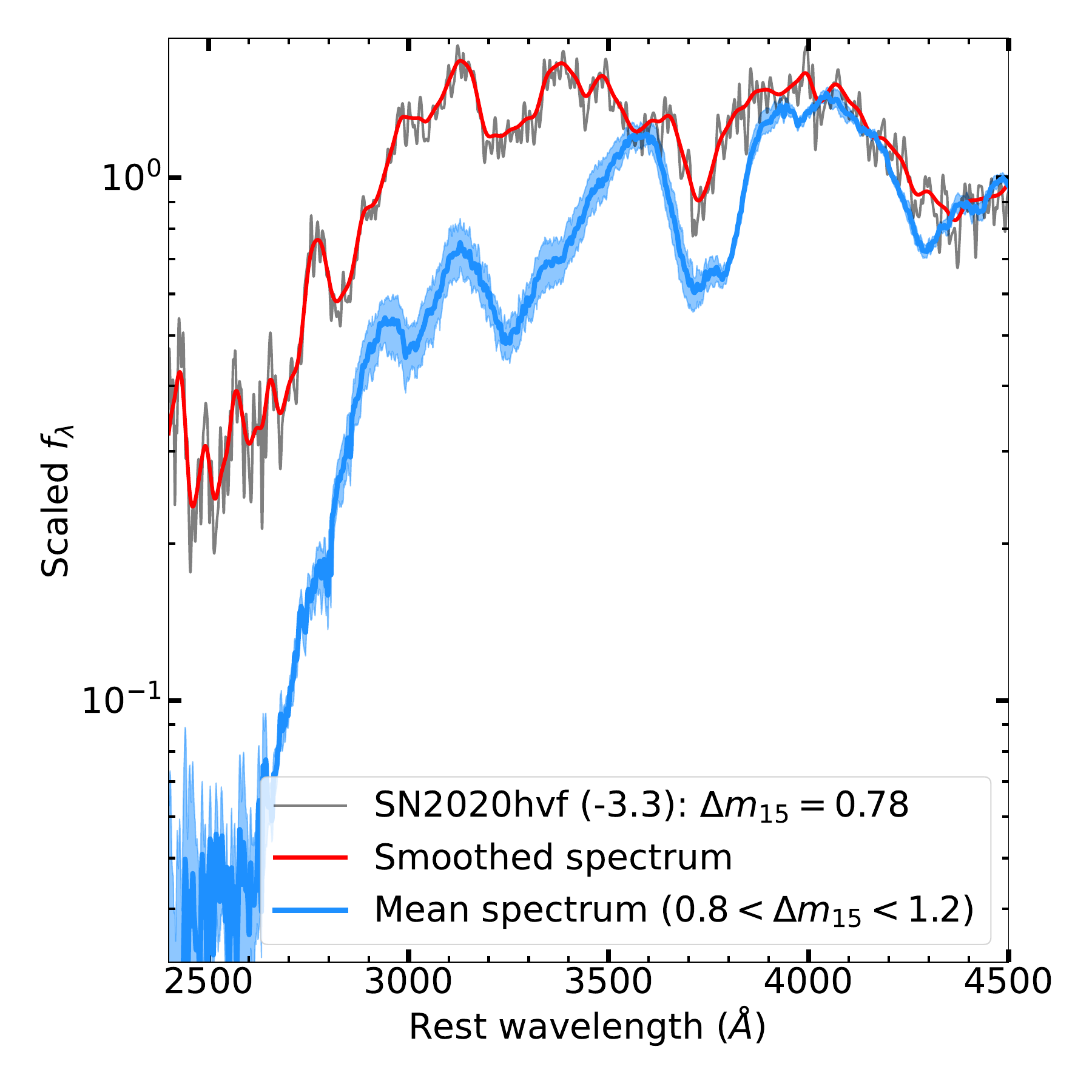}
\includegraphics[scale=0.19]{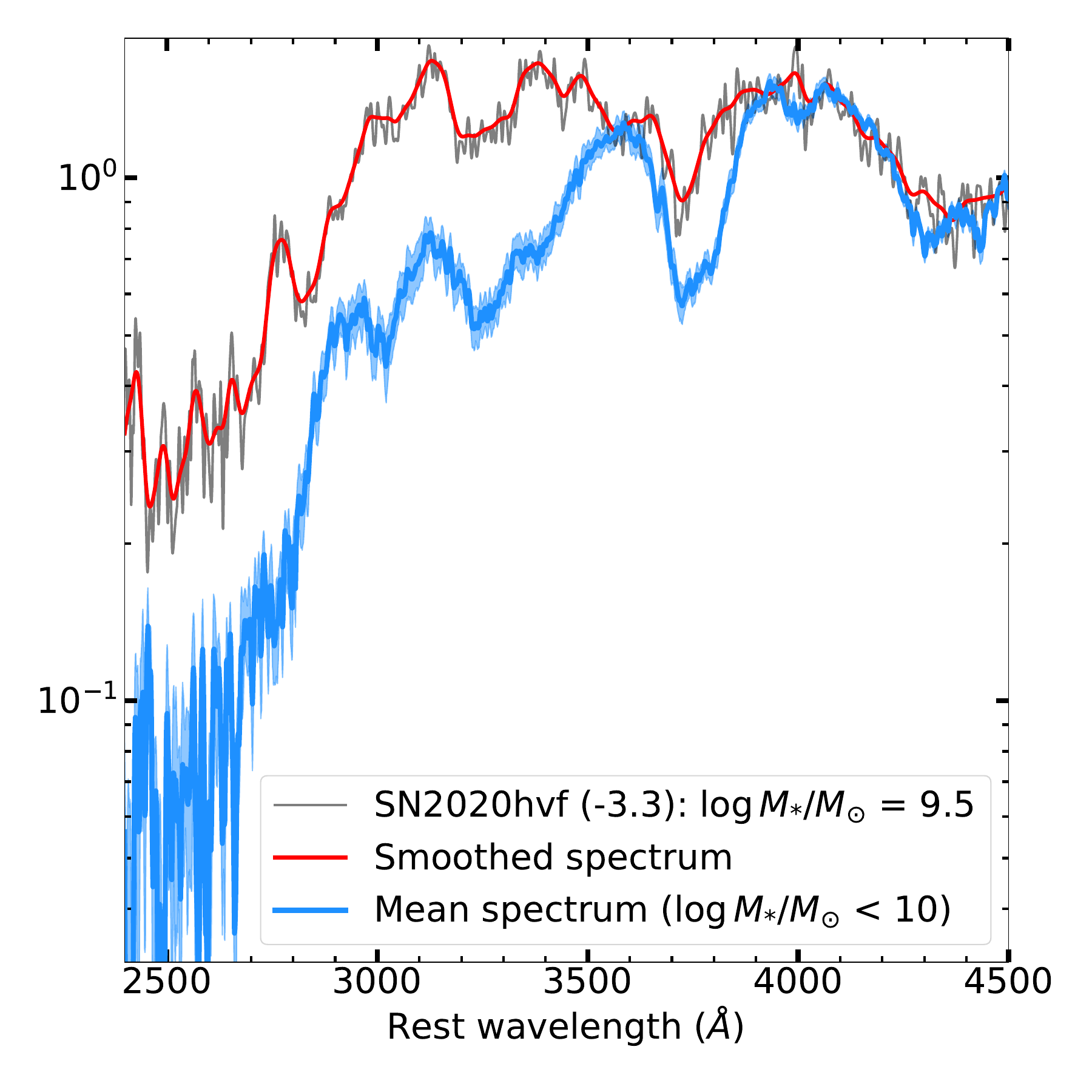}
\includegraphics[scale=0.19]{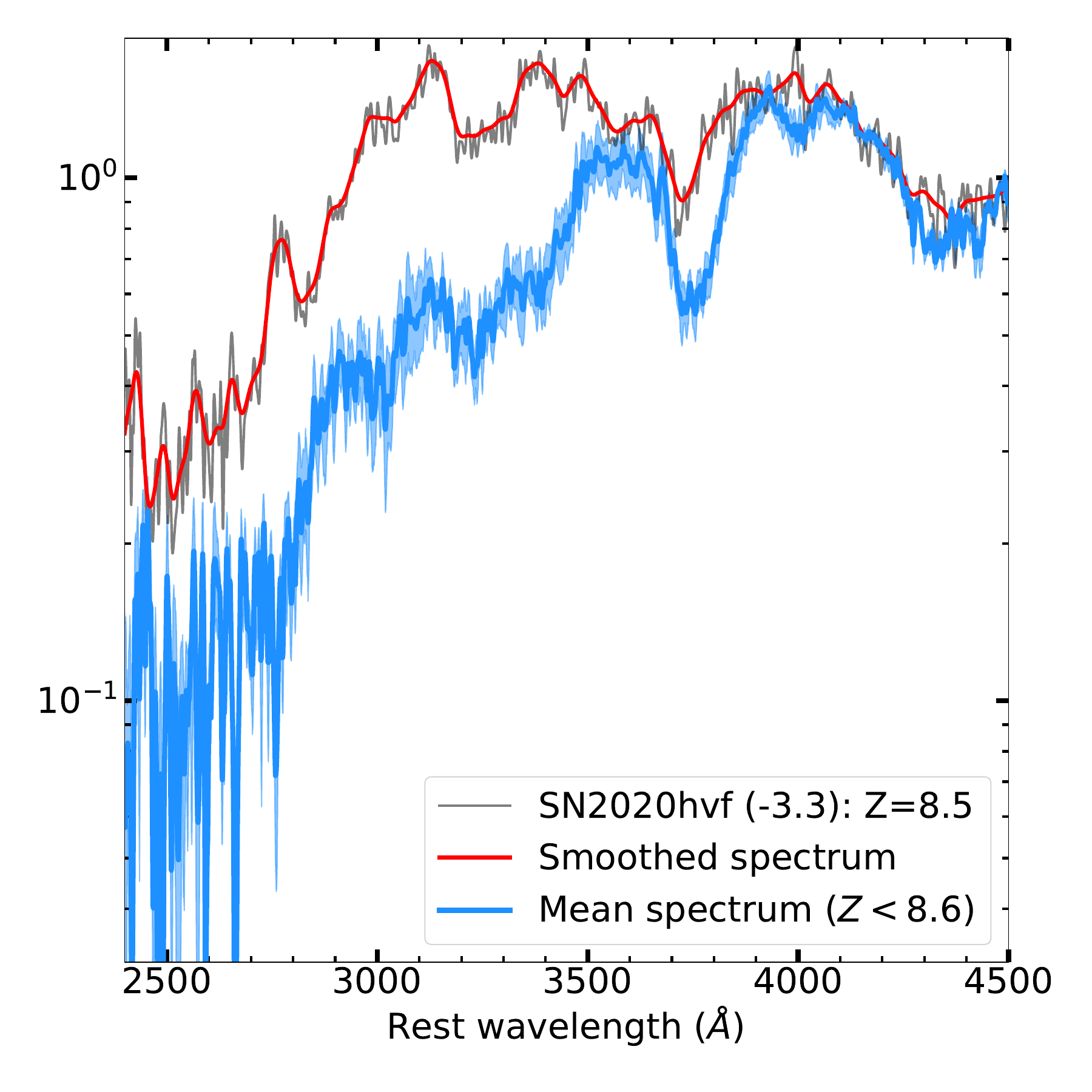}

\caption{\textit{Upper left}: The comparison between the near-peak UV spectra of SN~2012dn (in black) and the near-peak mean spectra of normal SNe~Ia with $0.8 < \Delta m_{15} < 1.2$ (in blue). The red spectrum represents the smoothed spectrum obtained using the inverse-variance Gaussian method \citep[e.g.,][]{blondin2006}. \textit{Upper 
 middle}: The same as the left panel, but compared to the mean spectra of normal SNe~Ia with host-galaxy stellar mass $\log (M_{*}/M_{\odot}) < 10$. \textit{Upper right}: The same as the left panel, but compared to the mean spectra of normal SNe~Ia with host-galaxy gas-phase metallicities $Z < 8.6$. \textit{Lower}: The same as upper panels but for SN~2020hvf.} \label{mean-spec}
\end{figure*}

The left panels of Figure~\ref{mean-spec} present a spectral comparison between SN~2012dn (upper left) and SN~2020hvf (lower left) and the mean spectra of normal SNe~Ia with decline rates in the range $0.8 < \Delta m_{15} < 1.2$, respectively. While SN~2012dn has a $\Delta m_{15}$ of 0.97\,mag, SN~2020hvf has a slightly lower $\Delta m_{15}$ of 0.78\,mag, just outside the range of the mean spectrum used for comparison. Since mean spectra for $\Delta m_{15}$ values below 0.8\,mag are not available in \citet{pan2020}, a more direct comparison is not possible. Nevertheless, we observe a clear trend: 03fg-like SNe tend to show significantly higher UV flux than the mean spectra of normal SNe~Ia at similar phases and decline rates, particularly at wavelengths below $\sim3500$\,\AA. This finding aligns with the UV flux ratios presented in Figure~\ref{ratio:2550_3025:dm15}.

The middle and right panels of Figure~\ref{mean-spec} show the same spectral comparisons, now using mean spectra produced with respect to the host-galaxy stellar mass ($M_{*}$) and gas-phase metallicity ($Z$), respectively. Since both SN~2012dn and SN~2020hvf have host-galaxy stellar mass $\log (M_{*}/M_{\odot}) < 10$, we compare them with the mean spectra of normal SNe~Ia of $\log (M_{*}/M_{\odot}) < 10$ from \citet{pan2020}. As direct measurements of host-galaxy metallicity are unavailable for our 03fg-like SNe, we estimate it from the host-galaxy stellar mass using the \textit{PP04 03N2} calibration \citep{pp2004} derived in \citet{kewley2008}. This yields gas-phase metallicities of $Z=8.6$ and $Z=8.5$ for SN~2012dn and SN~2020hvf, respectively. Accordingly, we compare their spectra with the mean spectra of normal SNe~Ia with $Z<8.6$ determined in \citet{pan2020}. Overall, we see a similar trend to that of the SN decline rate, in the sense that 03fg-like SNe tend to exhibit an excess in UV compared to normal SNe~Ia with similar host-galaxy stellar masses and metallicities.

\subsection{Comparison with the model templates of normal SNe~Ia}\label{sec:tem}

\citet{pan2020} generated empirical, data-driven spectral templates of normal SNe~Ia near the peak luminosity. These templates were constructed by fitting the smoothed flux for all the UV spectra at each wavelength using a linear function with various parameters, including the SN decline rate $\Delta m_{15}$, host-galaxy stellar mass $M_{*}$, and gas-phase metallicity $Z$. An advantage of these model templates is their ability to provide comparison spectra for specific parameter spaces, allowing for direct comparison between observed data and models. Since these templates are only available near the peak luminosity, we restrict our comparison to the UV spectra of SNe~2012dn and 2020hvf, for which the near-peak spectra are available.

\begin{figure*}
\centering
\includegraphics[scale=0.19]{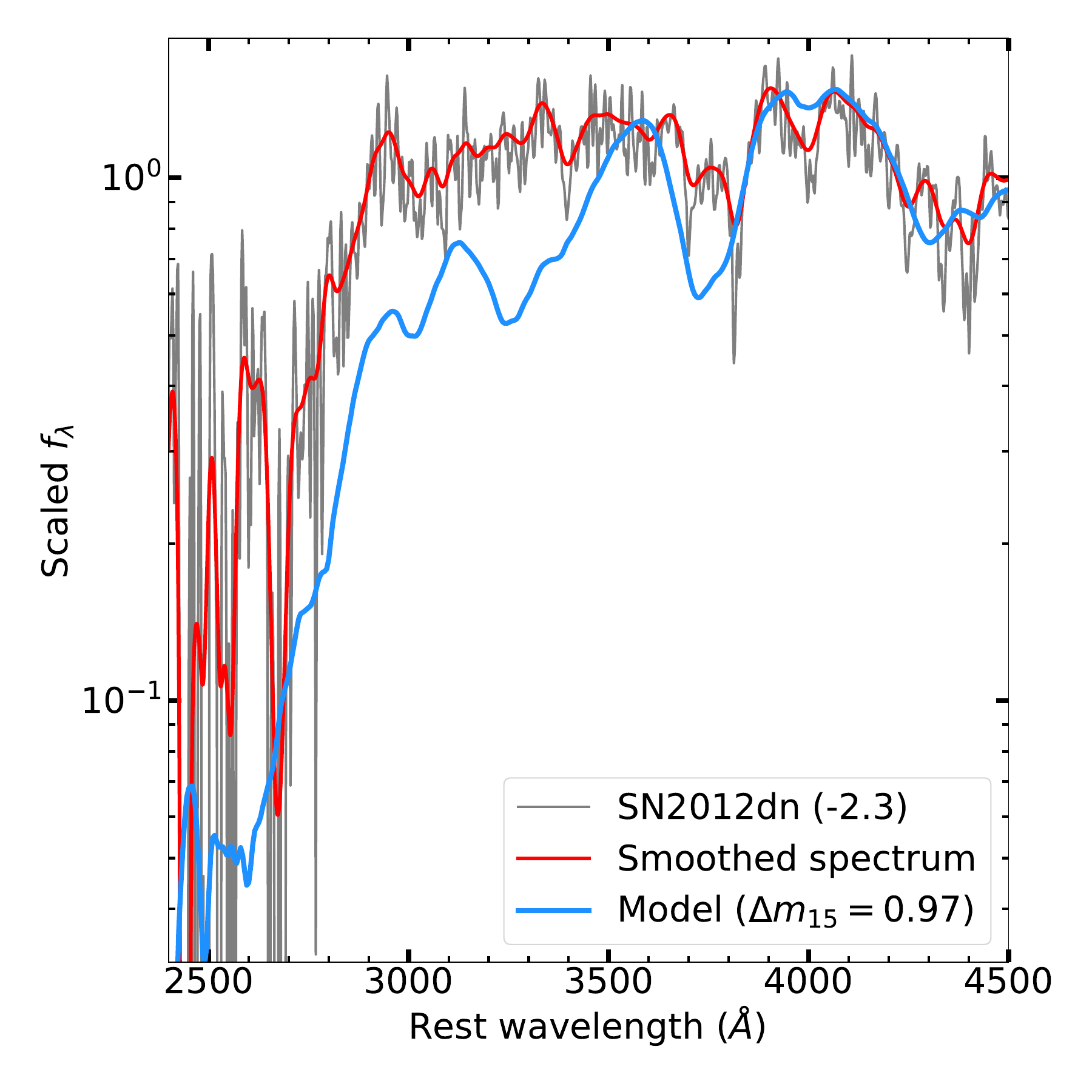}
\includegraphics[scale=0.19]{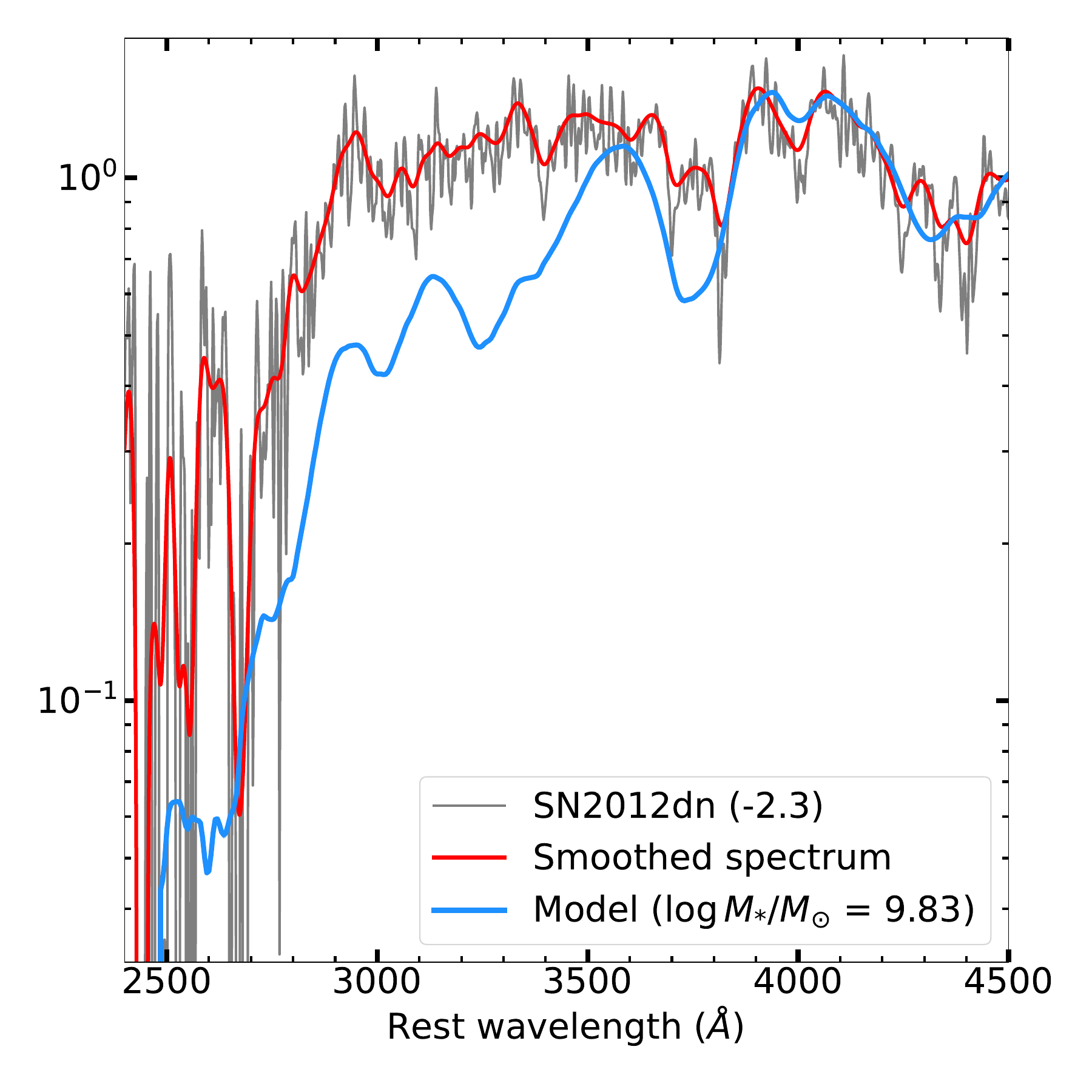}
\includegraphics[scale=0.19]{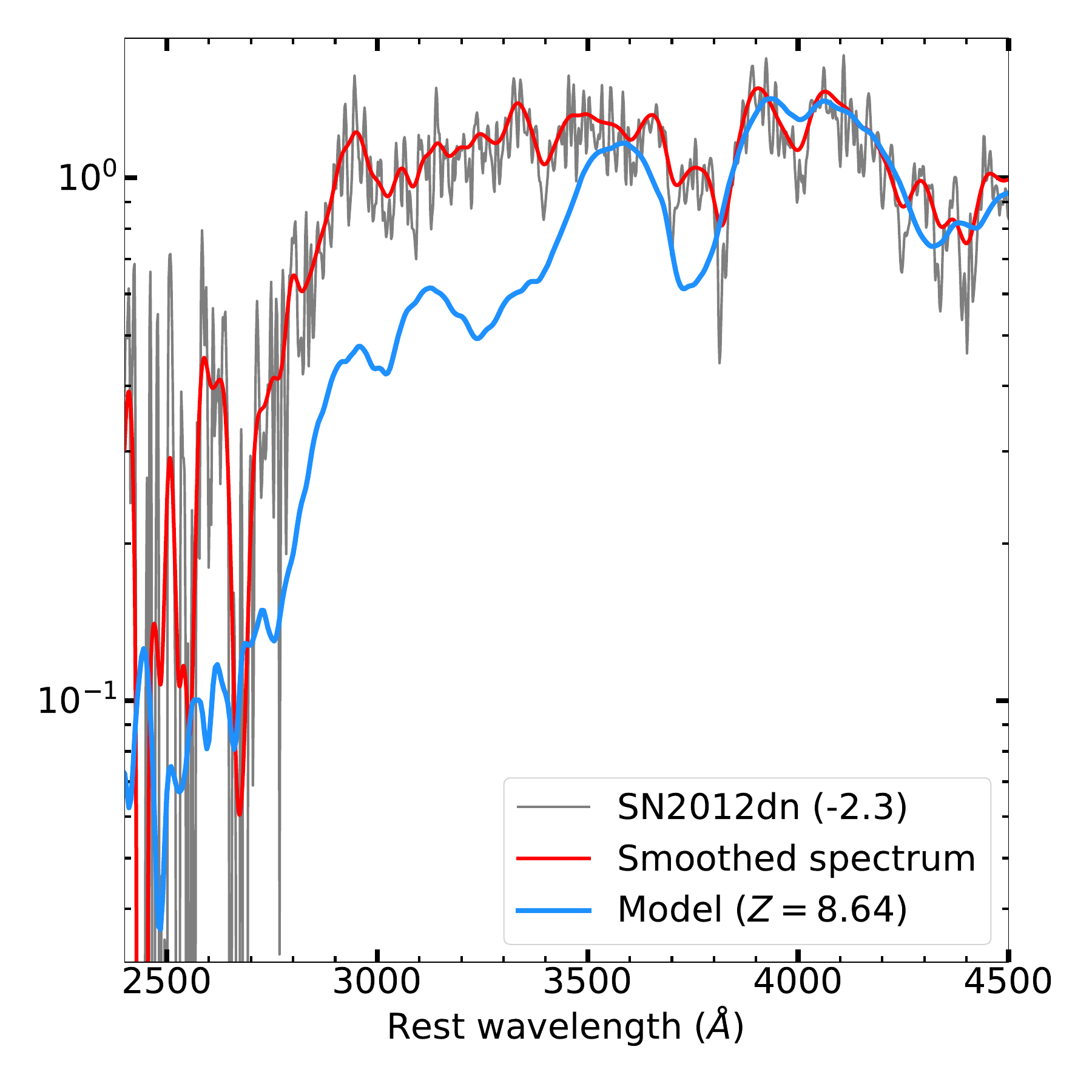}\\
\includegraphics[scale=0.19]{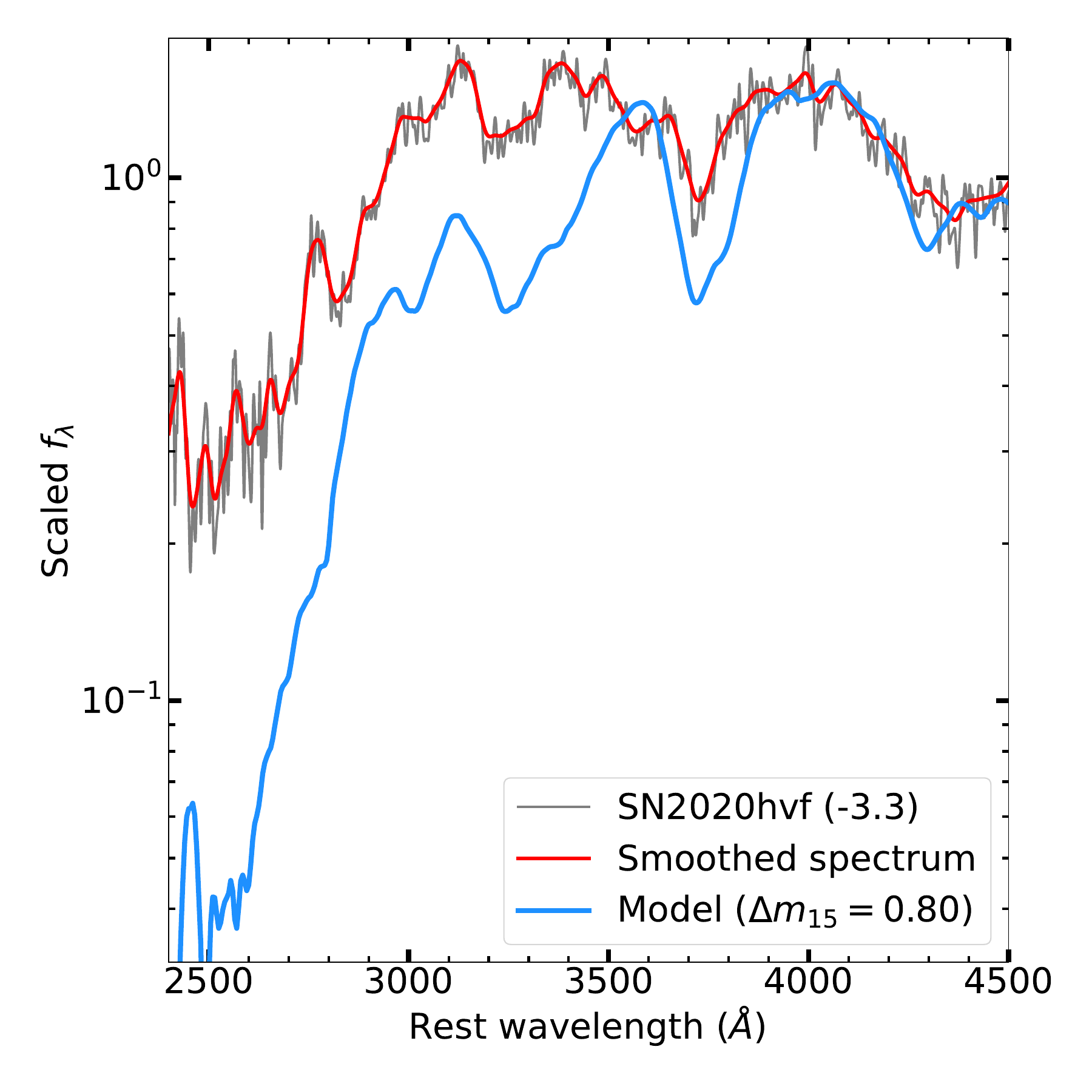}
\includegraphics[scale=0.19]{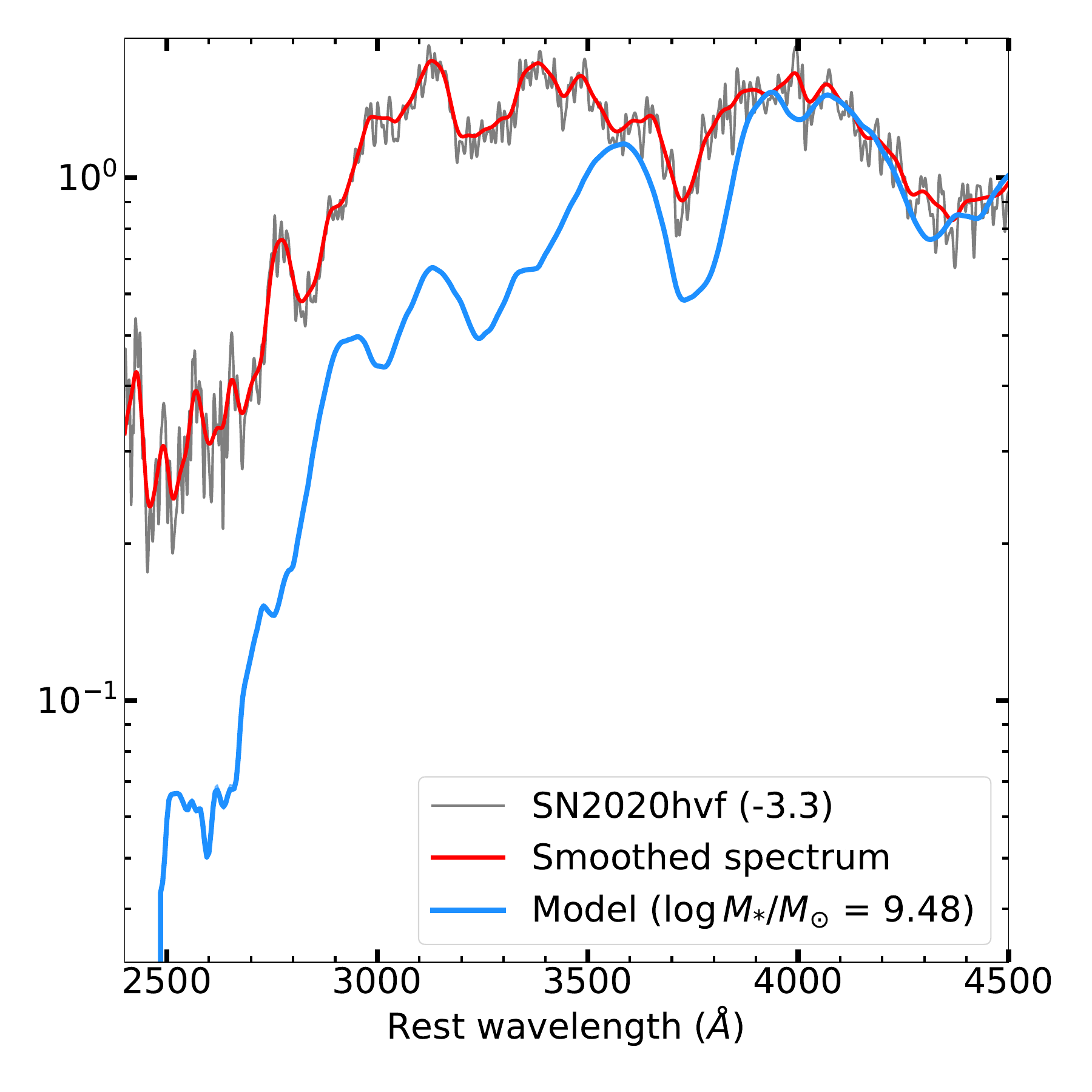}
\includegraphics[scale=0.19]{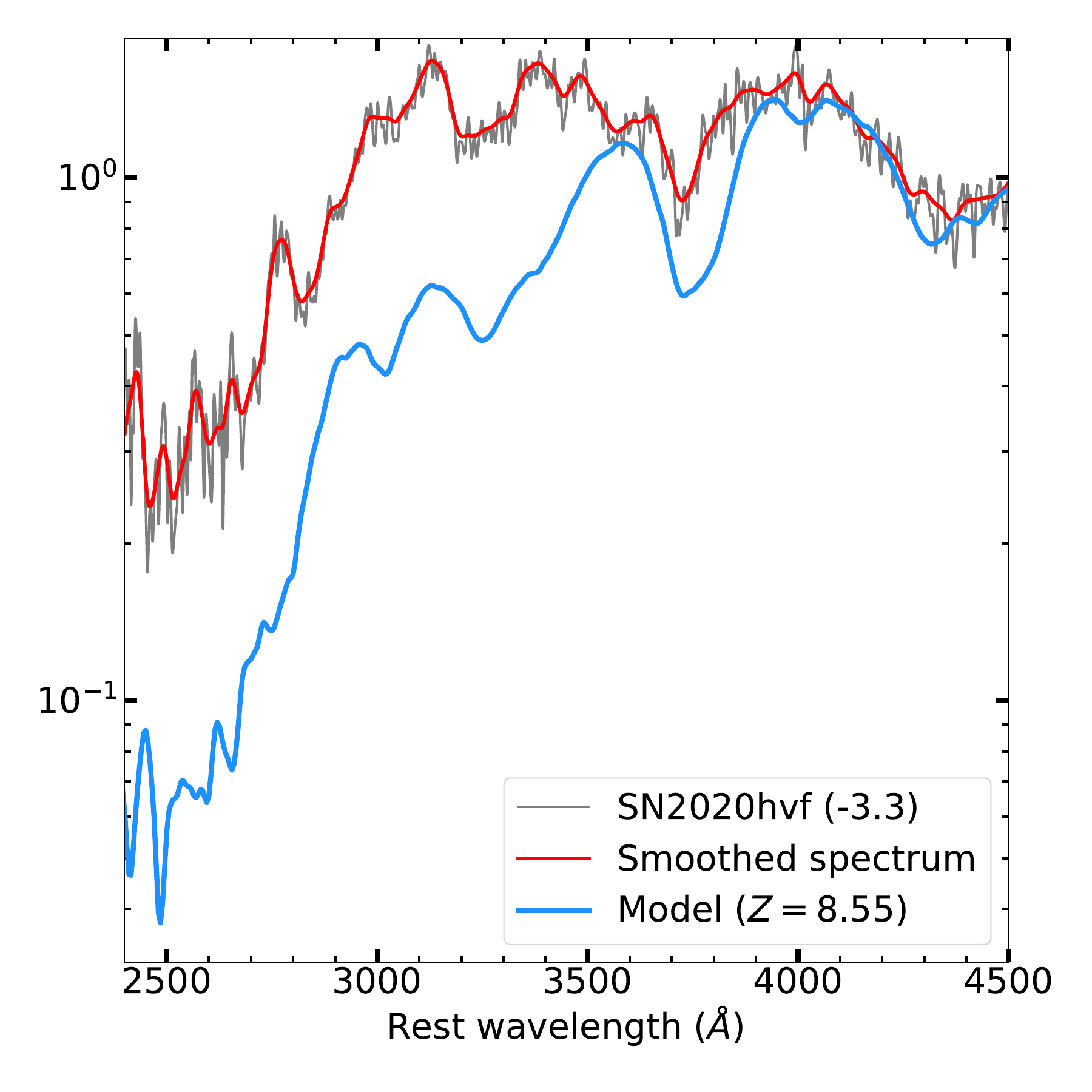}
 \caption{\textit{Upper left}: The comparison between the near-peak UV spectra of SN~2012dn (in black) and the near-peak model spectra of normal SNe~Ia with the same $\Delta m_{15}$ (in blue). The red spectra represent the smoothed spectra obtained using the inverse-variance Gaussian method. \textit{Upper middle}: The same as the left panel, but compared to the model spectra of normal SNe~Ia with the same host-galaxy stellar mass. \textit{Upper right}: The same as the left panel, but compared to the model spectra of normal SNe~Ia with the same host-galaxy gas-phase metallicities. \textit{Lower}: The same as upper panels but for SN~2020hvf.}\label{template}
\end{figure*}

The left panels of Figure~\ref{template} present a spectral comparison between SN~2012dn (upper) and SN~2020hvf (lower) with the spectral templates of the same $\Delta m_{15}$ as that of the normal SNe~Ia, respectively. The 03fg-like spectra are significantly bluer than the template spectra of normal SNe~Ia with the same $\Delta m_{15}$ in the UV ($\lesssim3500$\,\AA), consistent with the trend found with the mean spectra (Figure~\ref{mean-spec}). Notably, SN~2020hvf shows UV excess extending from a longer wavelength ($\lesssim4000$\,\AA) than that of SN~2012dn.

Similar trends are found in the middle and right panels of Figure~\ref{template}, where the UV spectra of SN~2012dn and SN~2020hvf are compared to spectral templates parametrized by the same host-galaxy stellar mass and gas-phase metallicity as those of the SNe, respectively. A significant discrepancy in UV colors remains between the template spectra of normal SNe~Ia and that of 03fg-like SNe, suggesting that the host-galaxy environment may not be the primary factor driving the difference between 03fg-like SNe and normal SNe~Ia in the UV. In addition to comparing models with parameters matched to those of 03fg-like SNe, we also explore the full parameter space of our data-driven models. Nevertheless, we are unable to reproduce the UV flux levels observed in SN~2012dn and SN~2020hvf near peak luminosity.

\begin{figure*}
\centering
\includegraphics[scale=0.24]{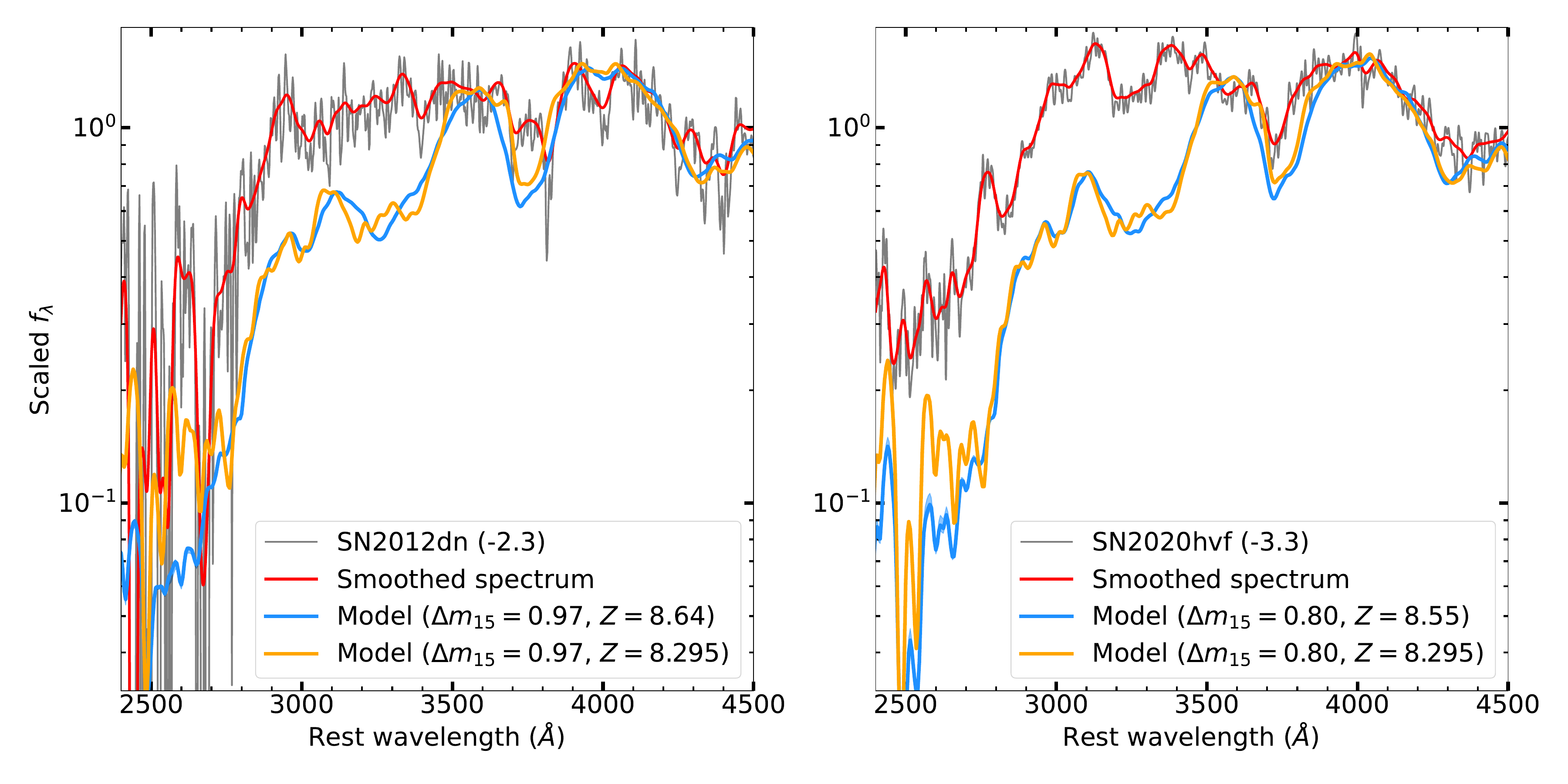}
\caption{The same as Figure~\ref{template}, but comparing with the 2-parameter ($\Delta m_{15}$ and $Z$) model spectra of normal SNe~Ia. 
}\label{template:metal-dm15}
\end{figure*}

Finally, Figure~\ref{template:metal-dm15} presents the spectral template parametrized by both $\Delta m_{15}$ and $Z$. \citet{pan2020} found that using both parameters yields a spectral template more consistent with observations, as it enables a more robust exploration of parameter space given that UV flux is influenced differently by each parameter. We apply the same $\Delta m_{15}$ and metallicity as the 03fg-like SN to the 2-parameter spectral model. However, a significant discrepancy in UV flux levels between the spectral template and the 03fg-like SNe remains. We further decrease $Z$ to $8.295$, the lower limit at which the spectral template remains valid in \citet{pan2020}, to assess if a normal SN~Ia in an extremely metal-poor galaxy could reach UV flux levels comparable to those of 03fg-like SNe. Despite this adjustment, we still find it difficult to reproduce the high UV flux levels observed in the 03fg-like SNe.

\subsection{Spectral modeling with \textsc{tardis}}\label{sec:tardis}

Here, we use the \tardis\ spectral modeling to investigate the UV spectral features of 03fg-like SNe and explore their differences from normal SNe~Ia in more detail. \tardis\ is an open-source, Monte Carlo radiative transfer code designed to model SNe ejecta in one dimension, given a specified density profile, elemental composition, atomic line data, and treatment of plasma interactions \citep{kerzendorf2014}. The code assumes spherical symmetry, homologous expansion of the ejecta, and a blackbody photosphere. \tardis\ is time-independent and, therefore, provides a snapshot of the spectral evolution at a chosen epoch. This makes it ideal for our analysis, as we aim to compare the \tardis\ synthetic spectra to the observed spectra of the 03fg-like SNe at specific epochs. However, \tardis\ has limitations, including its inability to model aspherical explosions \citep[e.g.,][]{siebert2023,nagao2024} and its inability to account for CSM interactions, which may be relevant for 03fg-like SNe (see Section~\ref{sec:discussion} for further details). Nonetheless, \tardis\ has yielded promising results in studies related to a diverse class of SNe \citep[e.g.,][]{magee2016,izzo2019,mulligan2019,gillanders2020,vogl2020,barna2021,williamson2021,kwok2022}.

We adopt the \texttt{branch85\_w7} \citep{branch1985} density model in our \tardis\ settings. For simplicity, we use uniform fractional element abundances, where the mass fractions are initially chosen randomly and then iteratively optimized along with other parameters, such as luminosities, ejecta velocities, and plasma temperatures, to improve the spectral fit. We list the key parameters of our \tardis\ setup in Table \ref{tab:tardis}. More details regarding our \tardis\ setup and \textsc{PYTHON} codes are available on \texttt{GITHUB}\footnote{\url{https://github.com/Bilton6/TARDIS_setups}.}. 

\begin{figure*}
\centering
\includegraphics[width=0.32\columnwidth]{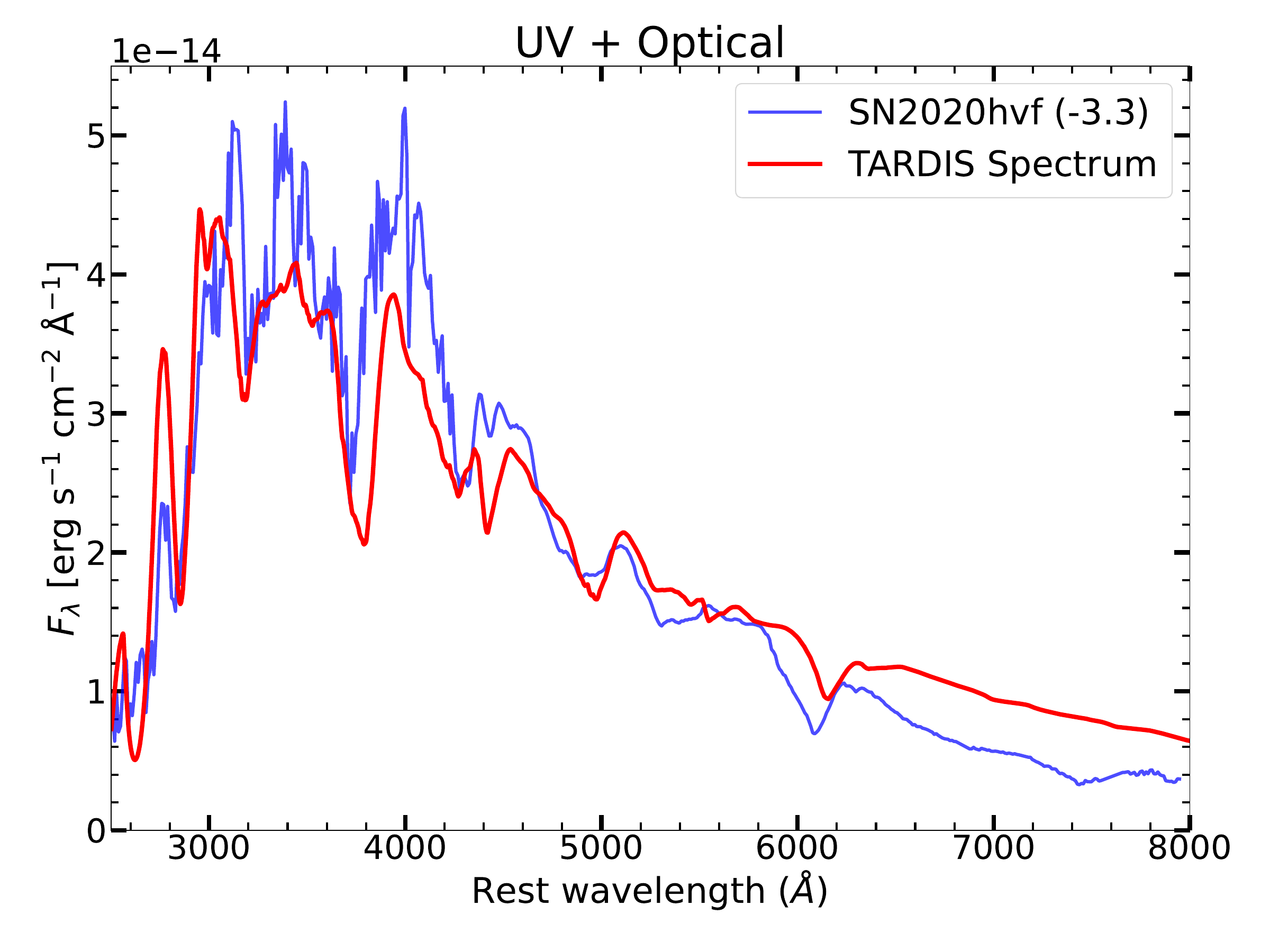}
\includegraphics[width=0.32\columnwidth]{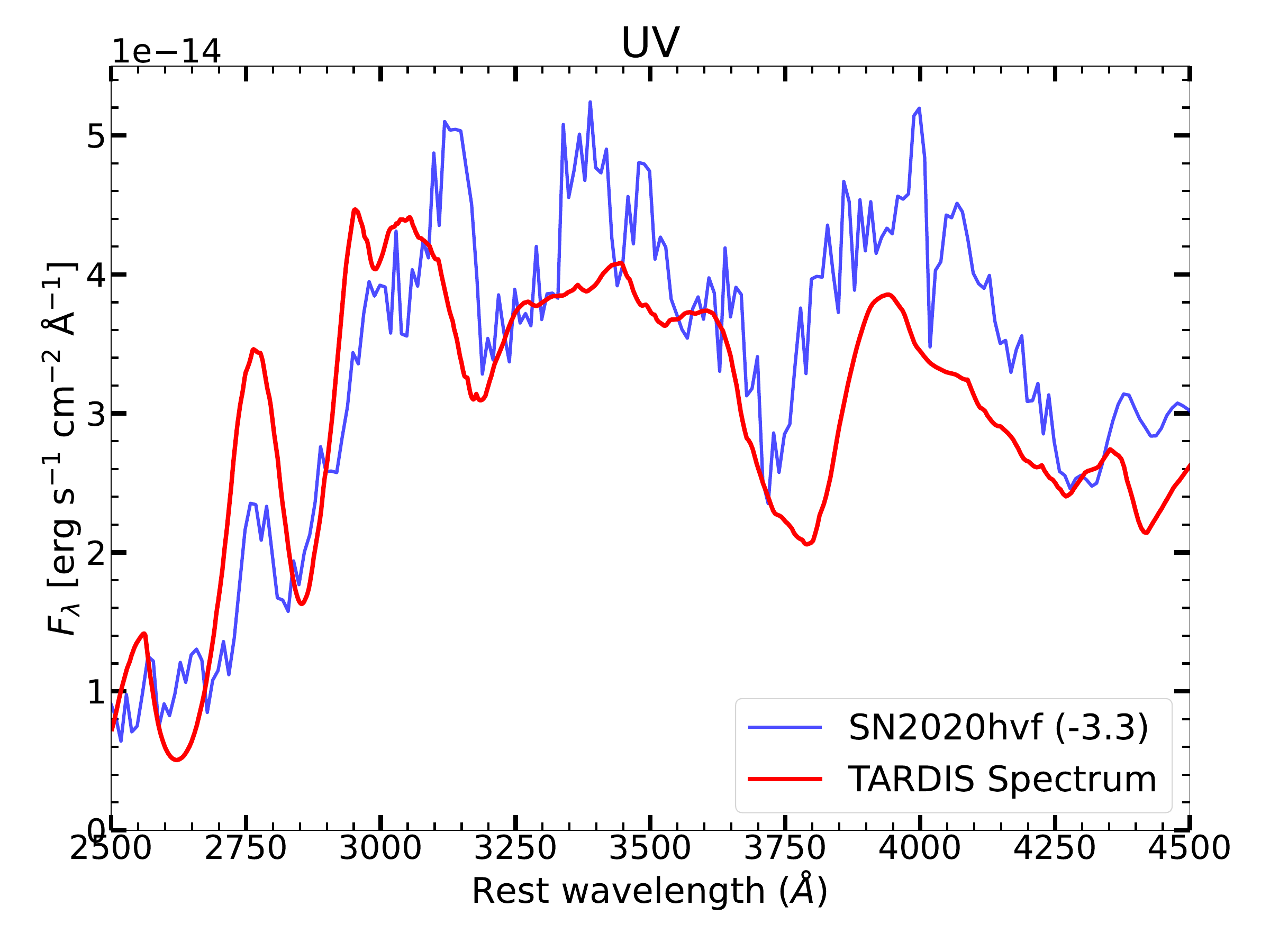}
\includegraphics[width=0.34\columnwidth]{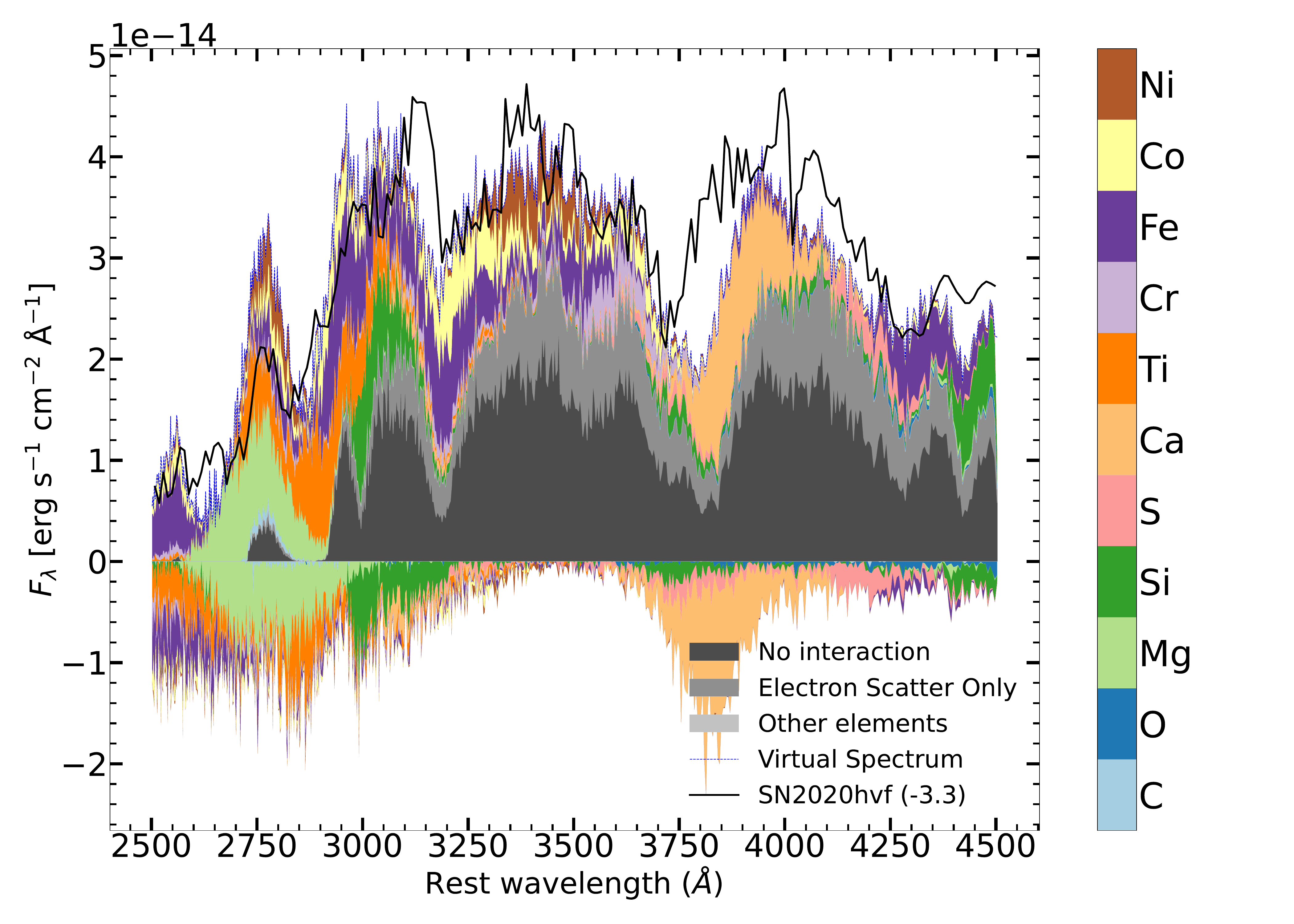}\\
\includegraphics[width=0.32\columnwidth]{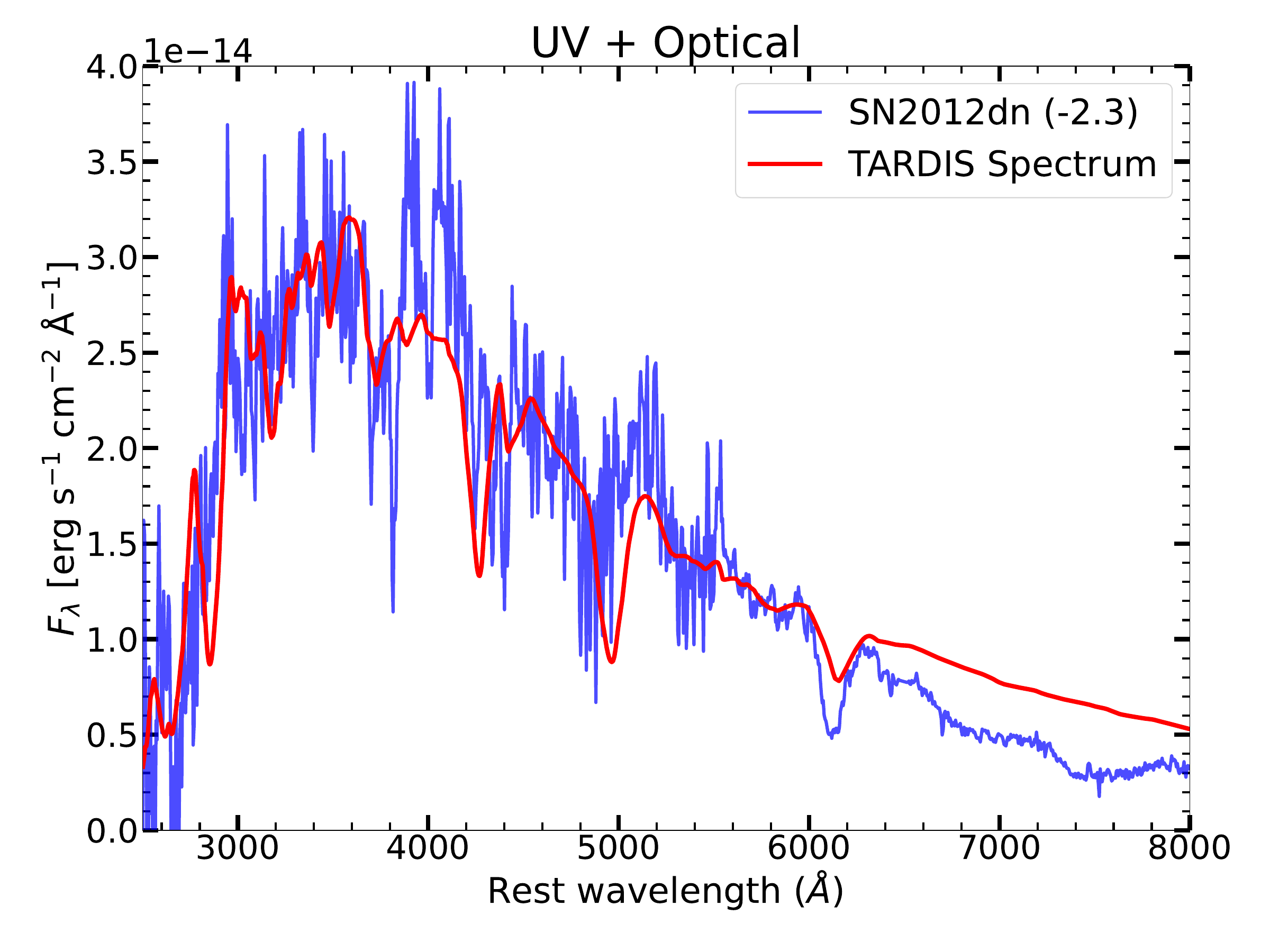}
\includegraphics[width=0.32\columnwidth]{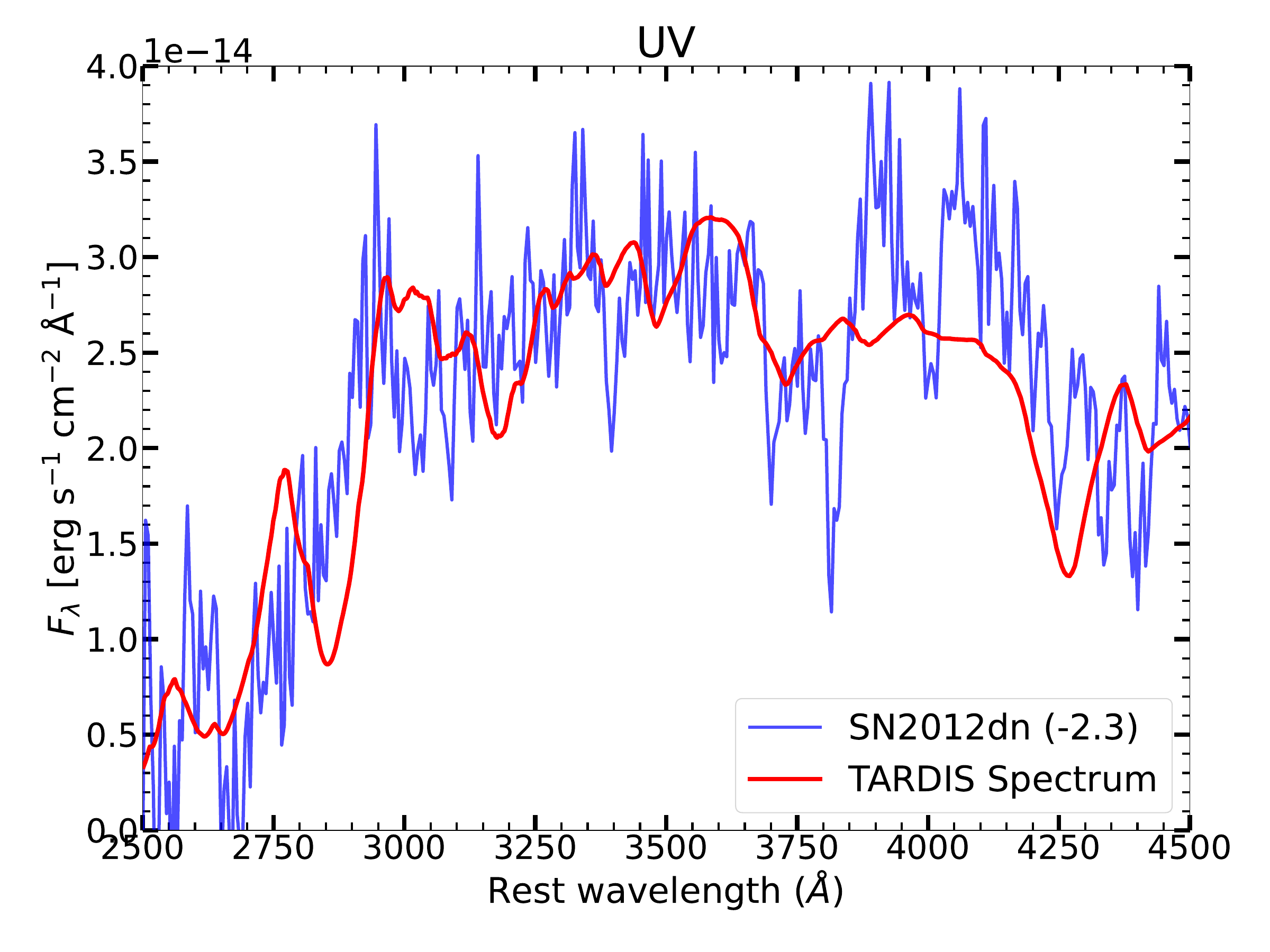}
\includegraphics[width=0.34\columnwidth]{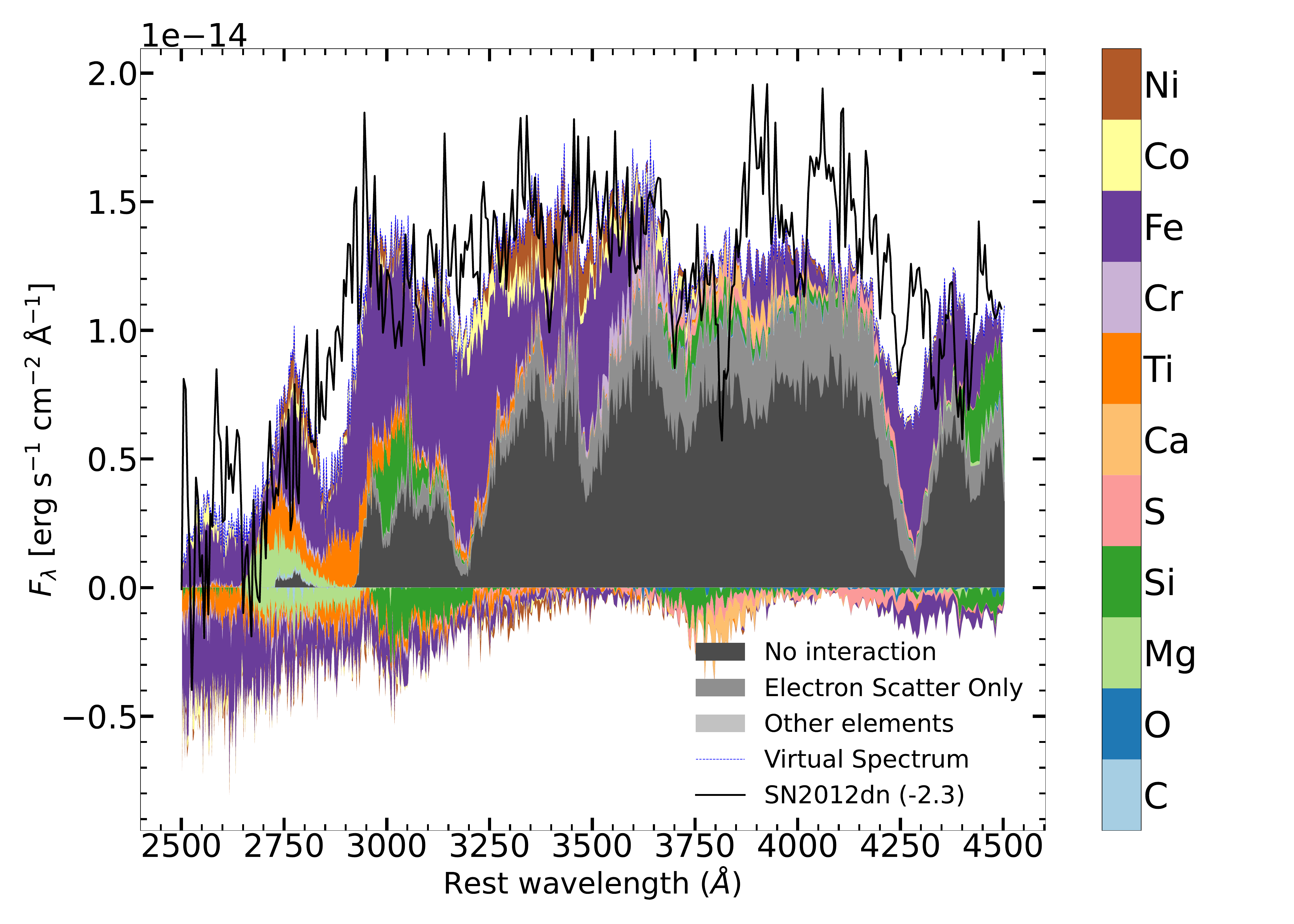}\\
\includegraphics[width=0.32\columnwidth]{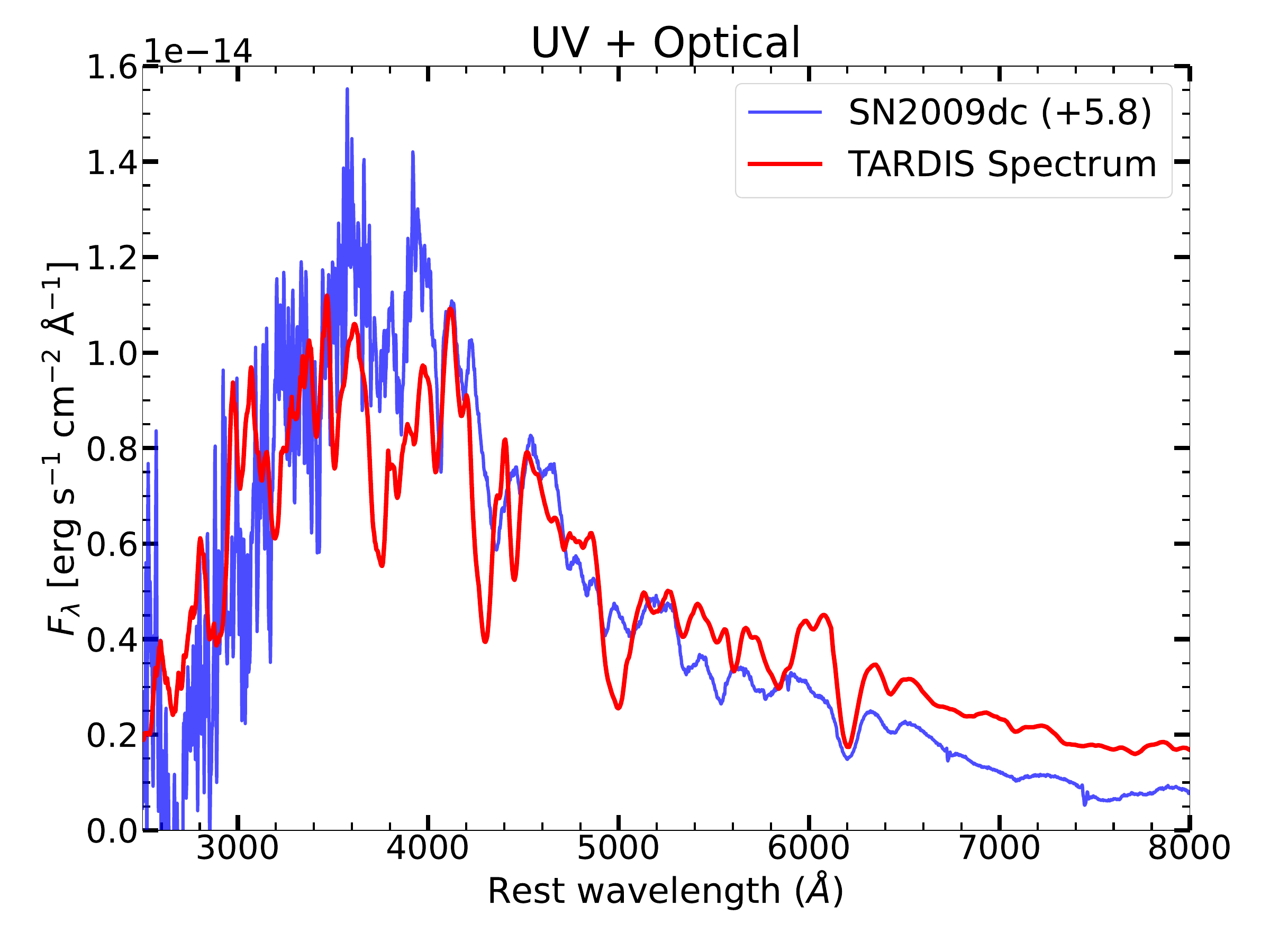}
\includegraphics[width=0.32\columnwidth]{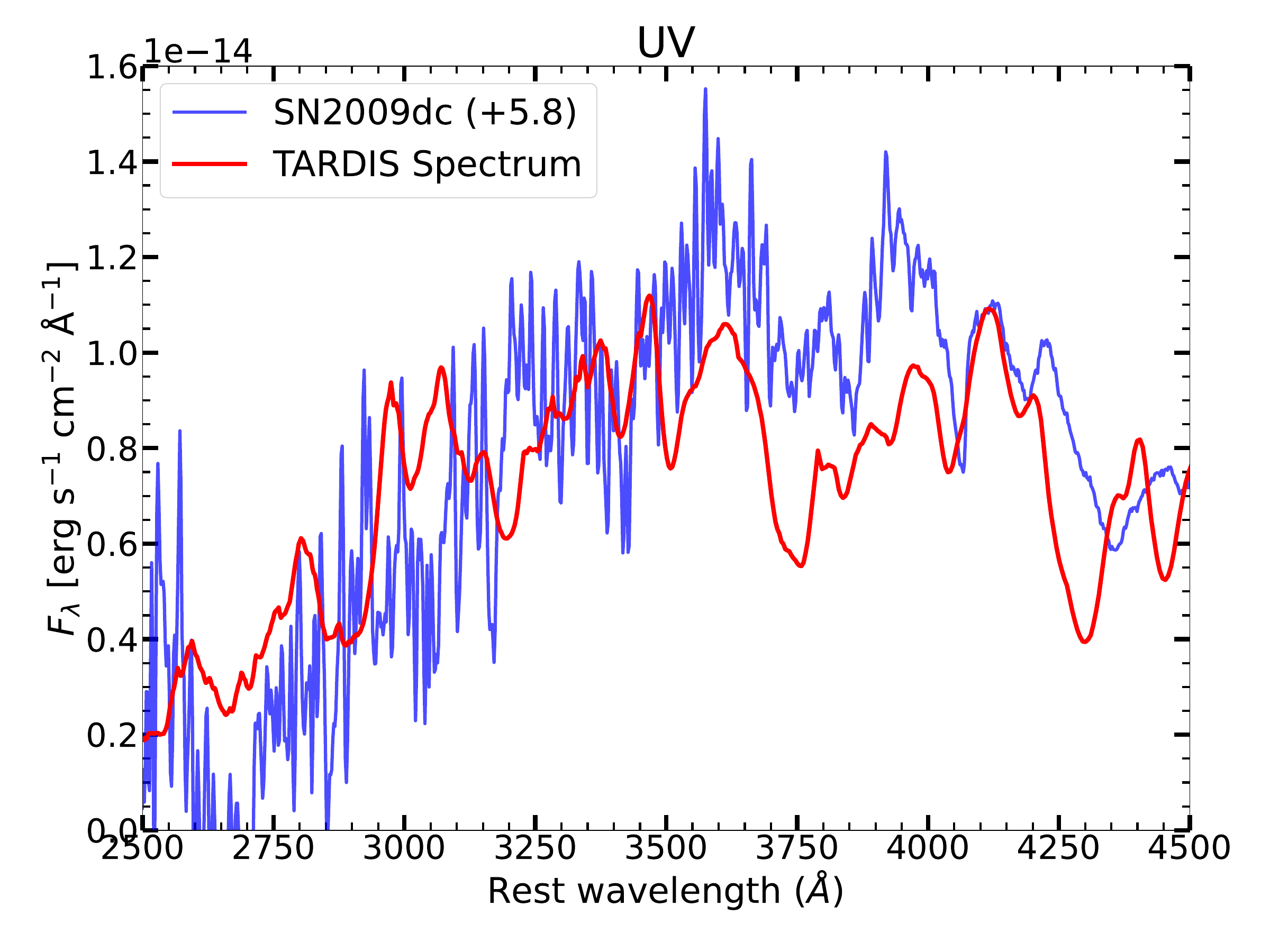}
\includegraphics[width=0.34\columnwidth]{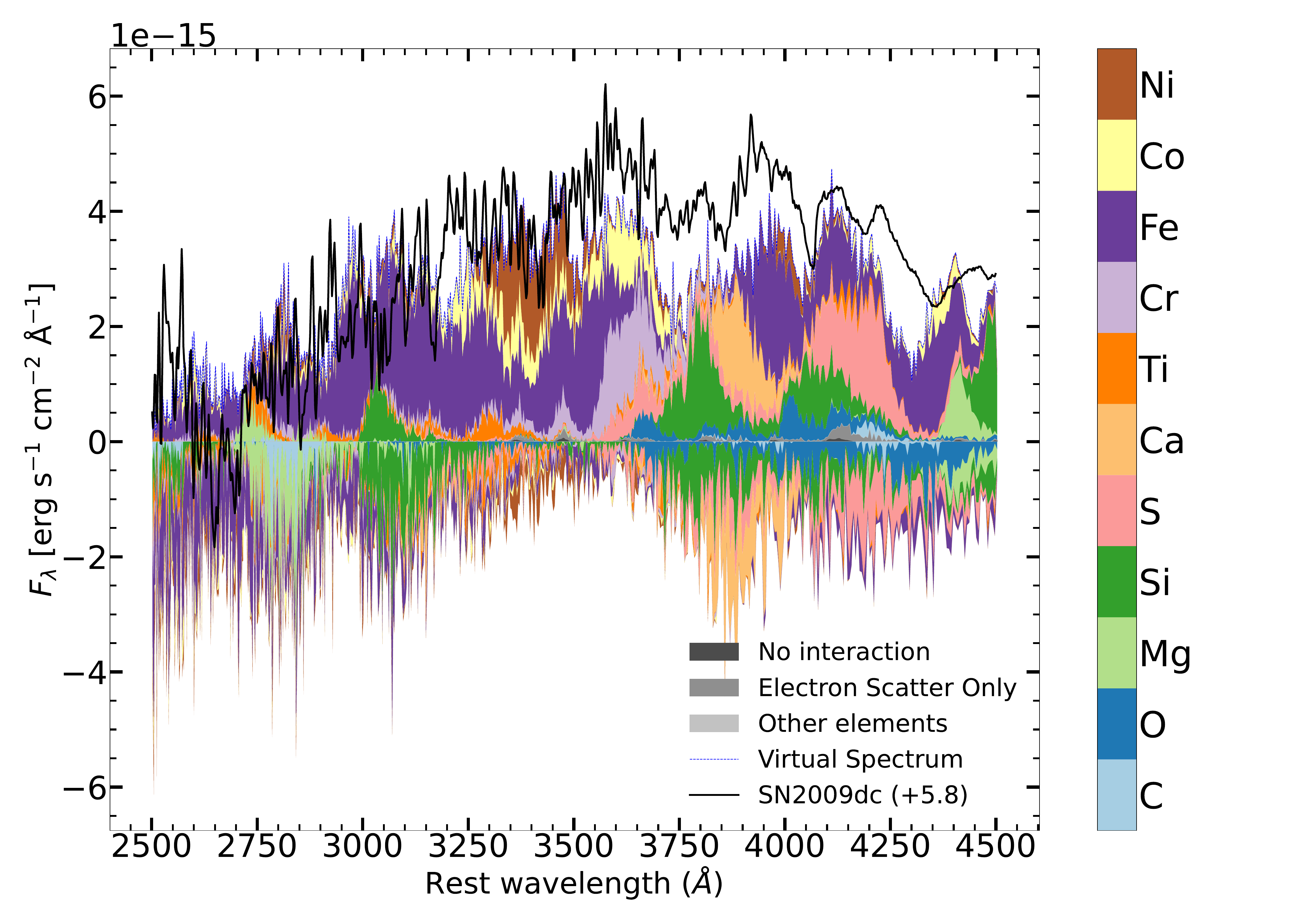}
\includegraphics[width=0.32\columnwidth]{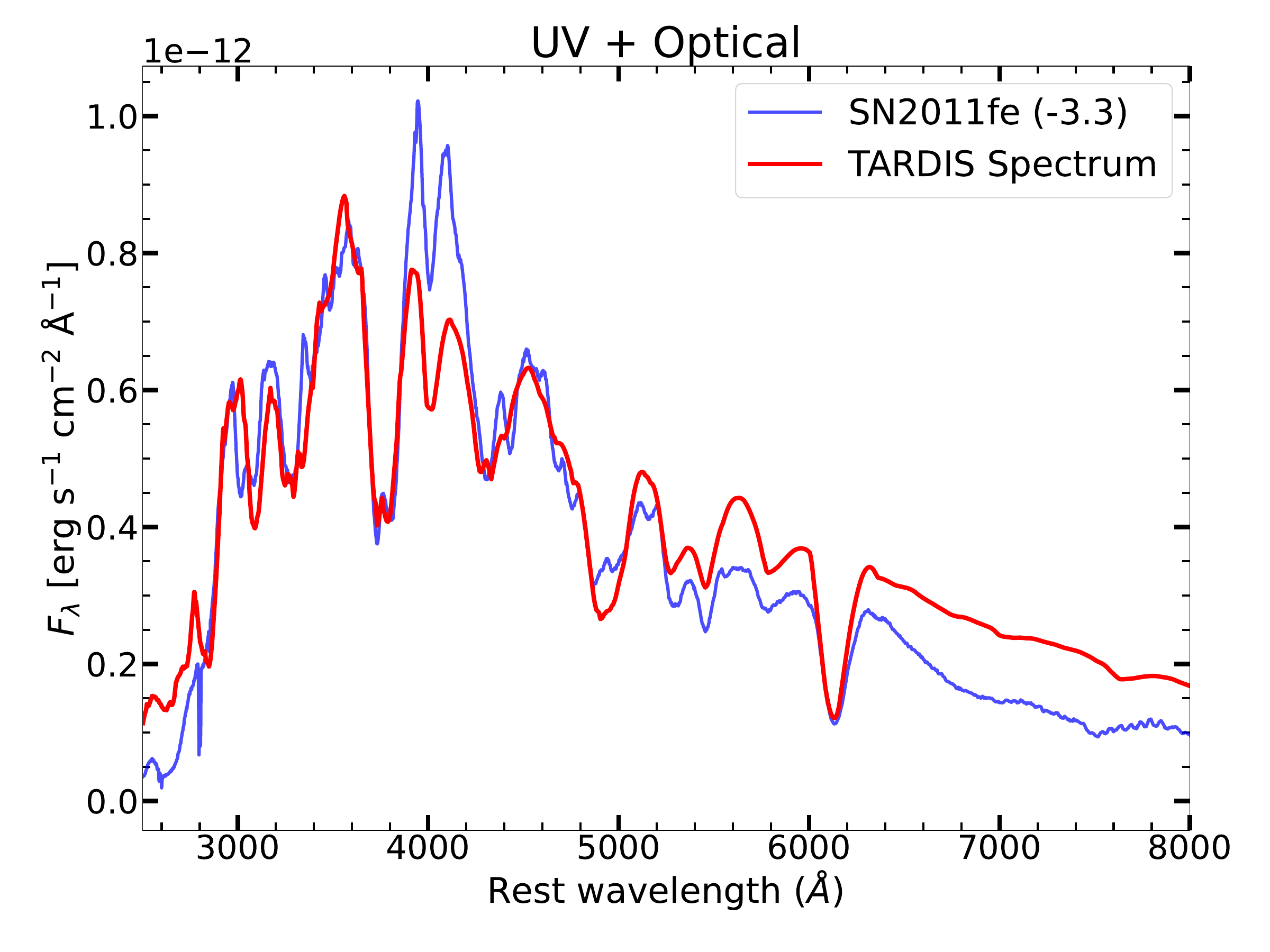}
\includegraphics[width=0.32\columnwidth]{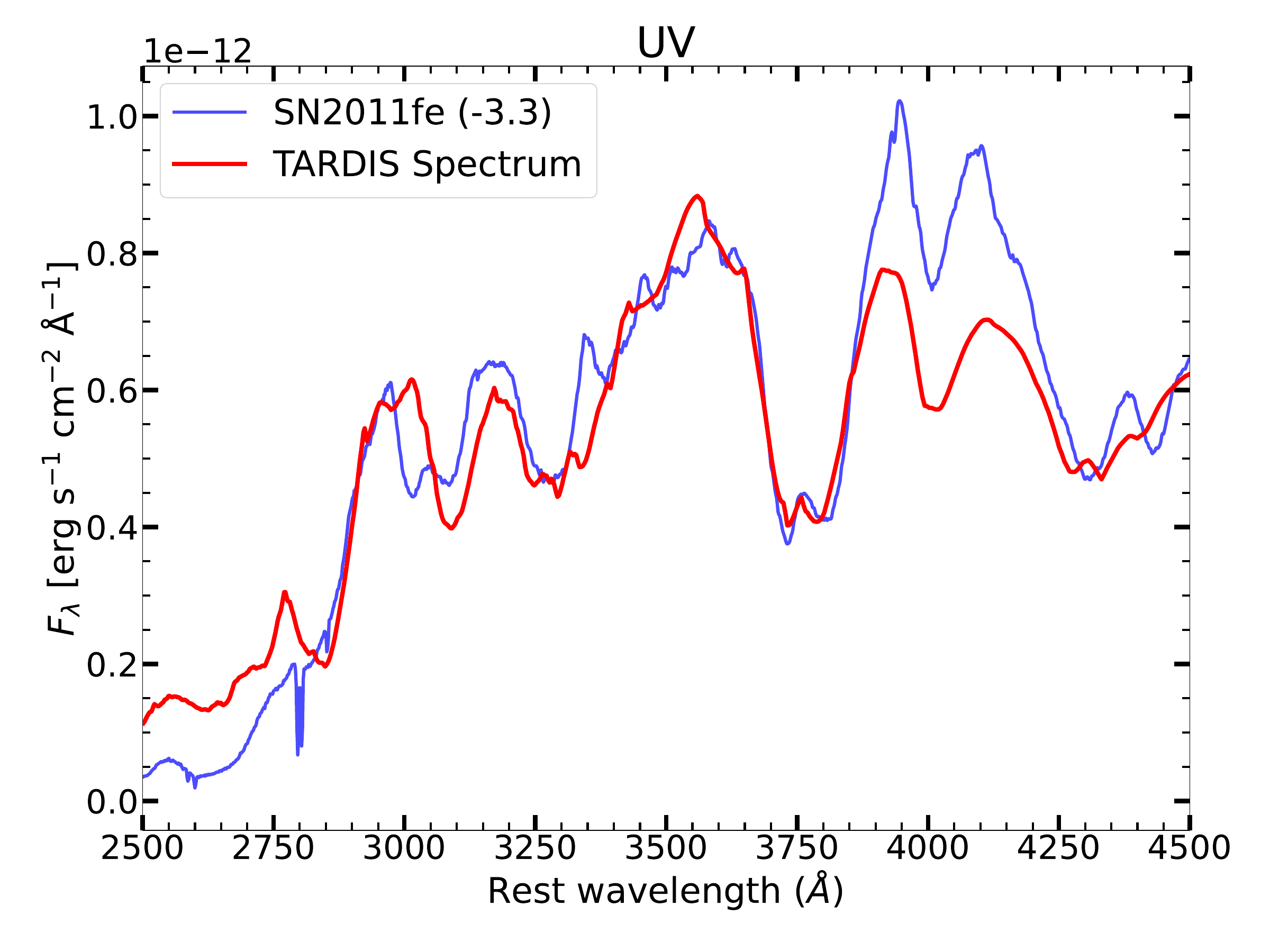}
\includegraphics[width=0.34\columnwidth]{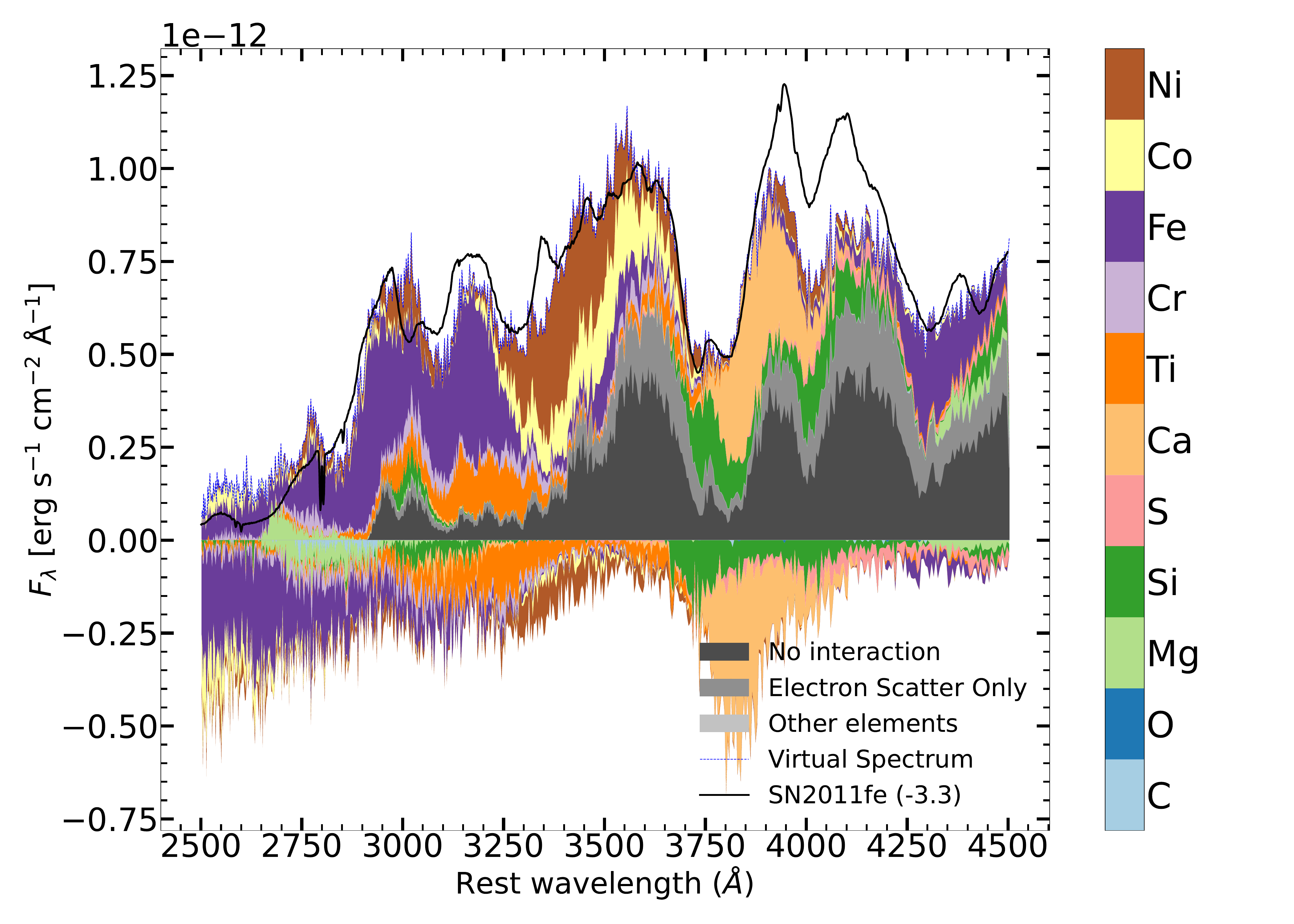}

\caption{\textit{Top row}: The \tardis\ fit to the spectrum of SN~2020hvf at $-3.3$\,d, covering the UV to optical wavelengths (left panel) and a zoom-in of the UV region (middle panel). The \tardis\ fit and observed spectrum are shown in red and blue, respectively. The right panel shows the \tardis\ spectral element decomposition for the UV region, with the colormap indicating the elements used in the fit. The dark gray region represents the contributions from the photons that do not undergo any interactions. The gray region shows the contributions from photons interacting via electron scattering, and the light gray region shows the contributions from elements not included in the fit. 
\textit{Second row}: The same as top panels, but for the spectrum of SN~2012dn at $-2.3$\,d. \textit{Third row}: The same as top panels, but for the spectrum of SN~2009dc at $+5.8$\,d. \textit{Bottom row}: The same as top panels, but for the spectrum of a spectroscopically normal SN~2011fe at $-3.3$\,d.}
\label{tardis:fits}
\end{figure*}

We combine each {\it Swift} UV spectrum with an optical spectrum of similar phase, obtained from the WISeREP database \citep{yaron2012}, and fit the resulting combined spectrum using \tardis. Since we focus on identifying the UV spectral features of 03fg-like SNe, a random scaling factor is applied to the \tardis\ spectrum to match the continuum of the observed spectrum (the UV wavelength in particular). Thus, direct comparison of the modeled and observed continuum flux levels would not be meaningful here. Figure~\ref{tardis:fits} shows our spectral modeling results for SN~2020hvf ($-$3.3\,d), SN~2012dn ($-$2.3\,d), and SN~2009dc ($+$5.8\,d). We do not show the fitting result for SN~2020hvf at $-$10.4\,d as we could not obtain a reasonable fit. This may be due to the simplified assumptions of \tardis\ fitting, which makes it challenging to fit our target at such an early phase. We did not perform \tardis\ fitting for SN~2011aa at $+$8.1\,d either, as the spectrum is noisy and without significant UV features.

The top panels of Figure~\ref{tardis:fits} show results of \tardis\ fitting for SN~2020hvf at $-$3.3\,d. Overall, \tardis\ successfully reproduces the UV spectral features of 03fg-like SNe. Although the overall spectral shapes of the model and observation do not match, our focus is primarily on matching the spectral features rather than the continuum, particularly in the UV region. The absorption feature between 3750\,\AA\ and 4000\,\AA\ is primarily attributed to Ca. The flux between 3000\,\AA\ and 3500\,\AA\ is mainly contributed by iron-group elements (IGEs), mostly Co and Fe. This aligns with the previous findings that the flux at this wavelength correlates with the decline rate of SN~Ia and, therefore, likely with Ni production \citep[e.g.,][]{foley2016,pan2020}. Additionally, the features between 2500\,\AA\ and 3000\,\AA\ are likely due to Mg and Ti. Our line identifications with \tardis\ are generally consistent with the previous spectral modeling of 03fg-like SNe \cite[e.g.,][]{hachinger2012}. 

Figure~\ref{tardis:fits} also shows the fitting results for SN~2012dn at $-$2.3\,d and SN~2009dc at $+$5.8\,d. Although the data quality for these spectra is lower than that for SN~2020hvf, both exhibit similar distributions of line species in the UV. The spectrum of SN~2012dn at $-$2.3\,d likely shows a greater contribution from Fe below 3750\,\AA\ compared to SN~2020hvf at a similar phase. However, the contribution of Ca between 3750\,\AA\ and 4000\,\AA\ appears weaker, possibly due to poor fitting in this region. Additionally, the UV spectrum of SN~2009dc, observed at a considerably later phase, shows stronger contributions from Si and S between 3750\,\AA\ and 4250\,\AA, along the IGEs.

For comparison, we perform the \tardis\ fitting on SN~2011fe (the bottom panels of Figure~\ref{tardis:fits}), a spectroscopically normal SN~Ia, using a spectrum at $-$3.3\,d. Our results indicate that 03fg-like SNe and SN~2011fe generally exhibit similar UV line species, although with different abundances. Notably, Mg and Ti contribute less between 2500\,\AA\ and 3000\,\AA\ in SN~2011fe compared to 03fg-like SNe. Furthermore, we attempt to perform \tardis\ fitting on SN~2011fe using the same setup as for 03fg-like SNe but find that it tends to produce weaker features of intermediate-mass elements (IMEs), such as Ca and Si. Additionally, it overestimates the UV flux and introduces extra features below 3000\,\AA.

\begin{table}
    \centering
    \caption{\tardis\ settings adopted in this work}
    \label{tab:tardis}
    \begin{tabular}{ll}
        \hline
        Configuration       & Value            \\
        \hline
        Disable Electron Scattering  & \texttt{no} \\
        Density                & \texttt{branch85\_w7}\\
        Abundances Type            & \texttt{uniform}\\
        Ionization             & \texttt{lte} \\
        Excitation             & \texttt{lte} \\
        Radiative Rate         & \texttt{detailed} \\
        Line Interaction       & \texttt{macroatom} \\
        \hline
    \end{tabular}
\end{table}

\section{Discussion}\label{sec:discussion}

\begin{figure*}
\centering
\includegraphics[scale=0.14]{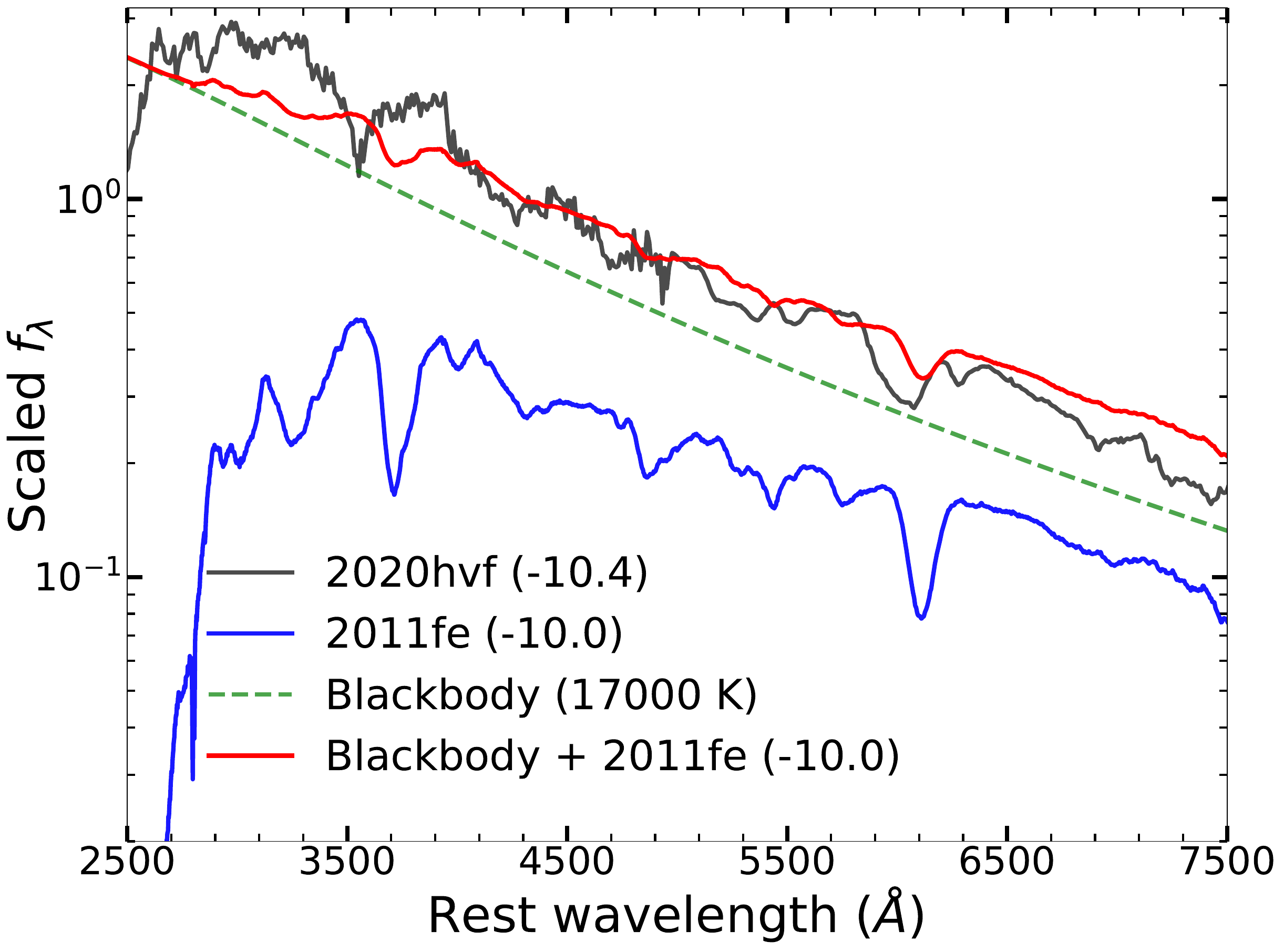}
\includegraphics[scale=0.14]{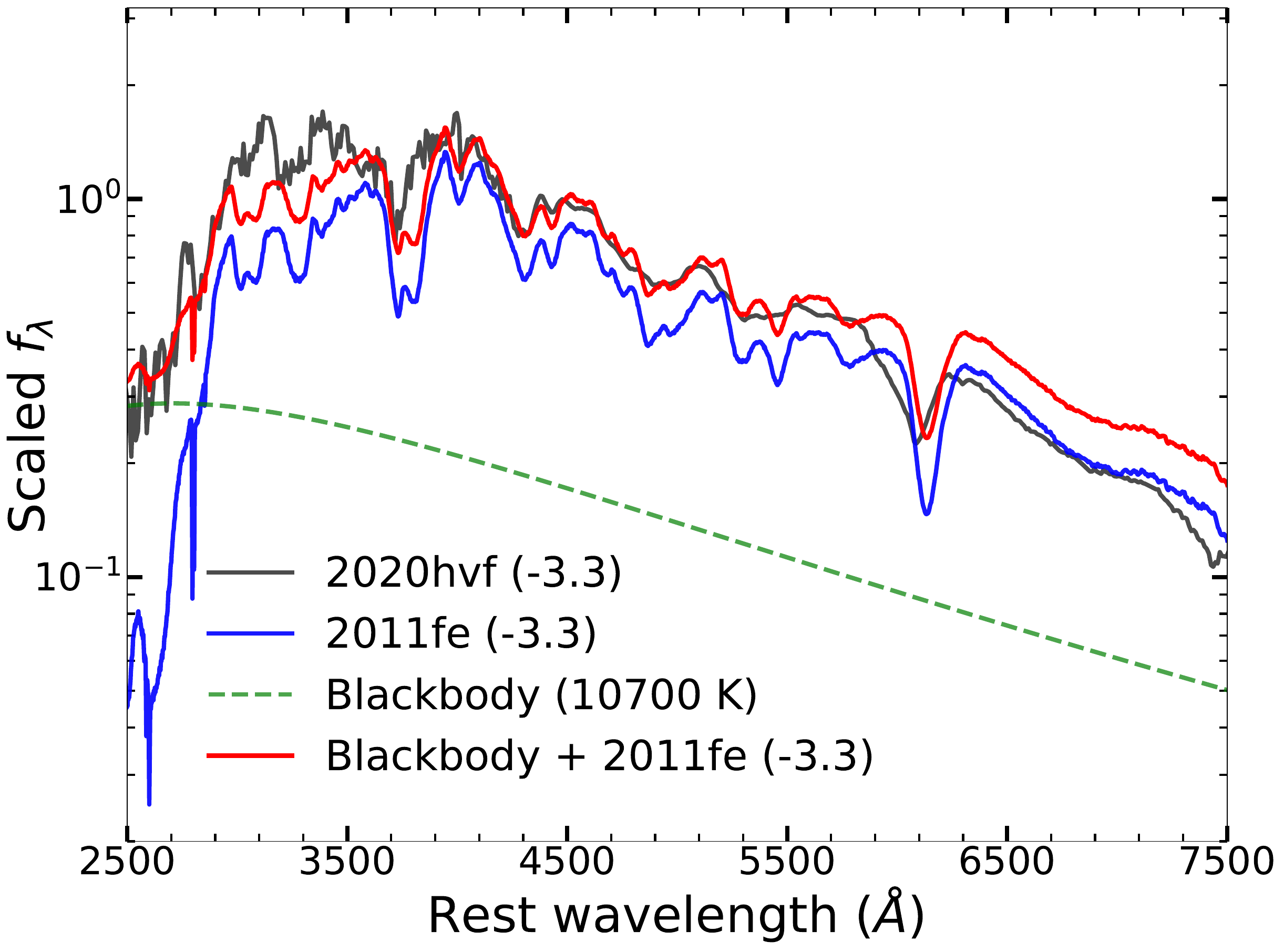}
\includegraphics[scale=0.14]{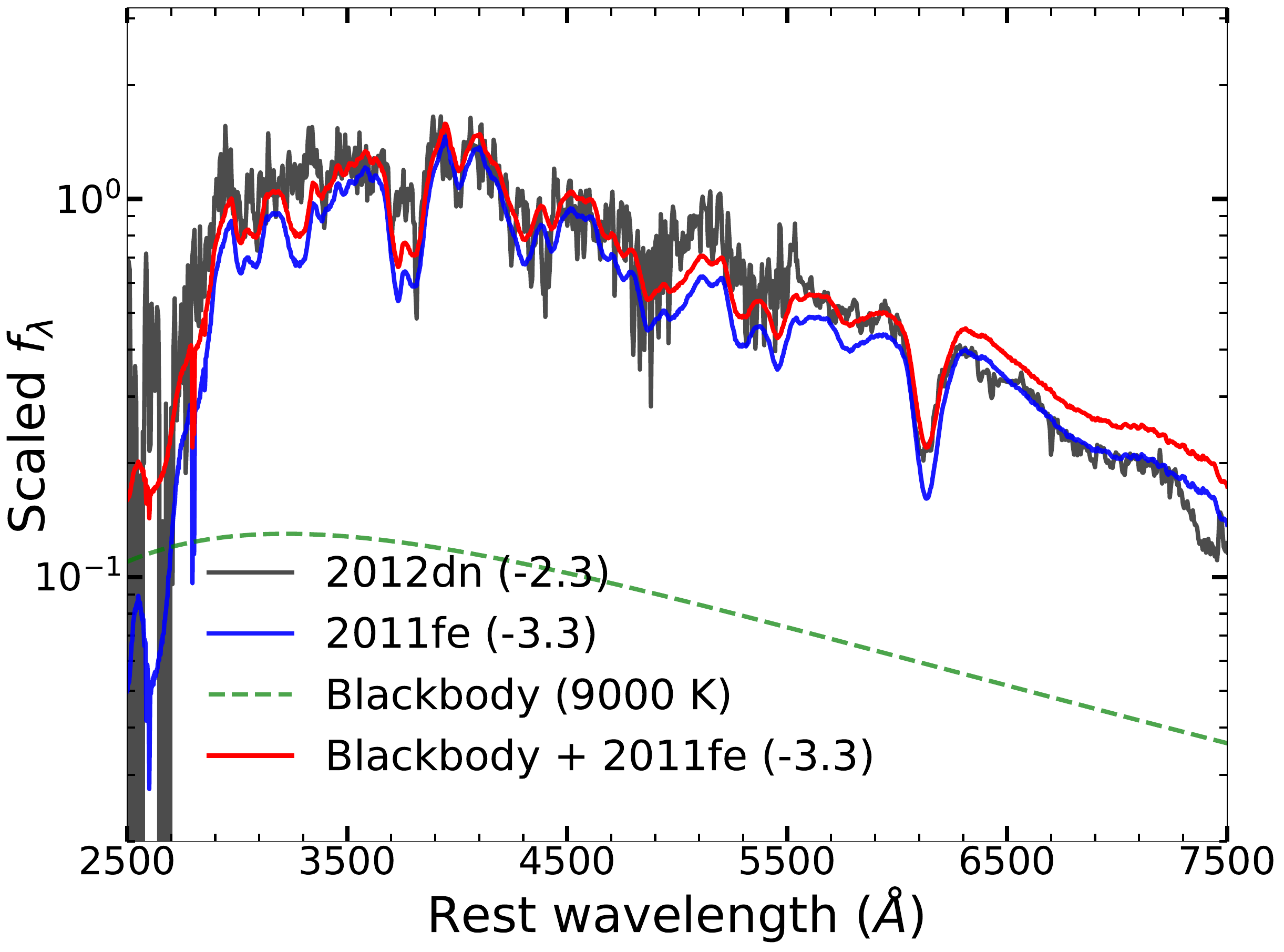}

\caption{\textit{Left}: The UV-through-optical spectrum of SN~2020hvf at $-10.4$\,d (shown in black) is compared (in red) with a combination of the spectrum of a normal SN~Ia, SN~2011fe, at $-10.0$\,d (in blue) and a blackbody spectrum of $T=17,000$\,K (in green). \textit{Middle}: The same as the left panel but for the spectrum of SN~2020hvf at $-3.3$\,d, using the spectrum of SN~2011fe at $-3.3$\,d and a blackbody spectrum of $T=10700$\,K. \textit{Right}: The same as the left panel but for the spectrum of SN~2012dn at $-2.3$\,d, using the spectrum of SN~2011fe at $-3.3$\,d and a blackbody spectrum of $T=9000$\,K.}
\label{bb}
\end{figure*}

03fg-like SNe is a peculiar subclass of SNe~Ia, characterized by their overluminous and broader optical light curves compared to the normal SNe~Ia \citep[e.g.,][]{brown2014,taubenberger2017,ashall2021}. These objects were generally thought to originate from explosions of super-$\rm M_{ch}$ WDs, given the observational evidence that they are likely to originate from a progenitor with a mass exceeding the Chandrasekhar limit \citep[e.g.,][]{howell2006,tanaka2010,scalzo2010,taubenberger2011}. Previous studies also suggested that 03fg-like SNe tend to be found in a unique host-galaxy environment, e.g., galaxies of low stellar masses, low metallicities, and high specific star-formation rates. \citep{howell2006,childress2011,khan2011,taubenberger2011,chakradhari2014,hsiao2020,ashall2021,lu2021}. Recent studies have also shown that these objects have unique UV properties: They appear to be bluer in UV and have a different UV color evolution at early times compared to those of normal SNe~Ia \citep[e.g.,][]{brown2014,hoogendam2024}.

\citet{pan2020} reported a mild correlation between the UV flux ratios and the host-galaxy metallicities, with SNe~Ia residing in more metal-poor galaxies tending to show higher UV flux levels. This motivated us to further investigate if the host-galaxy environment could drive the UV excess found for 03fg-like SNe. By comparing the UV spectra of 03fg-like SNe with those of normal SNe~Ia, we confirm that the 03fg-like SNe tend to be bluer in the UV than the majority of normal SNe~Ia, regardless of their phases. As seen in Figure~\ref{cdf} and \ref{ratio:2550_3025:dm15}, the 03fg-like SNe generally have higher UV flux ratios (for both $f_{2550}$ and $f_{3025}$) than the normal SNe~Ia, with the $f_{3025}$ of SN~2012dn being an exception. While SN~2012dn was classified as a 03fg-like SN, it has the largest $\Delta m_{15}$ (0.97) among all the 03fg-like SNe in our sample. It is probably not surprising that SN~2012dn exhibits relatively low $f_{3025}$, given that $f_{3025}$ is believed to be inversely proportional to the $\Delta m_{15}$ \citep{foley2016,pan2020}. 

This trend is also evident when comparing the UV spectra of 03fg-like SNe with the mean spectra of normal SNe~Ia near the peak luminosity. Examination of near-peak UV spectra reveals that SN~2012dn and SN~2020hvf show significant UV excess below $\sim4000$\,\AA, in contrast to the near-peak mean spectra of normal SNe~Ia with comparable $\Delta m_{15}$ and host-galaxy properties (e.g., low-mass galaxies). Moreover, this UV excess cannot be well described by the spectral templates of normal SNe~Ia at least for the parameter space (with respect to both SN and host parameters) investigated by \citet{pan2020}. Our spectral comparison and modeling also indicate that 03fg-like SNe share a similar spectral element composition with normal SNe~Ia in the UV, with flux levels being the major distinguishing factor. These results rule out the possibility that the host-galaxy environments of 03fg-like SNe~Ia primarily drive their UV excess.

The differences in methods used to measure the host-galaxy \( E(B-V)\) between 03fg-like SNe and the comparison sample are unlikely to affect this trend. If the host-galaxy reddening of our 03fg-like SNe is systematically underestimated, the difference in UV flux ratios would likely be even greater. Conversely, our conclusions remain robust even if the host-galaxy reddening for the 03fg-like SNe is systematically overestimated. To test this, we assume zero host-galaxy reddening for all the 03fg-like SNe and still find the trend to be significant. Our conclusions remain robust regardless of the models used for \texttt{SNooPy} LC fitting or the reddening laws applied to de-redden the spectra. The mean difference in flux ratios determined using the CCM and \citet{fm07} reddening laws is only $\sim0.01$, with a root-mean-square (RMS) scatter of $\sim0.07$. To assess the impact of different \texttt{SNooPy} fitting models on the host-galaxy \( E(B-V)\), we apply the \texttt{EBV2\_model} and find that the values remain consistent up to two decimal places. However, the trends observed in this study could be affected if the $R_{v}$ of 03fg-like SNe differs systematically from that of normal SNe~Ia. This possibility has not been explored in previous studies and is beyond the scope of this work.

Several scenarios have been proposed to explain the unique characteristics of the 03fg-like SNe, including a merger of two WDs with a total mass exceeding $\rm M_{ch}$ \citep{howell2006,scalzo2010,dimitriadis2023,ohora2024,kwok2024}, and an explosion of a super-$\rm M_{ch}$ WD supported by the fast rotation or magnetic fields \citep{yoon2005}. More recent studies suggested that some double degenerate scenarios, such as the merger of a CO WD with the core of an asymptotic giant branch (AGB) star, or a CO WD explosion within the carbon-enriched CSM (e.g., from the disrupted secondary CO WD) could be promising in explaining the bumps and UV excess seen in the early-time light curves and asymmetric explosions inferred by the spectral features in the nebular phase \citep[e.g.,][]{noebauer2016,hsiao2020,lu2021,ashall2021,jiang2021,dimitriadis2022,hoogendam2024,siebert2024}.

To explore if CSM interaction could account for the UV excess observed in 03fg-like SNe \citep[e.g.,][]{hachinger2012,siebert2024}, we add a blackbody spectrum to the spectrum of a normal SN~Ia to determine if this combination could reproduce the continua of the 03fg-like SN spectra. Here, the scaling parameters of normal SN~Ia and blackbody components are adjusted based on the visual inspection. Figure~\ref{bb} demonstrates that the UV-through-optical continua of SN~2020hvf at $-10.4$\,d and $-3.3$\,d can be reasonably described by a combination of SN~2011fe spectra at the same phases and blackbody spectra with temperatures of $T=17000$\,K and $T=10700$\,K, respectively. There is also a clear trend that the blackbody temperature decreases with time. We apply the same analysis to the spectrum of SN~2012dn at $-2.3$\,d, using a combination of the SN~2011fe spectrum and a blackbody spectrum with $T=9000$\,K. Since the UV excess of SN~2012dn is less significant, we obtain a blackbody component of lower temperature compared to that of SN~2020hvf at the same phase. These tests suggest that an additional blackbody radiation component could explain the UV excess observed in 03fg-like SNe. Further studies with a larger sample will be crucial to place more robust constraints on this hypothesis.

\section{Conclusions}\label{sec:conclude}

In this work, we study a sample of 03fg-like SNe~Ia using the \textit{Swift} UVOT Grism observations. Our sample comprises five UV spectra from four 03fg-like SNe: SNe~2009dc, 2011aa, 2012dn, and 2020hvf. We compare their UV spectra with a large sample of spectroscopically normal SNe~Ia. Our main findings are summarized below.

\begin{enumerate}
\item 03fg-like SNe~Ia are significantly bluer in the UV than the majority of normal SNe~Ia, regardless of the phase (at least until a week after the peak luminosity). Earlier epochs generally have higher UV flux levels than those of later epochs.

\item Using UV flux ratios at around 2550\,\AA and 3025\,\AA, 03fg-like SNe have significantly higher flux ratios than those of normal SNe~Ia, even compared with the SNe~Ia of similar $\Delta m_{15}$. Similar trends are found when compared with the mean spectra and spectral templates of normal SNe~Ia of similar phases, $\Delta m_{15}$, host-galaxy stellar mass, and metallicity (converted from stellar mass). 

\item The spectral comparison and modeling suggest that 03fg-like SNe~Ia and normal SNe~Ia could exhibit similar spectral element composition in the UV, with the primary difference being their UV flux levels.

\item Although the UV excess of normal SNe~Ia could potentially be attributed to the host-galaxy environment (e.g., low metallicity), we find it challenging to explain the UV properties of 03fg-like SNe with this explanation.

\item The UV excess observed in 03fg-like SNe~Ia can be explained by adding a hot blackbody component onto a normal SN~Ia spectrum. This supports the hypothesis that CSM interaction accounts for the unique UV properties of 03fg-like SNe~Ia. Future investigations with a larger sample of 03fg-like SNe~Ia will be crucial to shed light on their mysterious origins.

\end{enumerate}

\section*{ACKNOWLEDGEMENTS}
YCP is supported by the National Science and Technology Council (NSTC grant 112-2112-M-008-026-MY3). SB thanks Ryan Foley for valuable feedback and Paul Kuin for his assistance in installing and understanding the \texttt{UVOTPY} software. SB also thanks SNe~Ia for always being reliable cosmic candles, even when our deadlines are unpredictable. HYM acknowledges financial support from Agència de Gestió d’Ajuts Universitaris i de Recerca (AGAUR) under project PID2023-151307NB-I00 and 2021-SGR-01270, Consejo Superior de Investigaciones Científicas (CSIC) under PIE project 20215AT016, program JAEPRE23-18, and program Unidad de Excelencia Maria de Maeztu CEX2020-001058-M. This publication has made use of data collected at Lulin Observatory, partly supported by NSTC grant 113-2740-M-008-005.  C.D.K. gratefully acknowledges support from the NSF through AST-2432037, the HST Guest Observer Program through HST-SNAP-17070 and HST-GO-17706, and from JWST Archival Research through JWST-AR-6241 and JWST-AR-5441. 

This material is based upon work supported by the National Science Foundation Graduate Research Fellowship Program under Grant Nos. 1842402 and 2236415. Any opinions, findings, conclusions, or recommendations expressed in this material are those of the author(s) and do not necessarily reflect the views of the National Science Foundation.

\section*{DATA AVAILABILITY}
The UV spectra used in this study are available on GitHub.\footnote{\url{https://github.com/Bilton6/Swift_UV_spectra}}

\bibliographystyle{mnras}
\bibliography{example}

\appendix

\section{Cross-cuts and Host Images} \label{cross:cut}

\begin{figure*}
\centering
\includegraphics[width=0.3\columnwidth]{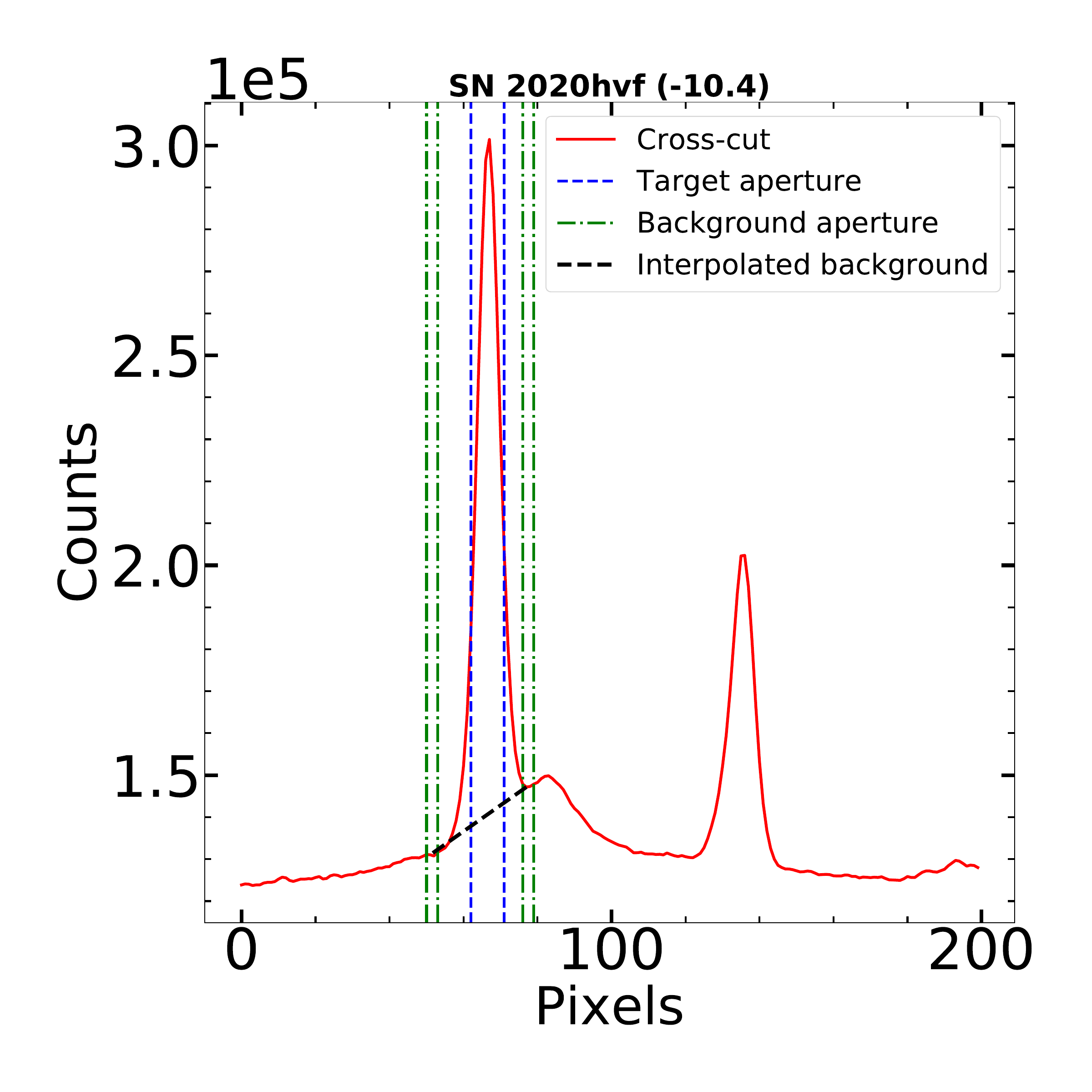}
\includegraphics[width=0.3\columnwidth]{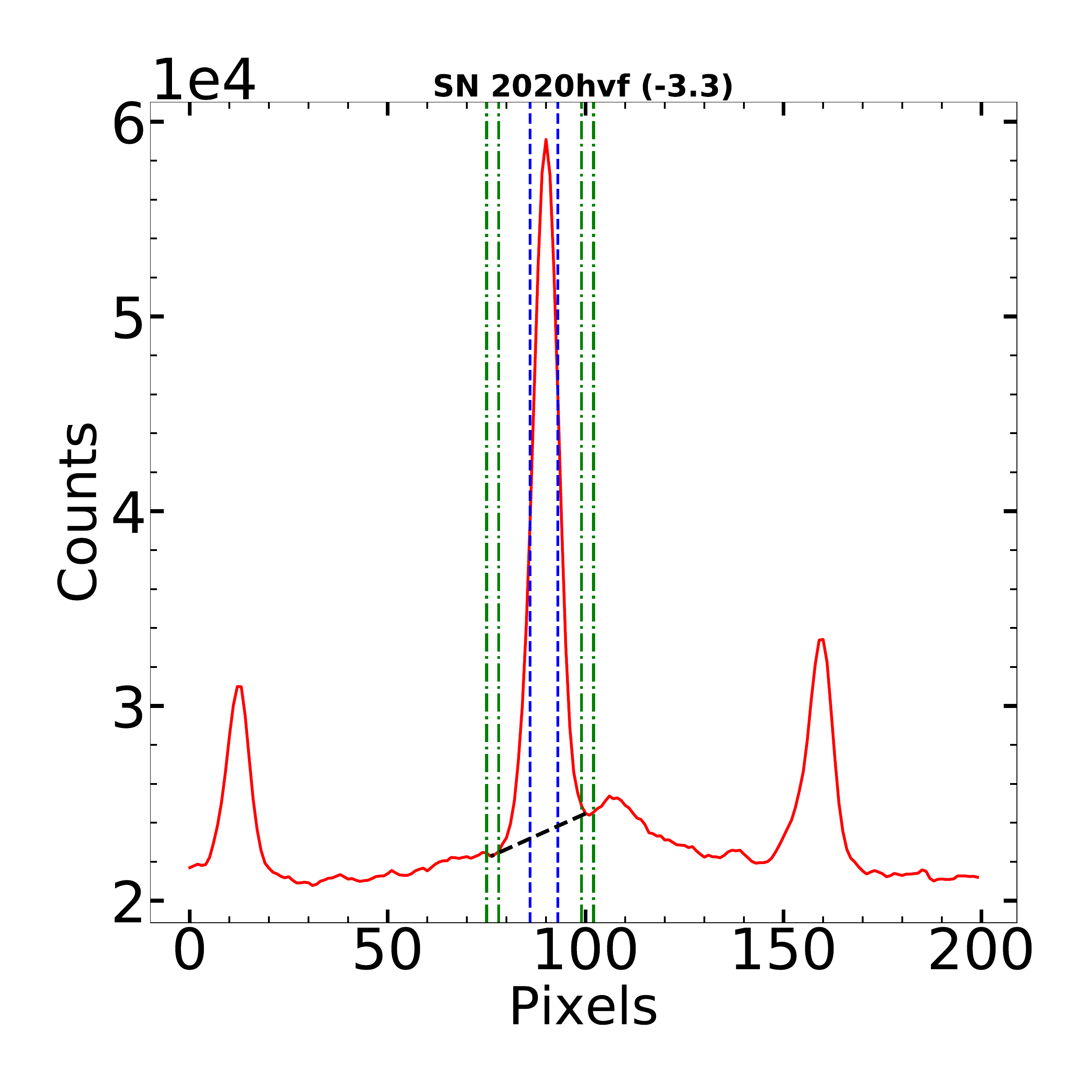}
\includegraphics[width=0.285\columnwidth]{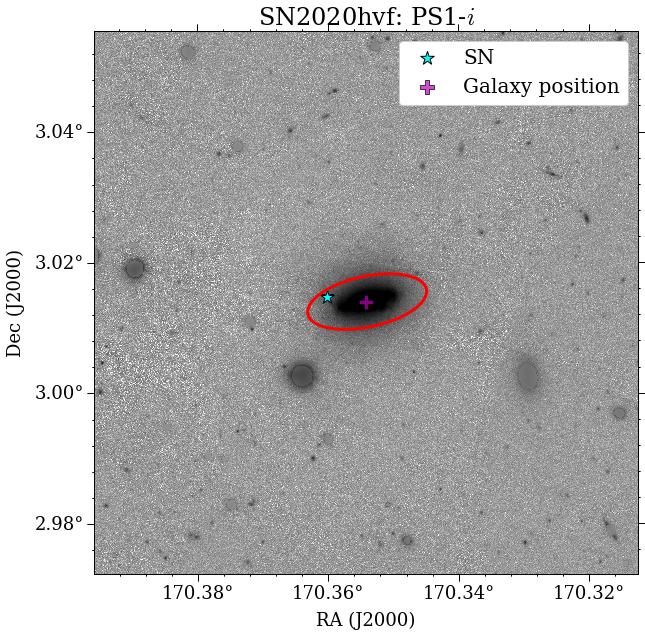}\\
\includegraphics[width=0.31\columnwidth]{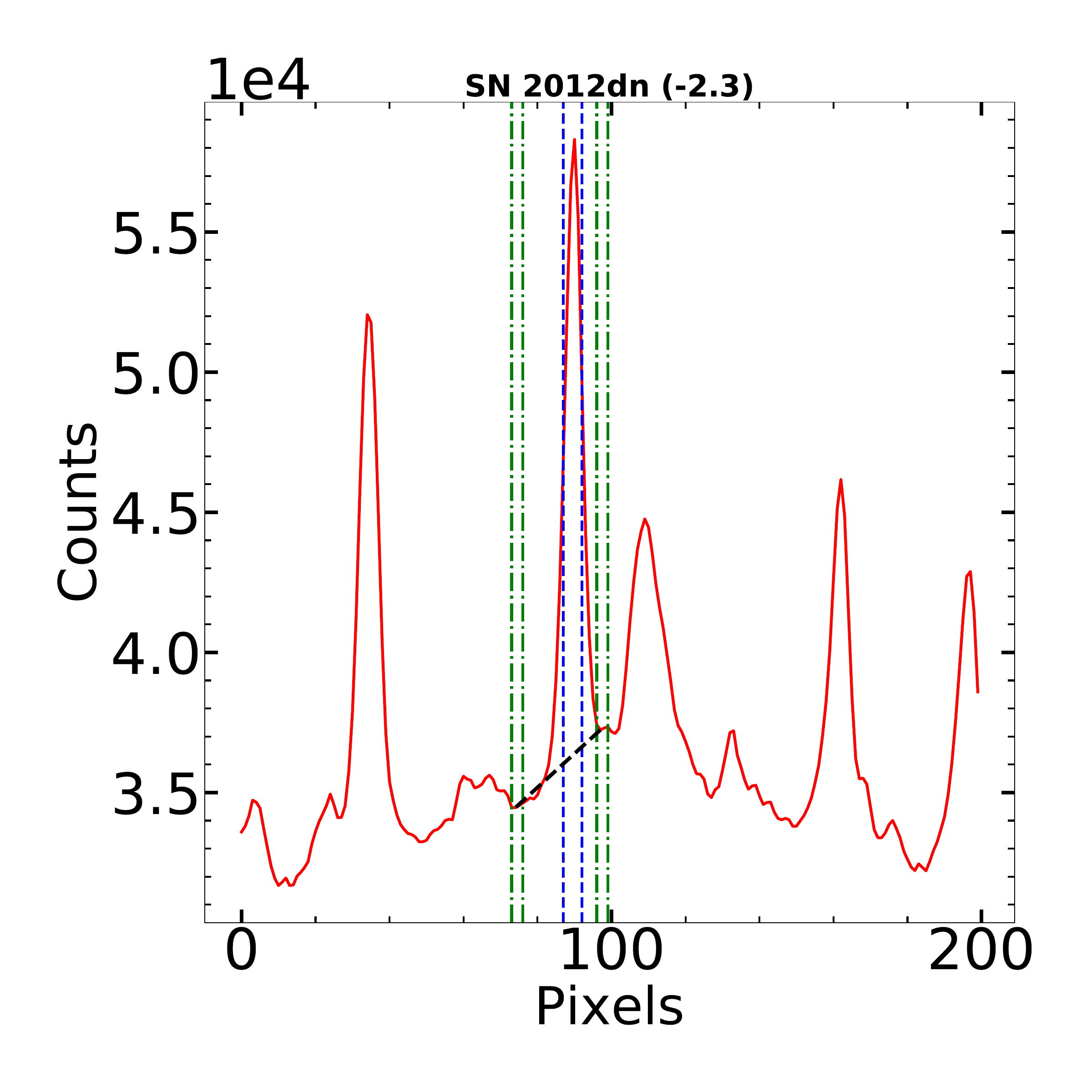}
\includegraphics[width=0.30\columnwidth]{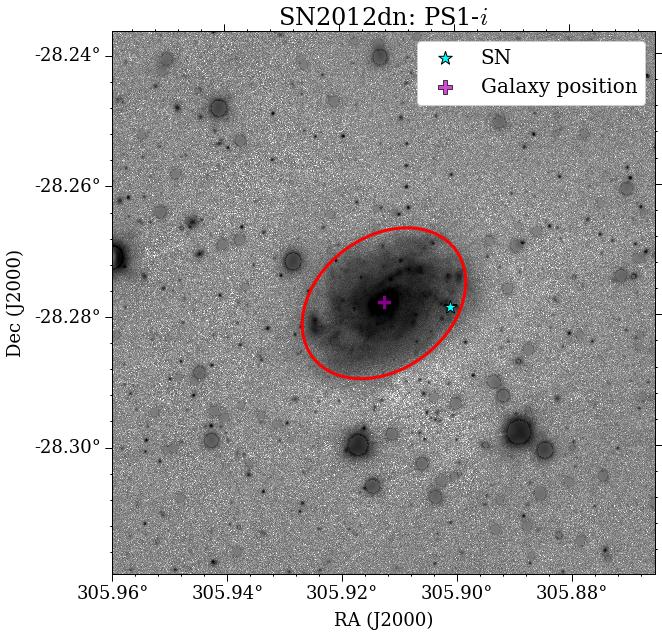}\\
\includegraphics[width=0.31\columnwidth]{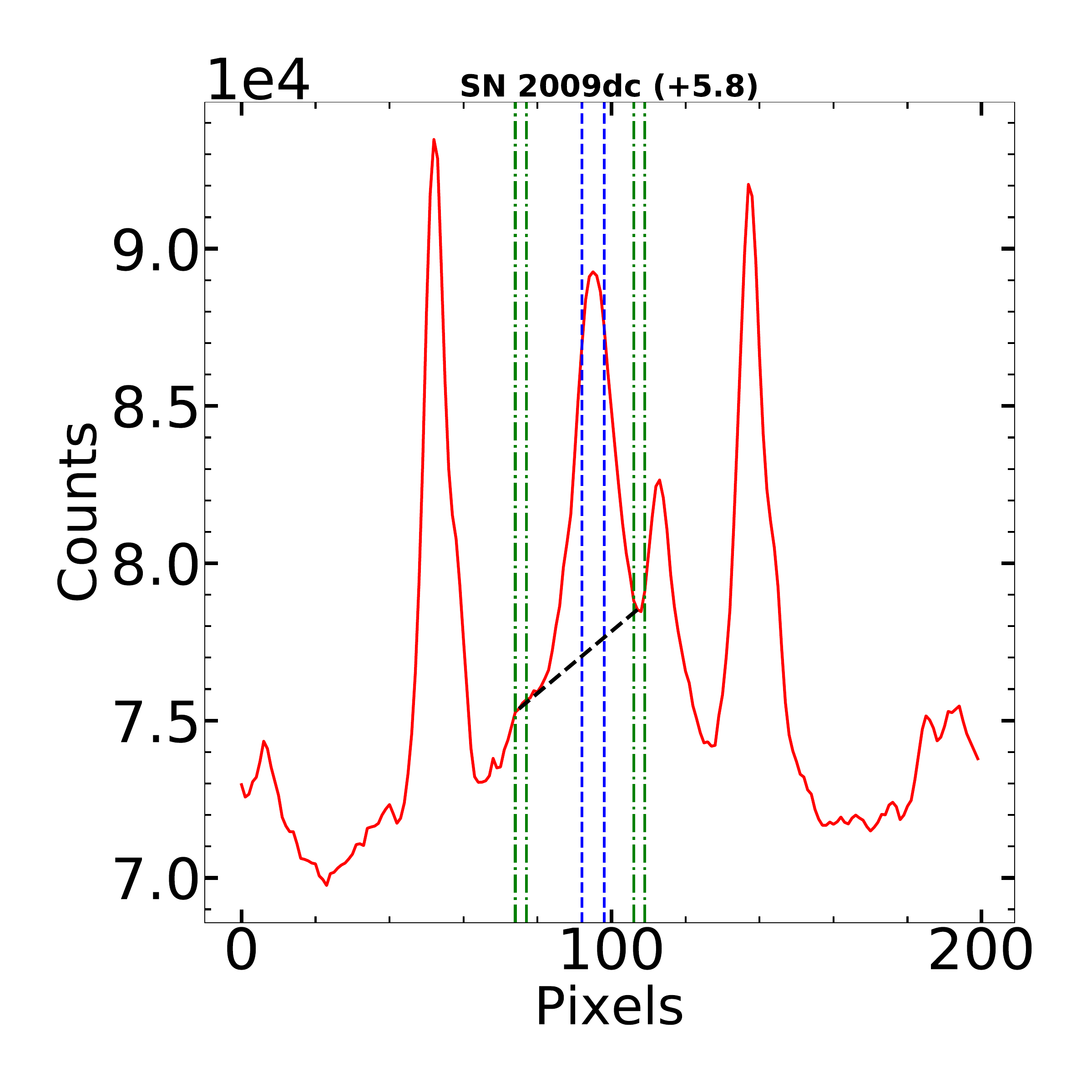}
\includegraphics[width=0.30\columnwidth]{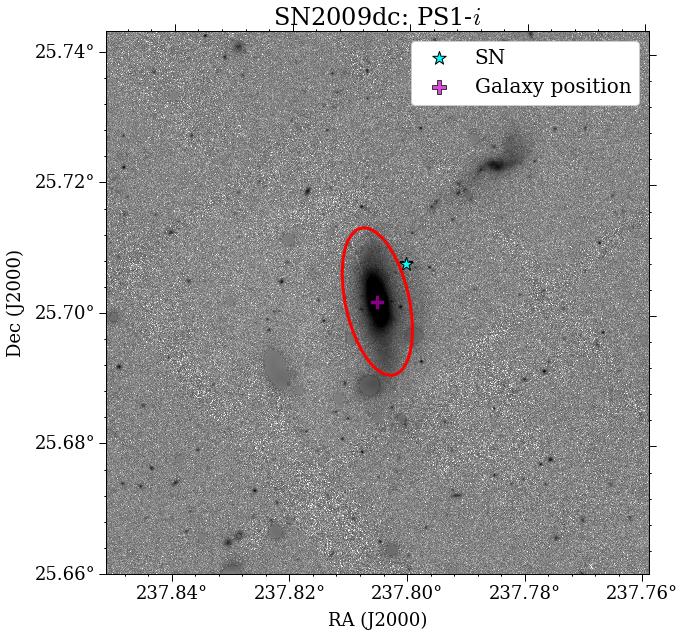}\\
\includegraphics[width=0.31\columnwidth]{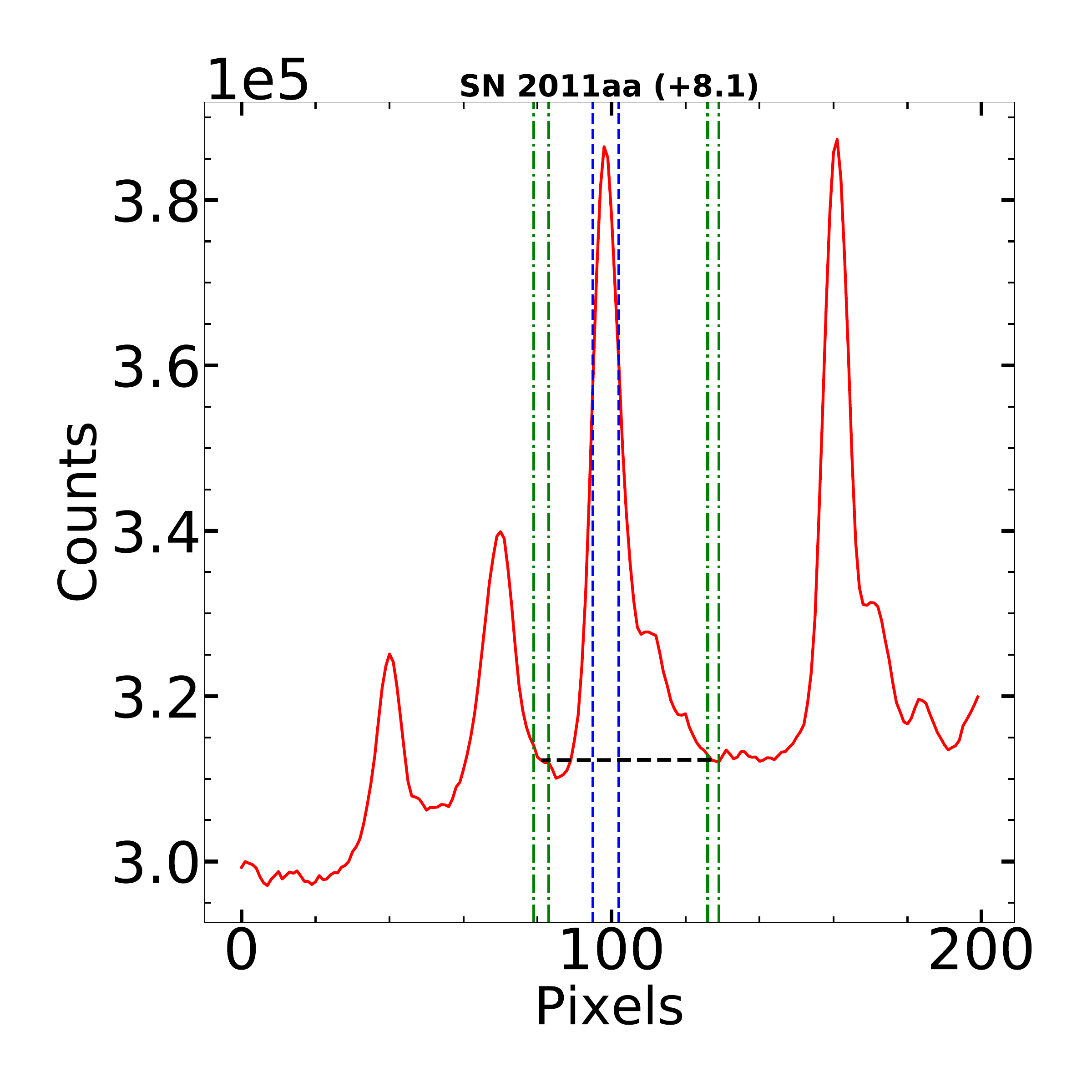}
\includegraphics[width=0.30\columnwidth]{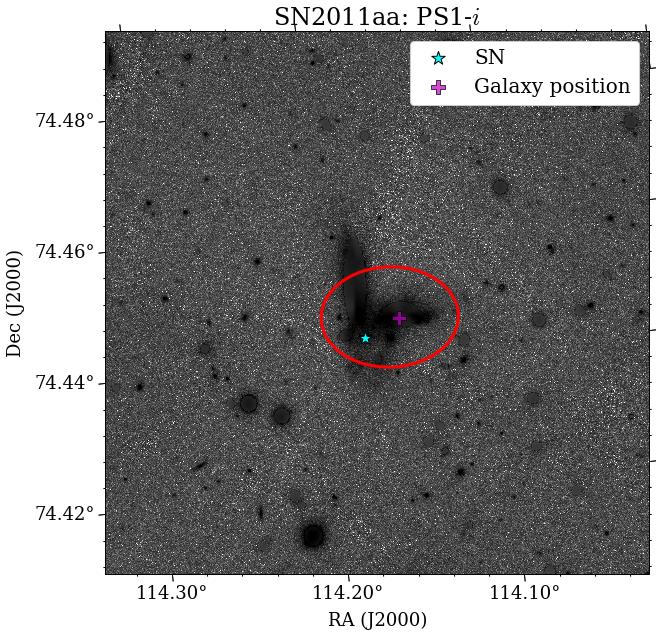}

\caption{\textit{Left:} The cross-cut profile from the {\it Swift} 2-D UVOT grism image for SN~2020hvf, SN~2012dn, SN~2009dc, and SN~2011aa (from top to bottom). The cross-cut profile is obtained by integrating the photon counts perpendicular to the SN trace. The apertures used for extracting the SN and background spectra are shown in blue and green, respectively. \textit{Right:} The PS1 $i$-band image of the 03fg-like SN host galaxies, with the cyan star and the ellipse indicating the SN location and host galaxy, respectively. The purple cross marks the coordinates of the host galaxy in the image.}
\label{crosscut}
\end{figure*}
\clearpage

\section{\texttt{SNooPy} Light Curve fitting}\label{snpy:app}

\begin{figure*}
\centering
\includegraphics[width=0.49\columnwidth]{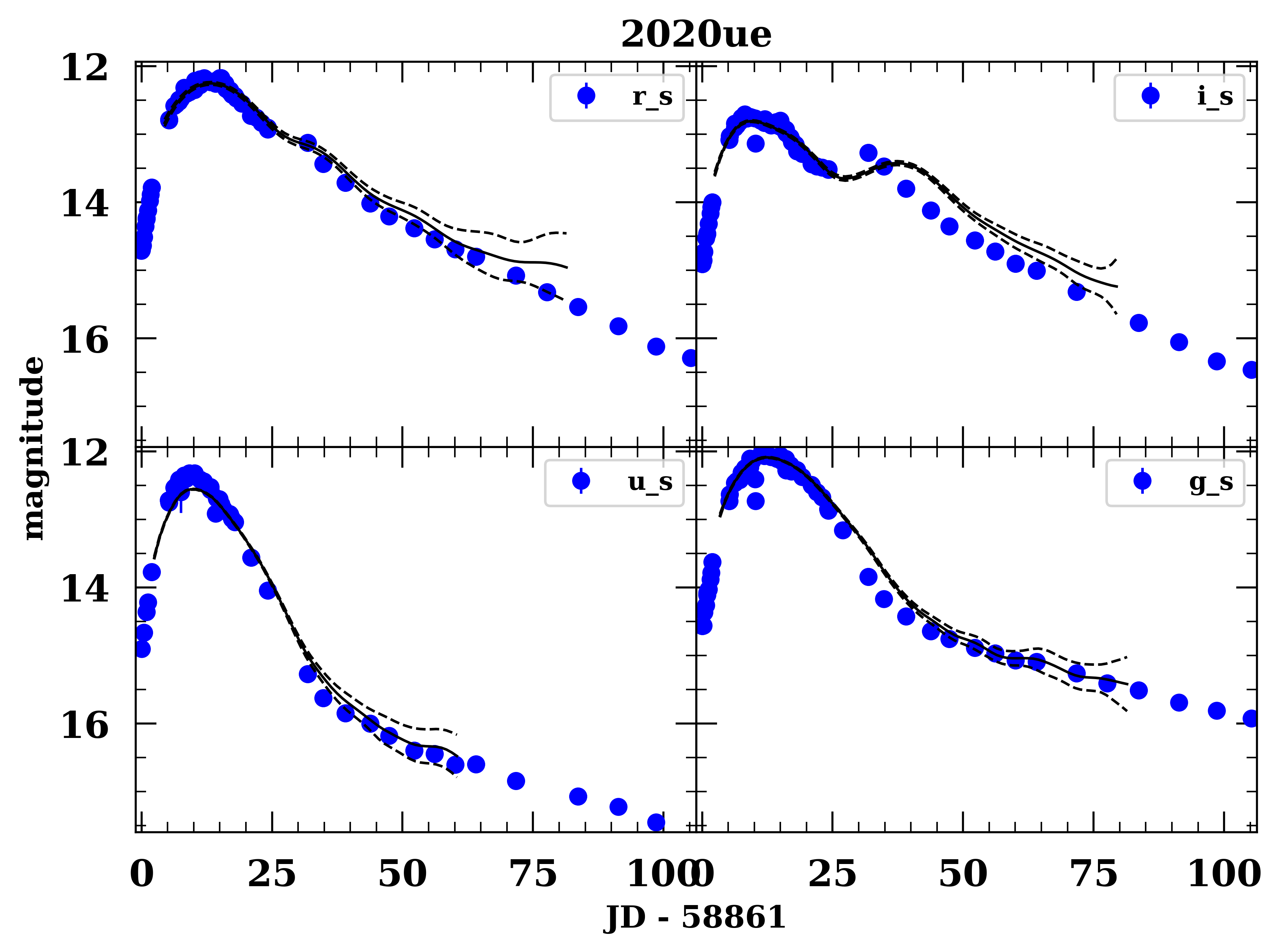}
\includegraphics[width=0.49\columnwidth]{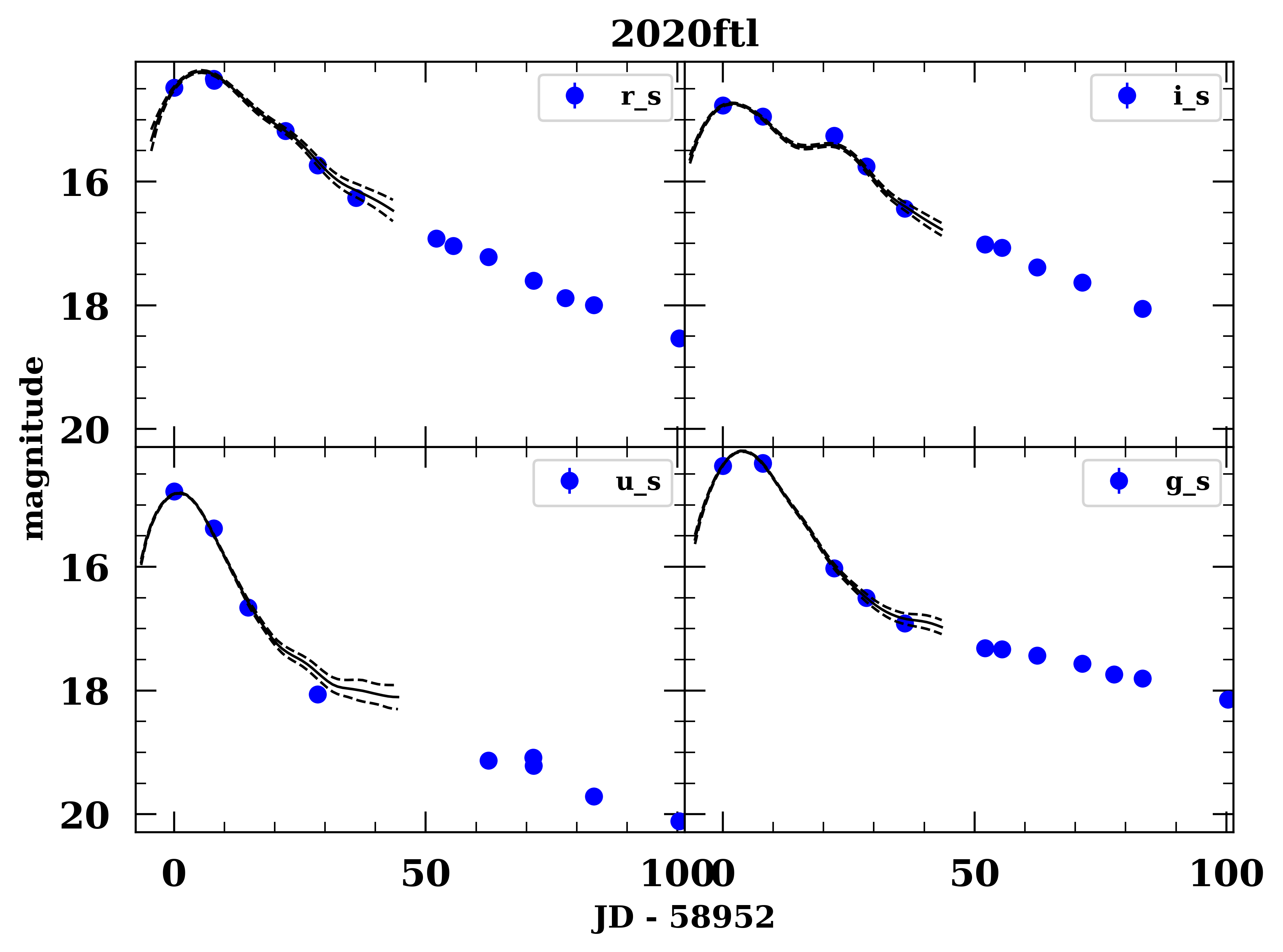}\\
\includegraphics[width=0.49\columnwidth]{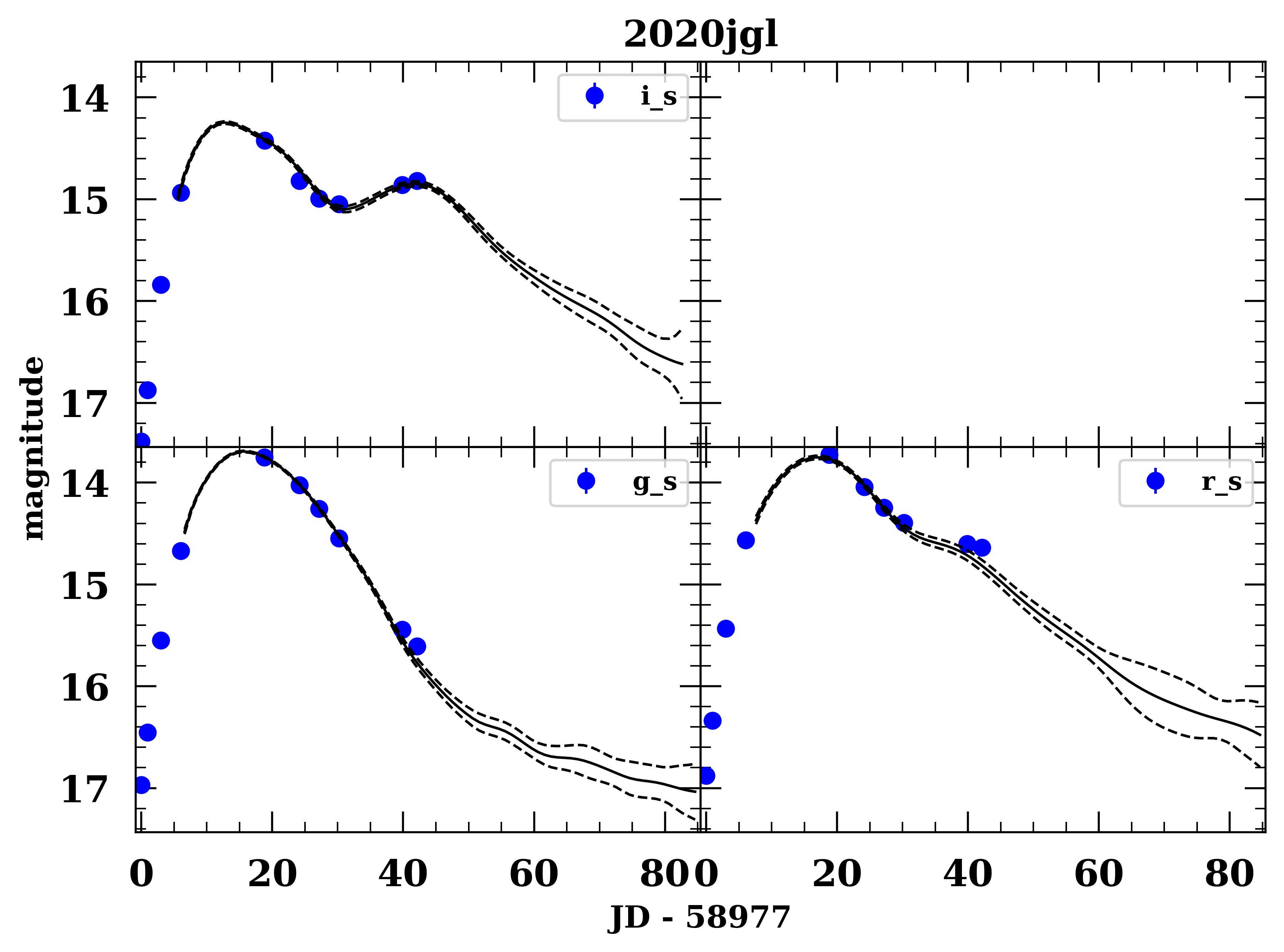}
\includegraphics[width=0.49\columnwidth]{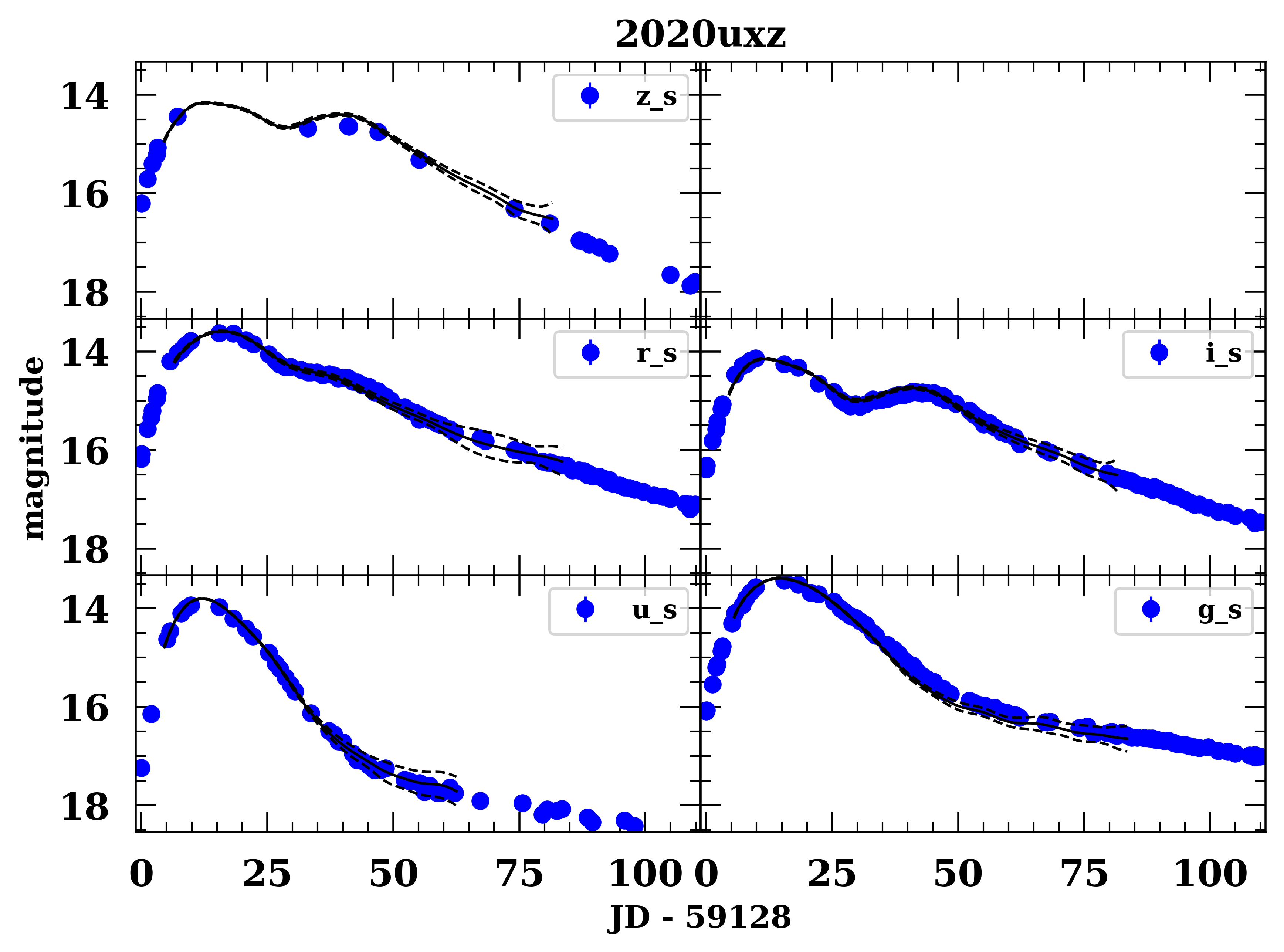}
\caption{\texttt{SNooPy} light-curve fitting for SN~2020ue (upper left), SN~2020ftl (upper right), SN~2020jgl (lower left), and SN~2020uxz (lower right). The \texttt{EBV\_model} method is adopted in the fitting. The observations and best fit from \texttt{SNooPy} are shown in blue solid circles and black curves, respectively. The $u_{s}$, $g_{r}$, $r_{s}$, $i_{r}$ and $z_{s}$ observations were obtained from the Lulin and Las Cumbres Observatory (LCO) one-meter telescopes.}
\label{snoopy:fits}
\end{figure*}

\bsp	
\label{lastpage}
\end{document}